\documentclass[10pt,a4paper,reqno]{amsart}
\usepackage{acronym}
\usepackage{amsfonts}
\usepackage{amsmath}
\usepackage{amssymb}
\usepackage{amsthm}
\usepackage{bm}
\usepackage{braket}
\usepackage{cite}
\usepackage{color}
\usepackage{geometry}
\usepackage{graphicx}
\usepackage{hyperref}
\usepackage{mathrsfs}
\usepackage{mathtools}
\usepackage{setspace}
\usepackage{stmaryrd}
\usepackage{subfigure}
\usepackage{tikz}

\usepackage{epsfig}
\usepackage{setspace}
\usepackage{booktabs}
\usepackage{threeparttable}
\usepackage{diagbox}
\usepackage{epstopdf}

\newgeometry{top=2cm,bottom=2cm,outer=1.5cm,inner=1.5cm}
\newcommand{\bse}{\begin{subequations}}
\newcommand{\ese}{\end{subequations}}

\usepackage{amssymb}
\usepackage[]{amsmath}
\usepackage{graphics}
\usepackage{graphicx}
\usepackage{amssymb}
\usepackage[]{amsmath}
\usepackage{amsthm}
\usepackage{setspace}
\usepackage{epsfig}

\usepackage{color}
\usepackage{pifont,bm}
\usepackage{mathrsfs}
\usepackage{booktabs}
\usepackage{diagbox}
\usepackage{makecell}

\numberwithin{equation}{section}

\title[Data-driven forward-inverse problems for the variable coefficients Hirota equation using deep learning method]
{Data-driven forward-inverse problems for the variable coefficients Hirota equation using deep learning method}
\author{Huijuan Zhou}
\address[HZ]{School of Mathematical Sciences, Shanghai Key Laboratory of Pure Mathematics and Mathematical Practice, and Shanghai Key Laboratory of Trustworthy Computing \\
East China Normal University \\ Shanghai 200241 \\ People's Republic of China}
\author{Juncai Pu}
\address[HZ]{School of Mathematical Sciences, Shanghai Key Laboratory of Pure Mathematics and Mathematical Practice, and Shanghai Key Laboratory of Trustworthy Computing \\
East China Normal University \\ Shanghai 200241 \\ People's Republic of China}
\author{Yong Chen}
\address[YC]{School of Mathematical Sciences, Shanghai Key Laboratory of Pure Mathematics and Mathematical Practice, and Shanghai Key Laboratory of Trustworthy Computing \\
East China Normal University \\ Shanghai 200241 \\ People's Republic of China}
\address[YC]{College of Mathematics and Systems Science \\ Shandong University of Science and Technology \\ Qingdao 266590 \\ People's Republic of China}
\address[YC]{Department of Physics \\ Zhejiang Normal University \\ Jinhua 321004 \\ People's Republic of China}
\email{ychen@sei.ecnu.edu.cn}

\begin{document}

\par
\begin{abstract}
Data-driven forward-inverse problems for the variable coefficients Hirota (VC-Hirota) equation are discussed in this paper. An improved physics-informed neural networks (IPINN) algorithm is used to recover the data-driven solitons, high-order soliton, as well as the data-driven parameters discovery for the VC-Hirota equation. We propose a PINN algorithm with sub-neural networks to learn the data-driven functions discovery of VC-Hirota equation with unknown functions under noise of different intensity. Numerical results are shown to demonstrate the facts: (i) data-driven soliton solutions and parameters discovery of the VC-Hirota equation are successfully learned by adjusting the network layers, neurons, the original training data, spatiotemporal regions and other parameters of the IPINN algorithm; (ii) the data-driven functions discovery of VC-Hirota equation can be trained stably and accurately via the PINN algorithm with sub-neural networks. The results achieved in this work verify that the forward-inverse problems including the data-driven function discovery of the variable coefficients equation can be solved based on deep learning method.
\end{abstract}

\maketitle

\section{Introduction}

As an important branch of nonlinear systems, solitons have been widely studied in many fields, such as plasma physics, fluid mechanics, optical fiber communication, condensed matter physics and so on \cite{zabusky-1965,ha1973,Gedalin-1997,Palacios-1999,Khaykovich-2002,Grelu-2012}. In the field of optical fiber communication, it is well known that a modified NLS equation, which is called Hirota equation, can be used to describe the propagation of subpicosecond or femtosecond optical pulse in fibers thanks to it taking into account higher-order dispersion and time-delay corrections to the cubic nonlinearity. Though there are extensive studies about Hirota equation and some types of exact soliton solutions in optical have been obtained, it is worth noting that these investigations of optical solitons have revolved mainly around homogeneous fibers. However, considering long-distance communication and manufacturing problems in realistic fiber transmission lines, the inhomogeneous variable coefficient model needs to be considered.   

Taking into account the higher-order effects influenced by spatial
variations of the fiber parameters, the inhomogeneous variable coefficient Hirota (VC-Hirota) equation which can be used to describe the certain ultrashort optical pulses propagating in a nonlinear inhomogeneous fiber is as follows \cite{kodama-1985}. 
\begin{equation}\label{VC-Hirotaeq}
iq_{z}+\alpha_{1}(z) q_{tt}-\frac{1}{3 \delta} i \alpha_{1}(z) q_{t t t}+\delta\alpha_{2}(z) q|q|^{2}-i\alpha_{2}(z)|q|^{2} q_{t}-i\alpha_{3}(z)q=0,
\end{equation}
where $\alpha_{3}(z)=\frac{\alpha_{1,z} \alpha_{2}-\alpha_{1} \alpha_{2, z}}{2 \alpha_{1} \alpha_{2}}$, $q=q(t,z)$ is complex-valued solution about the space $t$ and time $z$, $\delta$ is a real number, $\alpha_{1}(z)$ and $\alpha_{2}(z)$ are dispersion and nonlinear effects, respectively.
In order to understand the nonlinear phenomena of optical pulse propagation in inhomogeneous fiber media, it is necessary to analyze the analytical and numerical solutions of the VC-Hirota model. Due to the important application of the VC-Hirota model, there has been much research on this model, such as the exact bright and dark solitary wave solutions near the zero dispersion point is derived in \cite{Papaioannou-1996}. Three combined solitary wave solutions are given in the same expression, and the properties of bright and dark solitary waves are described respectively. Furthermore, the features of the solutions are analyzed, and numerically discuss the stability of these solitary waves under slight violations of the parameter conditions and finite initial perturbations \cite{Yang-2005}. It has also been extensively studied by many other authors and some types of localized waves solution have been obtained \cite{dai-2006-jpa,P-SAM-2010,tao-JNMP-2013,
Rajan-2015,Gao-2017,Yang-RC-2021}. Recently, we derive the exact form of $N$-soliton and high-order soliton solutions for the VC-Hirota equation by utilizing the Riemann-Hilbert approach \cite{zhj-arxiv-2022}.

In the past few decades, many methods and techniques have been presented to solve the nonlinear evolution equations, including Hirota bilinear method, Darboux transformation, Riemann-Hilbert method, physics-informed neural networks (PINN) deep learning method, etc. In scientific computing, the neural network (NN) method provides an ideal representation for the solution of differential equations due to its universal approximation properties. Recently, a PINN method which is controlled by mathematical physical systems based on the multi-layer NNs has been proposed and proved to be particularly suitable for dealing with both the forward problems and highly ill-posed inverse problems by obtaining the approximate solutions of governing equations and discovering parameters involved in the governing equation are inferred from the training data. Numerical results show that the PINN architecture is able to obtain remarkably accurate solutions with remarkably little data \cite{Raissi2019}. At the same time, this method also provides a better physical explanation for predicted solutions because of the underlying physical constraints, which are usually explicitly described by differential equations. Later on, using the PINN method to obtain data-driven solutions, parameters discovery and reveal the dynamic behavior of nonlinear partial differential equations with physical constraints has attracted extensive attention and raised a hot wave of research. Recently, PINN has played an important role in many physical applications \cite{KarniadakisGE2021}. Afterwards, global adaptive activation functions and locally adaptive activation functions are proposed to approximate smooth and discontinuous functions as well as solutions of linear and nonlinear partial differential equations by introducing a scalable parameters in the activation function and adding a slope recovery term based on activation slope to the loss function of locally adaptive activation functions. It demonstrates the locally adaptive activation functions further improve the training speed, performance and speed up the training process of NNs. \cite{Jagtap2020,JagtapA2020}. 

Due to the good properties of integrable systems, it is possible to solve the data-drive forward and inverse problems for abundant classical integrable nonlinear evolution equations via the PINN method. Since Chen's team first proposed the concept of integrable deep learning in 2020, a large number related literature has been published. 
In particular, Li and Chen construct abundant numerical solutions of second-order and third-order nonlinear integrable equations with different initial and boundary conditions by deep learning method based on the PINN model\cite{Li-2020-1,Li-2020-2}.
Pu, Peng et al. recover the solitons, breathers, rogue wave solutions and  rogue wave on the periodic background  of the nonlinear partial differential equation with the aid of the PINN model \cite{Pu2021, Peng2021}. 
Miao and Chen use the PINN method to high-dimensional system to solve the ($N$+1)-dimensional initial boundary value problem with 2$N$+1 hyperplane boundaries \cite{Miao2021}.
Lin and Chen devise a two-stage PINN method which is tailored to the nature of equations by introducing features of physical systems into NNs based on conserved quantities\cite{Lin2021}.  In addition, many important work on data-driven solutions and parameter discovery of different nonlinear systems have been doned by other scholars \cite{Fang2021,Wang2021,WuGZ2021,Ling-PLA2022,LB-ND-2022,Zhou-pla-2021}. 

In the previous literature, many soliton solutions of constant coefficients integrable systems have been accurately simulated numerically. The soliton solutions of variable coefficients integrable systems studied in this paper are quite different from the soliton solutions of constant coefficients equations. From the perspective of the solution dynamics, the center trajectory equation of the one-soliton solution of the constant coefficient equation is often a straight line, while the one-soliton solution of the variable coefficient equation presents a richer and more complex shape. For example, in this paper, we learned the case of the center trajectory equations of the one-solitons are parabola, "S-shape" and cosine wave type. For the two-soliton solution of variable coefficients equation, it can take on a more complex shape. To our knowledge, there is rare study about the data-driven soliton solutions, parameter and function discovery for the variable coefficient equation by the IPINN method. To fill this gap, the main purpose of this paper is to recover various solitons, high-order soliton, as well as parameters and functions discovery for the variable coefficient equation when only know the initial-boundary value conditions. It is note that we  use PINN algorithm with sub-neural networks to achieve data-driven function discovery of variable coefficient equations. 

 The rest of this paper is built up as follows. In Section 2, the IPINN method for the VC-Hirota equation is introduced. To provide a direct and precise description of the IPINN method, the algorithm flow schematic and algorithm steps are given. Section 3 provides the data-driven multi-soliton and high-order soliton solutions of the VC-Hirota equation using the IPINN approach, and related plots and dynamic analyses are revealed in detail. Section 4 presents experimental results of learning data-driven parameter discovery of VC-Hirota system. Furthermore, we learn the unknown function of the VC-Hirota via a PINN algorithm with sub-neural networks successfully.  The conclusions and discussions are given in Section 5.

\section{The improved physics-informed neural networks method for the variable coefficient Hirota equation}

An improved PINN (IPINN) approach with neuron-wise locally adaptive activation function
was presented to derive  data-driven localized waves and learn unknown parameters of integrable systems in complex space, and numerical results demonstrated the improved approach has faster convergence and better simulation effect than the classical PINN method \cite{PuJ2021,pu-manokov}. In the following, we will introduce the IPINN method in detail and extend it to the application of variable coefficient Hirota equations.

Considering the (1+1)-dimensional nonlinear time-dependent systems with unknown constant parameter $\delta$ and variable coefficients $\alpha_{1}(z),\alpha_{2}(z)$ and $\alpha_{3}(z)$ in complex space as below.
\begin{align}\label{E2}
&q_{z}+\mathcal{N}[q;\delta,\alpha_{1}(z),\alpha_{2}(z),\alpha_{3}(z)]=0,
\end{align}
where $\mathcal{N}[\cdot;\delta,\alpha_{1}(z),\alpha_{2}(z),\alpha_{3}(z)]$ is nonlinear differential operators in space. In order to simplify the structure of the complex-valued solutions $q(t,z)$ in Eq. \eqref{E2}, we decompose $q(t,z)$ into $q(t,z)=u(t,z)+\mathrm{i}v(t,z)$, where $u(t,z)$ and $v(t,z)$ are real-valued functions.  Then substituting $q(t,z)=u(t,z)+\mathrm{i}v(t,z)$ into Eq. \eqref{E2}, and letting the real and imaginary parts be equal to $0$, we have
\begin{align}\label{E3}
\begin{split}
&u_{z}+\mathcal{N}_u[u,v;\delta,\alpha_{1}(z),\alpha_{2}(z),\alpha_{3}(z)]=0,\\
&v_{z}+\mathcal{N}_v[u,v;\delta,\alpha_{1}(z),\alpha_{2}(z),\alpha_{3}(z)]=0,\\
\end{split}
\end{align}
where the $\mathcal{N}_u$ and $\mathcal{N}_v$ are nonlinear differential operators in space. Defining the physics-informed part $f_u(x,t)$ and $f_v(x,t)$ as
\begin{align}\label{E4}
\begin{split}
&f_u:=u_{z}+\mathcal{N}_u[u,v;\delta,\alpha_{1}(z),\alpha_{2}(z),\alpha_{3}(z)],\\
&f_v:=v_{z}+\mathcal{N}_v[u,v;\delta,\alpha_{1}(z),\alpha_{2}(z),\alpha_{3}(z)],
\end{split}
\end{align}
which play a role of regularization.

The IPINN of depth $D$ corresponding to the NN with an input layer, $D-1$ hidden-layers and an output layer. $N_d$ presents the number of neurons  in the $d$th hidden-layer, and each hidden-layer of the IPINN receives an output $\textbf{x}^{d-1}\in\mathbb{R}^{N_{d-1}}$ from the previous layer. Denote an affine transformation as follows:
\begin{align}\label{E-buchong}
\mathcal{L}_d(\textbf{x}^{d-1})\triangleq\textbf{W}^d\textbf{x}^{d-1}+\textbf{b}^d,
\end{align}
where the network weights $\textbf{W}^{d}\in\mathbb{R}^{N_d\times N_{d-1}}$ and bias term $\textbf{b}^d\in\mathbb{R}^{N_d}$ are associated with the $d$th layer. Then define neuron-wise locally adaptive activation function as
\begin{align}\nonumber
\sigma\left(na^d_i\left(\mathcal{L}_d\left(\textbf{x}^{d-1}\right)\right)_i\right),\,  n>1,\cdots d=1,2,\cdots,D-1,\,i=1,2,\cdots,N_d,
\end{align}
where $\sigma$ is the activation function, $n$ is a scaling factor and $\{a^d_i\}$ is additional $\sum\limits_{d=1}^{D-1}N_d$ parameter to be optimized. When $n$ greater than or equal to critical scaling factor $n_c$ in each problem set, the optimization algorithm will become sensitive. The neuron activation function acts as a vector activation function in each hidden layer, and each neuron has its own activation function slope.

The IPINN method with neuron-wise locally adaptive activation function can be expressed as:
\begin{align}\label{E5}
q(\textbf{x};\bar{\Theta})=\big(\left(\mathcal{L}_D\right)_{i'}\circ\sigma\circ na^{D-1}_{i}\left(\mathcal{L}_{D-1}\right)_{i}\circ\cdots\circ\sigma\circ na^1_i\left(\mathcal{L}_1\right)_i\big)(\textbf{x}),\,i'=1,2,
\end{align}
where $\textbf{x}$ represent the two inputs  and $q(\textbf{x};\bar{\Theta})$ represent the two outputs in the IPINN. The trainable parameters  set $\bar{\Theta}\in\bar{\mathcal{P}}$ consists of $\big\{\textbf{W}^d,\textbf{b}^d\big\}_{d=1}^{D}$ and $\big\{a_i^d\big\}_{d=1}^{D-1},\forall i=1,2,\cdots,N_d$, where the $\bar{\mathcal{P}}$ is a parameter space. In order to minimize the loss function below certain tolerance $\varepsilon$ until a prescribed maximum number of iterations,  seek a optimal values of weights $\textbf{W}$, biases $\textbf{b}$ and scalable parameter $a^d_i$ is necessary. Here, we initialize the scalable parameters in the case that $na_i^d=1,\forall n\geqslant1$ (we fixed n=10 in this paper).

Define the loss function as follows:
\begin{align}\label{E6}
\mathscr{L}(\bar{\Theta})=Loss=Loss_{u}+Loss_{v}+Loss_{f_u}+Loss_{f_v}+Loss_a,
\end{align}
where $Loss_{u}, Loss_{v}, Loss_{f_u}$, $Loss_{f_v}$ and $Loss_a$ are defined as below:
\begin{align}\label{E7}
\begin{split}
Loss_{u}&=\frac{1}{N_q}\sum^{N_q}_{j=1}\big|\hat{u}(t^j,z^j)-u^j\big|^{2},
\\
Loss_{v}&=\frac{1}{N_q}\sum^{N_q}_{j=1}\big|\hat{v}(t^j,z^j)-v^j\big|^{2},
\\
Loss_{f_u}&=\frac{1}{N_f}\sum^{N_f}_{l=1}\big|f_u(t_f^l,z_f^l)\big|^{2}, \\
Loss_{f_v}&=\frac{1}{N_f}\sum^{N_f}_{l=1}\big|f_v(t_f^l,z_f^l)\big|^{2},\\
Loss_a&=\frac{N_a}{\frac{1}{D-1}\sum\limits_{d=1}^{D-1}\mathrm{exp}\Bigg(\frac{\sum\limits_{i=1}^{N_d}a_i^d}{N_d}\Bigg)}.
\end{split}
\end{align}
$\{t^j,z^j,u^j,v^j\}^{N_q}_{j=1}$ denote the inputs data of initial-boundary value on Eqs. \eqref{E3} and \eqref{E4}. $\hat{u}(t^j,z^j)$ and $\hat{v}(t^j,z^j)$ represent the optimal training outputs data through the IPINN. Furthermore, $\{t_f^l,z_f^l\}^{N_{f}}_{l=1}$ represent the collocation points on networks $f_u(t,z)$ and $f_v(t,z)$. $N_a$ is the hyper-parameter for slope recovery term $Loss_a$, and we take $N_a=\frac{1}{100}$ for dominating the loss function to ensure that the final loss value is not too large in this paper. $Loss_{u}$ and $Loss_{v}$ correspond to the loss on the initial and boundary data. $Loss_{f_u}$ and $Loss_{f_v}$ penalize the collocation points which not satisfy the VC-Hirota equation. The $Loss_a$ changes the topology of $Loss$ function forces the NN to increase the activation slope value quickly, which ensures the non-vanishing gradient of the loss function and improves the convergence speed and network optimization ability. Therefore, the loss function is evaluated using the contribution from the NN part as well as the residual from the governing equation given by the physics-informed part. The resulting optimization algorithm will attempt to find the optimized parameters including the weights,  biases and additional coefficients in the activation to minimize the new loss function.
In addition, the $\mathbb{L}_2$ norm error is introduced to measure the training error better.  $\mathbb{L}_2$ norm error is defined as follows:
\begin{align}\nonumber
\mathrm{Error}=\frac{\sqrt{\sum\limits_{k=1}^{N}\big|q^{\mathrm{exact}}(\textbf{x}_{k})-q^{\mathrm{predict}}(\textbf{x}_{k};\bar{\Theta})\big|^{2}}}{\sqrt{\sum\limits_{k=1}^{N}\big|q^{\mathrm{exact}}(\textbf{x}_{k})\big|^{2}}},
\end{align}
where $q^{\mathrm{predict}}(\textbf{x}_{k};\bar{\Theta})$ represent the model training prediction solution and $q^{\mathrm{exact}}(\textbf{x}_{k})$ represent the exact analytical solution at point $\textbf{x}_{k}=(t_k,z_k)$.

The physics-informed parts of the IPINN for VC-Hirota equation \eqref{VC-Hirotaeq} can be defined as:
\begin{align}\label{E-pi}
\begin{split}
&f_u:=-v_z+\alpha_{1}(z)u_{tt}+\frac{1}{3\delta}\alpha_{1}(z)v_{ttt}+\delta\alpha_{2}(z)(u^{3}+uv^{2})+\alpha_{2}(z)(u^{2}v_{t}+v^{2}v_{t})+\alpha_{3}(z)v,\\
&f_v:=u_z+\alpha_{1}(z)v_{tt}-\frac{1}{3\delta}\alpha_{1}(z)u_{ttt}+\delta\alpha_{2}(z)(vu^{2}+v^{3})-\alpha_{2}(z)(u^{2}u_{t}+v^{2}u_{t})-\alpha_{3}(z)u.
\end{split}
\end{align}
The physics-informed parts of the IPINN for VC-Hirota equation \eqref{VC-Hirotaeq} is integrable due to the equation \eqref{VC-Hirotaeq} have the lax pair as follows:
\begin{equation}
 U=\left[ \begin {array}{cc} \lambda&\sqrt {{\frac {\delta\alpha_{2}(z) }{2\alpha_{1}(z)}}}q(z,t) \\ \noalign{\medskip}-q^{*}(z,t)\sqrt {{\frac {\delta\alpha_{2}(z) }{2\alpha_{1}(z) }}}&-\lambda\end {array} \right] ,
 \end{equation}
  \begin{equation}
V=\left(\begin{matrix}
A&B\\
C&-A
\end{matrix}
\right),
\end{equation} 
where
$$A=\frac{4\alpha_{1}(z)}{3\delta}\lambda^{3}+2i\alpha_{1}(z){\lambda}^{2}+\frac{\delta\alpha_{2}(z)|q(z,t)|^{2}}{3\delta}\lambda+
 \frac{\delta\alpha_{2}(z)(q^{*}(z,t)q_{t}(z,t)-q(z,t)q^{*}_{t}(z,t))}{6\delta}
+\frac{i\delta\alpha_{2}(z)q(z,t) q^{*}(z,t)}{2},$$
$$B= \sqrt{\frac{\delta\alpha_{2}(z)}{2\alpha_{1}(z)}}(
\frac{4\alpha_{1}(z)q(z,t)}{3\delta}{\lambda}^{2}+2({\frac {\alpha_{1}(z)q_{t}(z,t)}{3\delta}}+i\alpha_{1}(z)q(z,t))\lambda +i\alpha_{1}(z)q_{t}(z,t) +{\frac {
\alpha_{1}(z)}{3\delta}(q_{tt}(z,t) +{\frac {\delta\alpha_{2}(z) 
|q(z,t)|^{2}q(z,t)}
{\alpha_{1}(z)}}) })  $$
and $$C=\sqrt{{\frac {\delta\alpha_{2}(z)}{2\alpha_{1}(z)}}}(-{\frac{4\alpha_{1}(z)q^{*}(z,t) }{3\delta}{\lambda}^{2}}+2({\frac {\alpha_{1}(z)
q^{*}_{t}(z,t) }{3\delta}}-i\alpha_{1}(z)q^{*}(z,t))\lambda 
+i\alpha_{1}(z)q^{*}_{t}(z,t)-\frac{\alpha_{1}(z)}{3\delta}(q^{*}_{tt}(z,t) +
\frac {\delta\alpha_{2}(z) |q(z,t)|^{2}q^{*}(z,t) }{\alpha_{1}(z)})).$$

\begin{figure}[htbp]
\centering
\begin{minipage}[t]{0.99\textwidth}
\centering
\includegraphics[height=9cm,width=15cm]{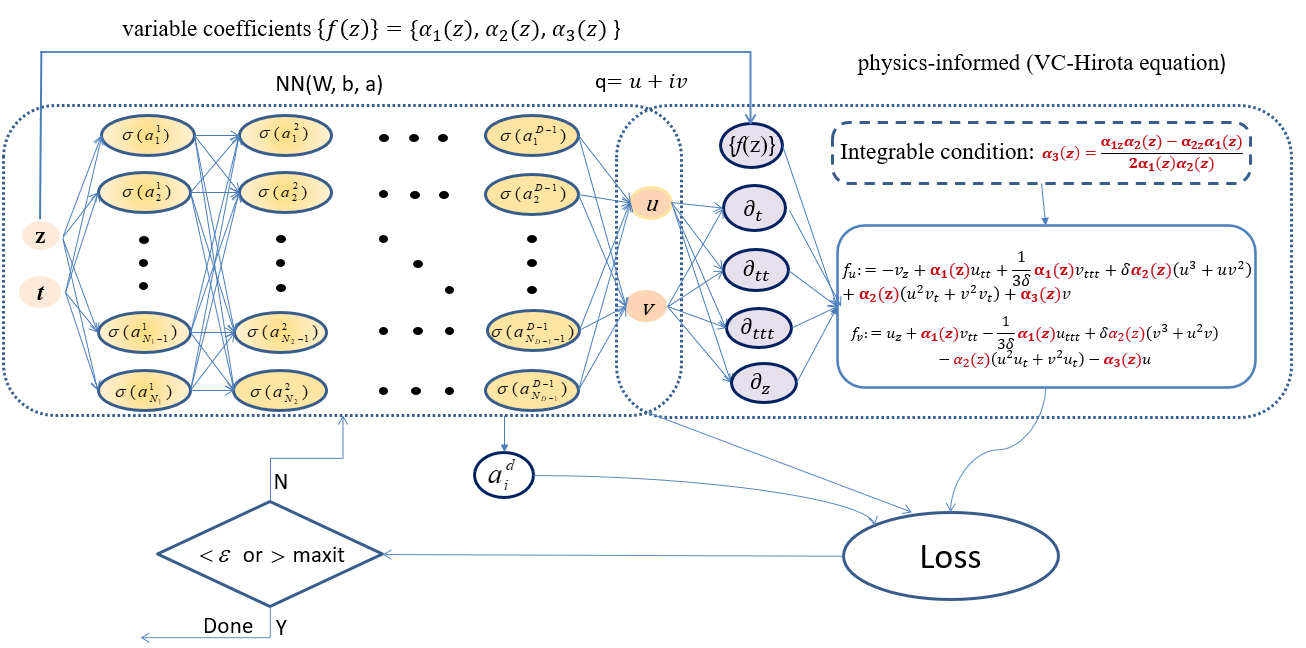}
\end{minipage}
\centering
\caption{(Color online) Schematic of IPINN for the VC-Hirota equation. The left NN is the universal approximation network while the right one induced by the governing equation is the physics-informed network. The two NNs share hyper-parameters and they both contribute to the loss function.}
\label{F1}
\end{figure}

In order to describe the IPINN algorithm for the VC-Hirota equation more intuitively, we give the flow chart of the algorithm in Fig. \ref{F1}. 
Since the complex function $q$ is decomposed into $u+iv$, one can see that the ``NN" part has two output functions $\{u,v\}$, and there are two nonlinear equation constraints in the physics-informed part. What need to note is that in the physics-informed part, we introduce the constant parameter $\delta$ and variable coefficients $\alpha_{1}(z),\alpha_{2}(z)$, $\alpha_{3}(z)$ into the governing equation \eqref{E4}. Note ${f(z)}$=${\alpha_{1}(z),\alpha_{2}(z), \alpha_{3}(z)}$ and $\alpha_{3}(z)=\frac{\alpha_{1,z} \alpha_{2}-\alpha_{1} \alpha_{2, z}}{2 \alpha_{1} \alpha_{2}}$ is a necessary constraint for the integrability of the equation \eqref{E4}. In addition, we also show the corresponding procedure steps of the IPINN algorithm for the VC-Hirota equation in 
Tab.\ref{Tab:bookRWCal}.

\begin{table}[htbp]
  \caption{IPINN algorithm of the VC-Hirota equation.}
  \label{Tab:bookRWCal}
  \centering
  \begin{tabular}{p{15cm}}
  \toprule[2pt]
  \quad \textbf{Step one}: Introduce constant parameter $\delta$ and variable coefficients $\alpha_{1}(z),\alpha_{2}(z)$, $\alpha_{3}(z)$ into the governing equation \eqref{E4} and take $\alpha_{3}(z)=\frac{\alpha_{1,z} \alpha_{2}-\alpha_{1} \alpha_{2, z}}{2 \alpha_{1} \alpha_{2}}$ insure Eq. \eqref{E4} completely integrable.\\
  \quad \textbf{Step two}: Specification of training set in computational domain:\\
  \quad \emph{Training data}: $\{t^j,z^j,u^j,v^j\}^{N_q}_{j=1}$, \emph{Residual training points}: $\{t_f^l,z_f^l\}^{N_{f}}_{l=1}.$\\
  \quad \textbf{Step three}: Construct NN $q(\textbf{x};\bar{\Theta})$ with random initialization of parameters $\bar{\Theta}$.\\
  \quad \textbf{Step four}: Construct the residual NN $\{f_u,f_v\}$ by substituting surrogate $q(\textbf{x};\bar{\Theta})$ into the governing equations using automatic differentiation and other arithmetic operations.\\
  \quad \textbf{Step five}: Specification of the loss function $\mathscr{L}(\bar{\Theta})$ that includes the slope recovery term.\\
  \quad \textbf{Step six}: Find the best parameters $\bar{\Theta}^*$ using a suitable optimization method for minimizing the loss function $\mathscr{L}(\bar{\Theta})$ as\\
  \qquad\qquad\qquad\qquad\qquad\qquad\qquad\quad $\bar{\Theta}^*=\mathop{\mathrm{arg\,min}}\limits_{\bar{\Theta}\in\bar{\mathcal{P}}}\mathscr{L}(\bar{\Theta})$.\\
  \bottomrule[2pt]
  \end{tabular}
\end{table}

Adam and L-BFGS algorithms are used to optimize all loss functions in the IPINN method. The Adam optimization algorithm is a variant of the traditional stochastic gradient descent algorithm, while the L-BFGS optimization algorithm is a full-batch gradient descent optimization algorithm based on the quasi-Newton method \cite{Kingma2014,Liu1989}. In particular, unless otherwise stated, the scalable parameters in the adaptive activation function are typically initialized as $n=10$, $a_i^d=0.1$. In addition, we chose feedforward NNs with Xavier initialization and hyperbolic tangent ($\tanh$) as activation functions. All the code in this article is based on Python 3.7 and Tensorflow 1.15. All the numerical experiments reported in this article were run on a DELL Precision 7920 Tower computer. The computer is equipped with a 2.10 GHz 8-core Xeon Silver 4110 processor, 64 GB memory and an 11 GB NVIDIA GeForce GTX 1080 Ti video card.

\section{Data-driven forward problems of the VC-Hirota equation}

In this section, we will focus on the data-driven forward for the VC-Hirota equation with Dirichlet boundary conditions and initial conditions by means of IPINN method. The VC-Hirota equation with initial–boundary value conditions is as follows.
\begin{align}\label{E1}
\begin{split}
\begin{cases}
iq_{z}+\alpha_{1}(z) q_{tt}-\frac{1}{3 \delta} i \alpha_{1}(z) q_{t t t}+\delta\alpha_{2}(z) q|q|^{2}-i\alpha_{2}(z)|q|^{2} q_{t}-i\alpha_{3}(z)q=0,\\
q(L_0,z)=q^{\mathrm{lb}}(z),\quad q(L_1,z)=q^{\mathrm{ub}}(z),\\
q(t,T_0)=q^0(t),\quad t\in[L_0,L_1],\quad z\in[T_0,T_1],\\
\end{cases}
\end{split}
\end{align}
where $``\mathrm{i}"$ is an imaginary number, the subscripts denote the partial derivatives of the complex fields $q(t,z)$ with respect to the space $t$ and time $z$, while the $L_0$ and $L_1$ represent the lower and upper boundaries of $t$ respectively. Similarly, $T_0$ and $T_1$ represent the initial and final times of $z$ respectively. Moreover, the $q^0(t)$ represents initial value of the $q(t,z)$ at $z=T_0$, the $q^{\mathrm{lb}}(z)$ and $q^{\mathrm{ub}}(z)$ are the lower and upper boundaries of the $q(t,z)$ corresponding to $t=L_0$ and $t=L_1$ respectively.

The $N$-solitons of the VC-Hirota have been derived by the Riemann-Hilbert method in \cite{zhj-arxiv-2022}, which can be expressed as follows:
 \begin{equation}\label{VC-Hirotas}
 q=-2i\frac{|F|}{|M|}f(z)e^{ig(t, z)},
\end{equation}
where $F$ is the following $(N +1)\times(N +1)$ matrix
\begin{equation}
\left(\begin{matrix}
0&e^{-\theta_{1}^{*}}&...&e^{-\theta_{N}^{*}}\\
c_{1}e^{\theta_{1}}&M_{11}&...&M_{N1}\\
.&.&.&.\\
.&.&.&.\\
.&.&.&.\\
c_{N}e^{\theta_{N}}&M_{1N}&...&M_{NN}
\end{matrix}
\right),
\end{equation}
the elements of the $N \times N$ matrix $M$ are given by
\begin{equation}\notag
M_{jk}=\frac{e^{-(\theta_{k}+\theta_{j}^{*})}
+c_{j}^{*}c_{k}e^{\theta_{k}+\theta_{j}^{*}}}{\zeta_{j}^{*}-\zeta_{k}},
\end{equation}
\begin{equation}\notag
\theta_{k}=-i\zeta_{k}T-(4i\beta \zeta_{k}^{3}+2i\gamma \zeta_{k}^{2})Z,
\end{equation}
\begin{equation}\notag
\begin{gathered}
f(z)=\sqrt{\frac{\alpha_{1}(z)}{\alpha_{2}(z)}},\enspace 
Z=-\frac{\sqrt{2 \delta}}{12 \beta} \int \alpha_{1}(z) d z, \enspace
T=\frac{\sqrt{2 \delta}}{2}( t-(\frac{\gamma^{2}
}{36 \beta^{2}}- \delta)\int \alpha_{1}(z) dz), \\ and \enspace
g(t, z)=-\frac{6 \beta \delta+\gamma \sqrt{2 \delta}}{6 \beta} t-\frac{216 \delta^{2} \beta^{3}+54 \gamma \beta^{2} \delta \sqrt{2 \delta}-\gamma^{3} \sqrt{2 \delta}}{324 \beta^{3}} \int \alpha_{1}(z) dz,
\end{gathered}
\end{equation}
and the symbol $*$ represents complex conjugate, $\beta$ and $\gamma$ are real numbers.

Note that we can obtain the $N$-solitons from formula Eq. \eqref{VC-Hirotas}. As the number $N$ gets larger, the form and dynamic behavior of $N$-solitons are more complex, with more parameters and stronger adjustability. In there, we only consider the case of $N=1$ and $2$ as examples to solve the data-driven one-solitons and two-solitons of the VC-Hirota equation.

\subsection{Data-driven one-solitons of the VC-Hirota equation}

In order to get the one-soliton solution, we set  $N=c_{1}=1$ and $\zeta_{1}=\xi+i\eta$ in the above formulae \eqref{VC-Hirotas}, the one-soliton solution of VC-Hirota equation is shown as

\begin{equation}\label{q1}
q_{1}=2\sqrt{\frac{\alpha_{1}(z)}{\alpha_{2}(z)}}\eta e^{\frac {A}{324\,{\beta}^{3}}}sech(B),
\end{equation}
\begin{equation}\notag
\begin{split}
A&=(\xi\,\beta+\frac{\gamma}{6})(216\,i\sqrt {2\delta}( {\xi}^{2}{\beta}^{2}-3\,{\eta}^{2} {\beta}^{2}+\frac{\gamma\,\beta\,\xi}{3}+\frac{\,{\gamma}^{2}}{36})-324\,i\sqrt {2}
  {\beta}^{2}{\delta}^{\frac{3}{2}}-216\,i{\beta}^{3}{\delta}^{2} ) \int \!\alpha_{1}(z) \,{\rm d}z\\&-324\,i{\beta}^{2}
( \sqrt{2\delta}( \xi\,\beta+\frac{\gamma}{6})
+\beta\,\delta)t,\\
B&=\eta\sqrt {2\delta} ((\frac{2}{3}\,{\eta
}^{2}-2\,{\xi}^{2}+\delta-\frac{2\gamma\xi}{3\beta}-\frac{{\gamma
}^{2}}{18{\beta}^{2}}) \int \!\alpha_{1}(z) \,{\rm d}z+t).
\end{split}
\end{equation}

For simplicity, we take  $\gamma=0$ in the following unless otherwise stated.

$\bullet$ \textbf{Data-driven one-soliton solution $\rm\uppercase\expandafter{\romannumeral1}$}

Taking $\alpha_{1}(z)=\alpha_{2}(z)=z$ and $\xi$=$\eta$=1, then the center trajectory equation of the solution \eqref{q11} is $t=\frac{1}{2}(\frac{4}{3}-\delta)z^{2}$. It shows that the value of $\delta$ has a great influence on the propagation path of the solution. In order to adjust the curvature of the curve in order to get better training results, here we take $\delta=10$, then the explicit one-soliton solution is as follows: 
\begin{equation}\label{q11}
q_{11}(t,z)=\frac{4e^{\frac{\sqrt{5}}{3}((6-6i)t+(26-34i)z^{2})-\frac{100}{3}iz^{2}-10it}}{e^{\frac{4}{3}\sqrt{5}(13z^{2}+3t)}+1}.
\end{equation}

Then we will focus on the data-driven one-solitons with the initial conditions $q^0(t)$ and Dirichlet boundary conditions $q^{\mathrm{lb}}(z)$ and $q^{\mathrm{ub}}(z)$ of the VC-Hirota equation. Taking $[L_0,L_1]$= $[-3.0,1.0]$ and $[T_0,T_1]$=$[-1.0,1.0]$ in Eq. \eqref{E1}, then the corresponding initial condition and Dirichlet boundary condition are shown as follows:
\begin{align}\label{E15}
\begin{split}
&q^0(t)=q_{\mathrm{11}}(t,-1.0),\quad t\in[-3.0,1.0],
\end{split}
\end{align}
\begin{align}\label{E16}
q^{\mathrm{lb}}(z)=q_{\mathrm{11}}(-3.0,z),\quad q^{\mathrm{ub}}(z)=q_{\mathrm{11}}(1.0,z),\quad z\in[-1.0,1.0].
\end{align}

We employ the traditional finite difference scheme on even grids in Matlab to simulate one-solitons which contains the initial data \eqref{E15} and boundary data \eqref{E16} to acquire the original training datas. Specifically, we divide spatial region $[-3.0,1.0]$ and temporal region $[-1.0,1.0]$ into 1000 points, respectively. Accordingly, the one-solitons \eqref{q11} is discretized into 1000 snapshots. Using the Latin Hypercube Sampling method (LHS) \cite{Stein1987}, a smaller training dataset containing initial-boundary data by randomly extracting $N_q=1500$ from original dataset and $N_f = 20000$ collocation points is generated. After giving a dataset of initial and boundary points, the data-driven one-soliton solution $\rm\uppercase\expandafter{\romannumeral1}$ have been successfully learned by tuning all learnable parameters of the 9 layers IPINN with 40 neurons per layer and regulating the loss function, in which we utilize the 20000 steps Adam firstly and subsequently use 19108 steps L-BFGS optimizations for minimizing the loss function \eqref{E6}. 
The relative $\mathbb{L}_2$ errors of the IPINN is 1.168275$\rm e^{-2}$ with the train time of 10732.7923 seconds and  39108 times iterations.
  
\begin{figure}[htbp]
\centering
\includegraphics[height=7.5cm,width=15cm]{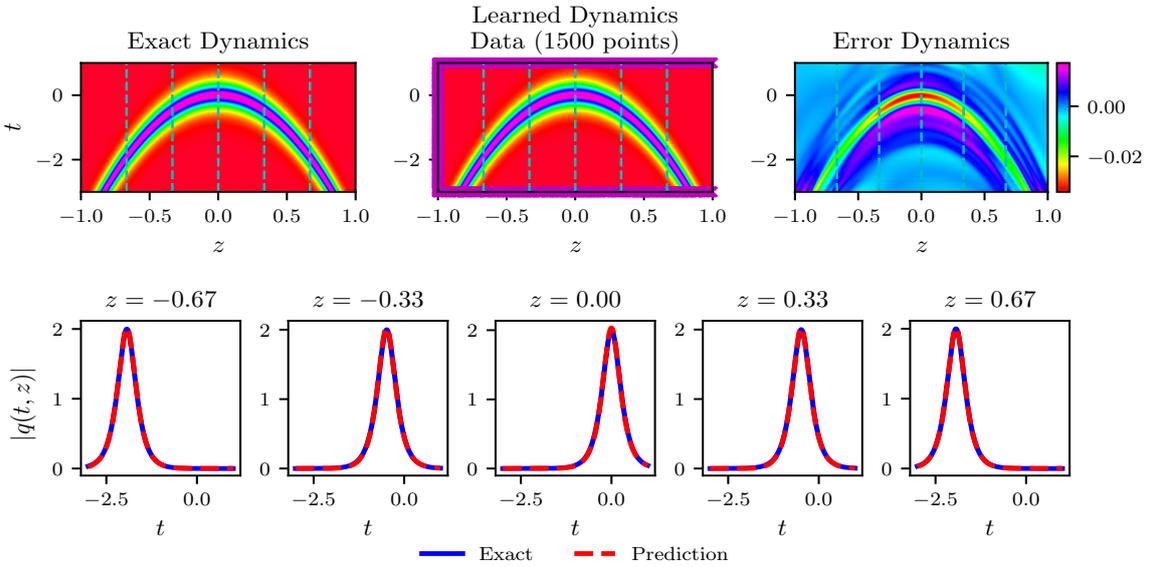}
\centering
\caption{(Color online) The density plots and sectional drawings for the one-soliton $q_{11}(t,z)$: The one-solitons $q_{11}(t,z)$ resulted from the IPINN with the randomly chosen initial and boundary points $N_q=1500$ which have been shown by using mediumorchid $``\times"$ in learned dynamics, and $N_f = 20000$ collocation points in the corresponding spatiotemporal region. The exact, learned and error dynamics density plots for the one-solitons $q_{11}(t,z)$ with five distinct training moments $t=-0.67, -0.33, 0.00, 0.33$ and $0.67$ (darkturquoise dashed lines), and the sectional drawings which contain the learned and explicit one-solitons $q_{11}(t,z)$ at the aforementioned five distinct moments.  }
\label{F2}
\end{figure}

\begin{figure}[htbp]
\centering
\subfigure[]{\label{F3}
\begin{minipage}[t]{0.3\textwidth}
\centering
\includegraphics[height=5cm,width=5cm]{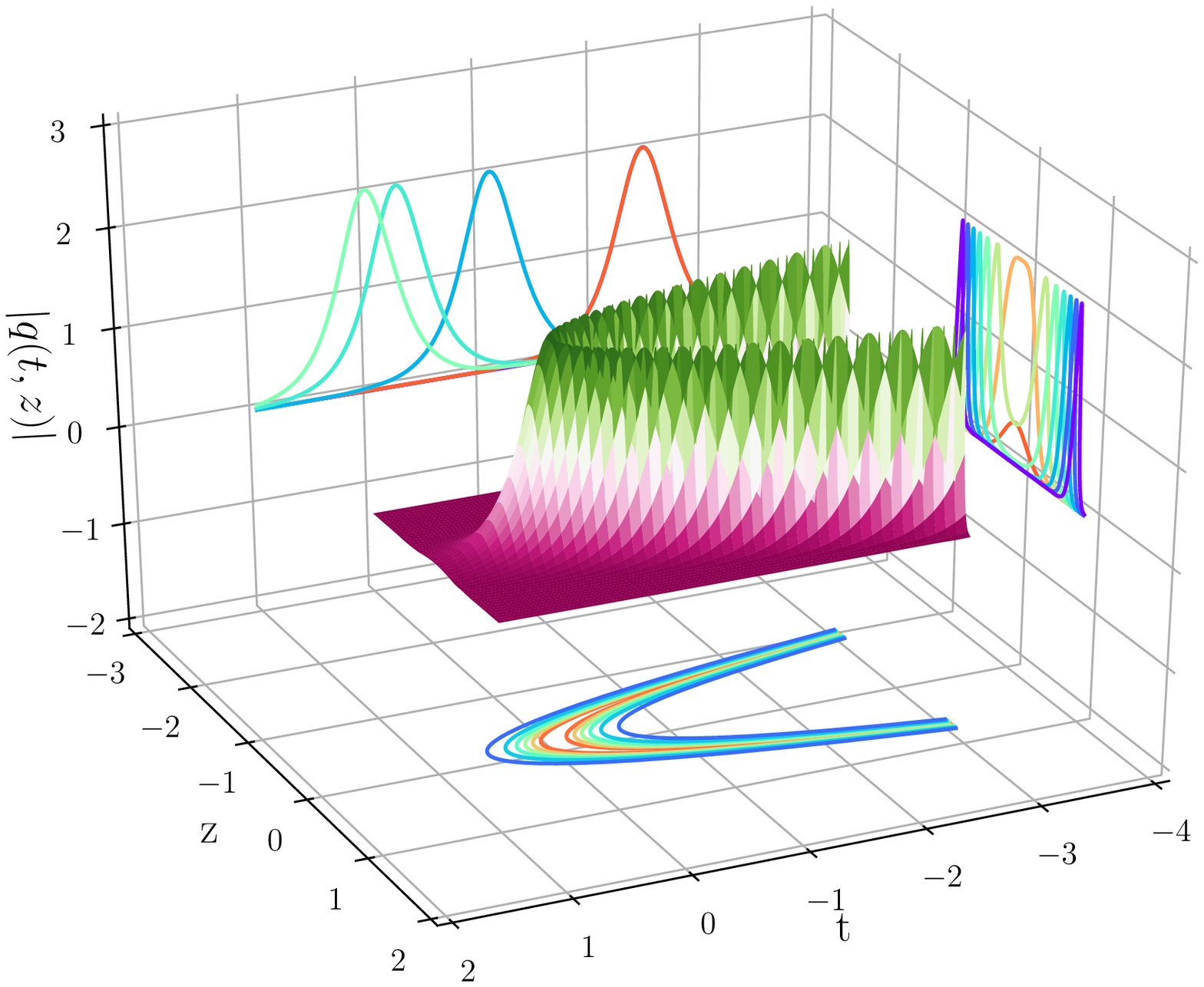}
\end{minipage}%
}%
\subfigure[]{\label{F41}
\begin{minipage}[t]{0.3\textwidth}
\centering
\includegraphics[height=4.5cm,width=4.5cm]{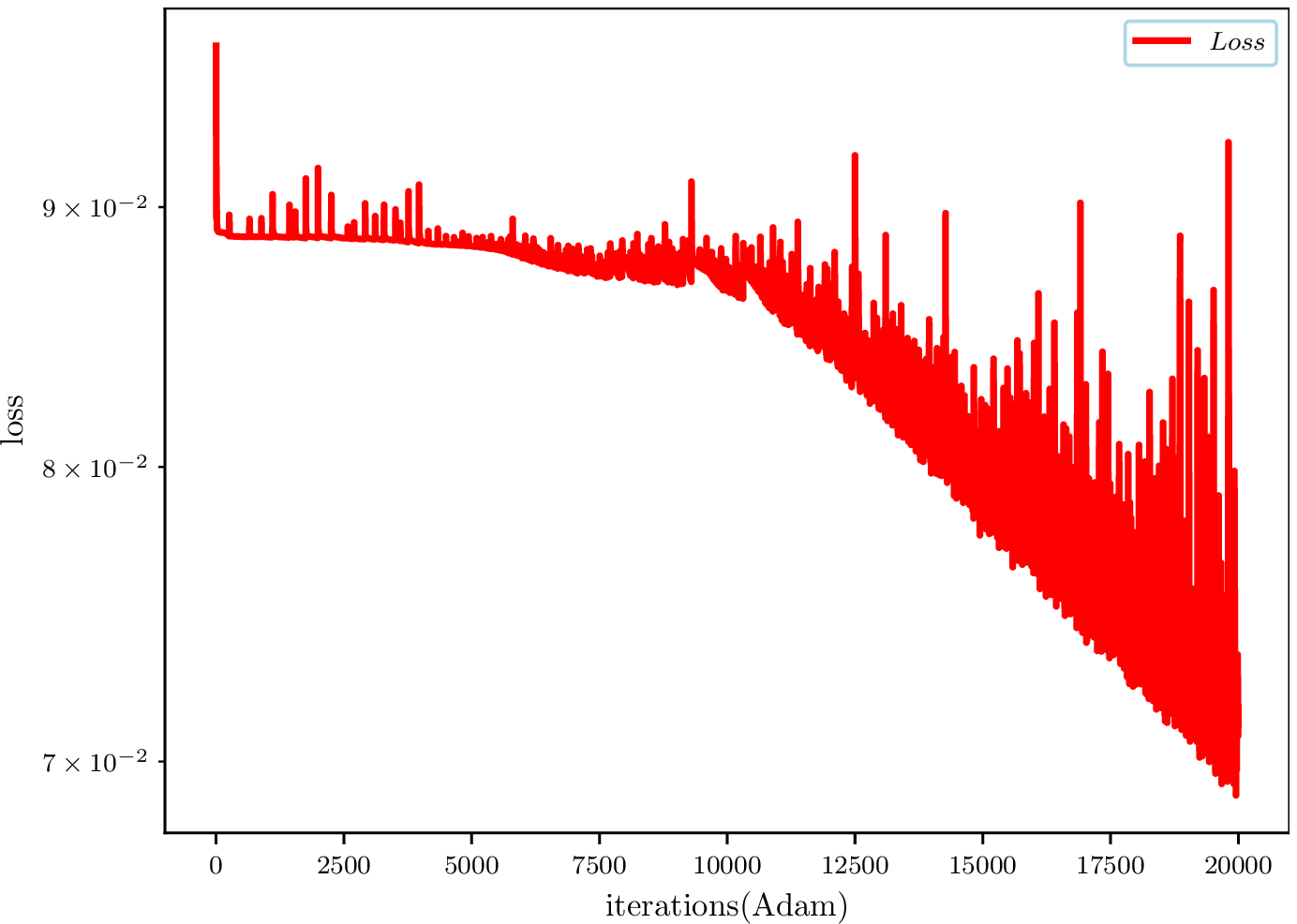}
\end{minipage}
}%
\subfigure[]{\label{F42}
\begin{minipage}[t]{0.3\textwidth}
\centering
\includegraphics[height=4.5cm,width=4.5cm]{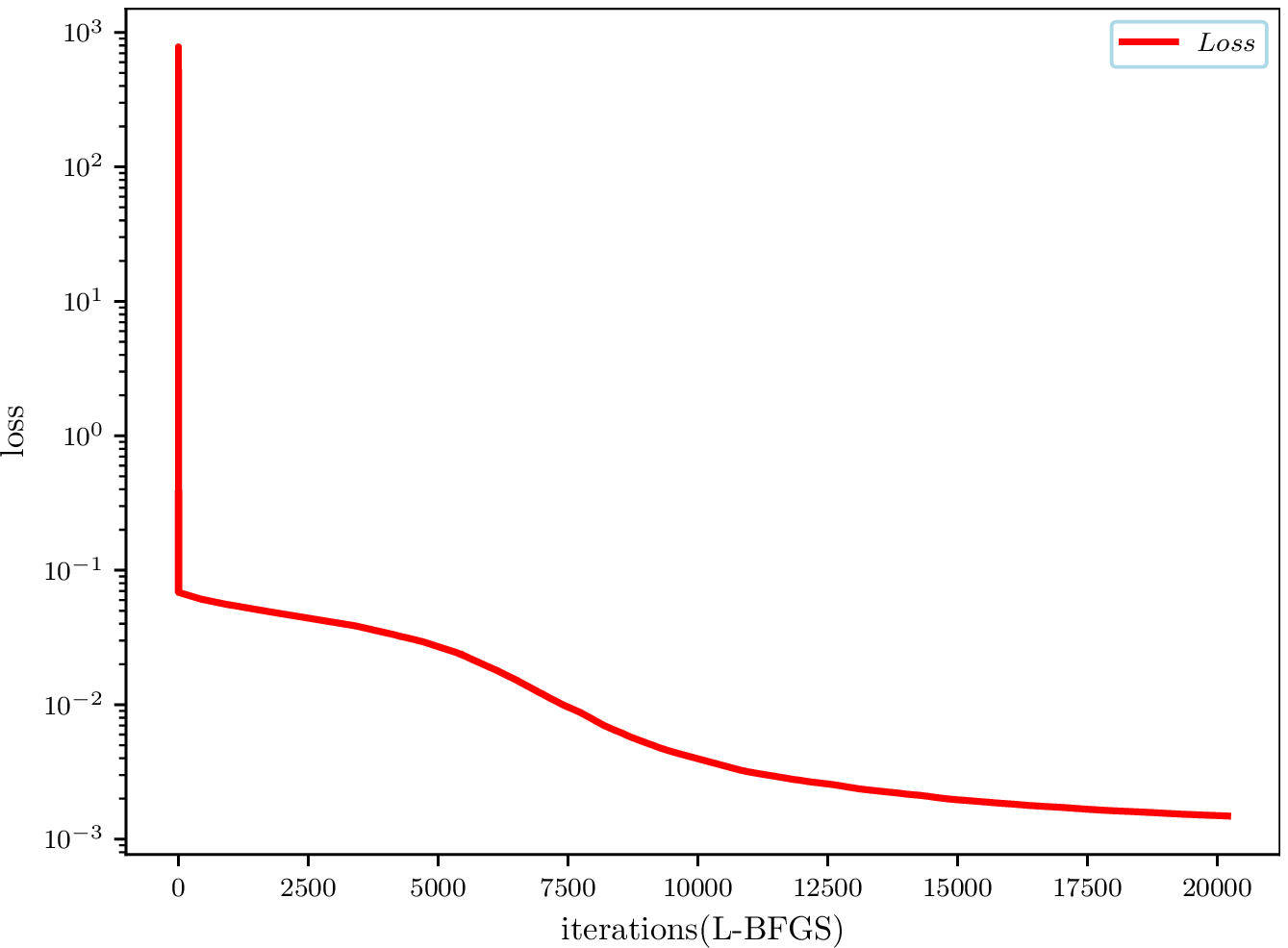}
\end{minipage}%
}%
\centering
\caption{(Color online) (a) three-dimensional plot and corresponding contour maps of the predicted one-soliton solution $\rm\uppercase\expandafter{\romannumeral1}$ stemmed from the IPINN; (b) The loss function curve of the predicted one-soliton solution $\rm\uppercase\expandafter{\romannumeral1}$  with 20000 Adam optimization iterations; (c) The loss function curve of the predicted one-soliton solution $\rm\uppercase\expandafter{\romannumeral1}$  with 19108 L-BFGS optimization iterations.}
\label{F34}
\end{figure}

Figs. \ref{F2}-\ref{F34} display the deep learning results of data-driven one-solitons of the VC-Hirota equation \eqref{E1} based on the IPINN method related to the initial boundary value problem \eqref{E15} and \eqref{E16}. Specifically, the density plots with the corresponding peak scale for diverse dynamics which contain exact dynamics, learned dynamics and error dynamics have been exhibited in detail, and the sectional drawings of exact solitons and their corresponding data-driven solutions at different moments arising from the IPINN are displayed in Fig. \ref{F2}. From the bottom panels of Fig. \ref{F2}, one can indicate one-solitons \eqref{q11} propagate from right to left along the $t$-axis, and show that the amplitudes of solitons are constant $2$ with the development of time $z$. It can be seen that the predicted solution agrees well with the exact solution. Fig. \ref{F3} exhibits the three-dimensional plot with contour map on three planes of the predicted one-solitons based on the IPINN. From the three-dimensional  plot  Fig. \ref{F3}, we can see the center trajectory of $q_{11}(t,z)$ takes the shape of a parabola, which is also intuitively reflected in the $(z,t)$ plane of a soliton prediction graph. The loss function curve figures account for the one-solitons $\rm\uppercase\expandafter{\romannumeral1}$ have also been given out, where Fig. \ref{F41} demonstrates that Adam is an optimizer with oscillatory loss function curves, Fig. \ref{F42} illustrates L-BFGS is an optimization algorithm with linear loss function curves.  It is worth mentioning that the $N_q=1500$ training data points involved in the initial-boundary condition are marked by mediumorchid symbol $``\times"$ in the learned density graphs.

$\bullet$ \textbf{Data-driven one-soliton solution  $\rm\uppercase\expandafter{\romannumeral2}$}

Letting coefficient $\alpha_{1}(z)=\alpha_{2}(z)=z^{2}$, $\xi$=$\eta$=1 and $\delta=10$, then the explicit one-soliton solution is as follows:
\begin{equation}\label{q12}
q_{12}= \frac{4e^{\frac{1}{9}\sqrt{5}((18-18i)t+(52-68i)z^{3})-\frac{200}{9}iz^{3}-10it}}{e^{\frac{4}{9}\sqrt{5}(26z^{3}+9t)}+1}.
\end{equation}

For recovering the data-driven one-solitons $\rm\uppercase\expandafter{\romannumeral2}$ of the VC-Hirota equation in the 9-layer IPINN with 40 neurons per layer, we consider the initial value and  Dirichlet boundary conditions:
\begin{align}\label{E28}
&q^0(t)=q_{\mathrm{12}}(t,-0.75),\quad t\in[-3.0,3.0],
\end{align}
\begin{align}\label{E29}
q^{\mathrm{lb}}(z)=q_{\mathrm{12}}(-3.0,z),\quad q^{\mathrm{ub}}(z)=q_{\mathrm{12}}(3.0,z),\quad z\in[-0.75,0.75].
\end{align}
With the aid of Matlab, the sampling points are selected by means of the finite difference method with spatial-temporal region $[-3.0,3.0]\times[-0.75,0.75]$, and the one-soliton \eqref{q12} is discretized into $1000\times1000$ data points which contain initial-boundary value condition \eqref{E28} and \eqref{E29}. Then the train dataset is obtained by randomly extracting $N_q = 1500$ from original initial-boundary value condition dataset and $N_f= 20000$ collocation points produced via the LHS in spatial-temporal region. Through 20000 Adam iterations and 16150 L-BFGS iterations to optimize loss function $\mathscr{L}(\bar{\Theta})$, the data-driven one-soliton  has been successfully recovered by using the IPINN method. The relative $\mathbb{L}_2$ error is 1.466678$\rm e^{-2}$, and spend 9521.1221 seconds doing 36150 times  iterations.

\begin{figure}[htbp]
\centering
\includegraphics[height=7.5cm,width=15cm]{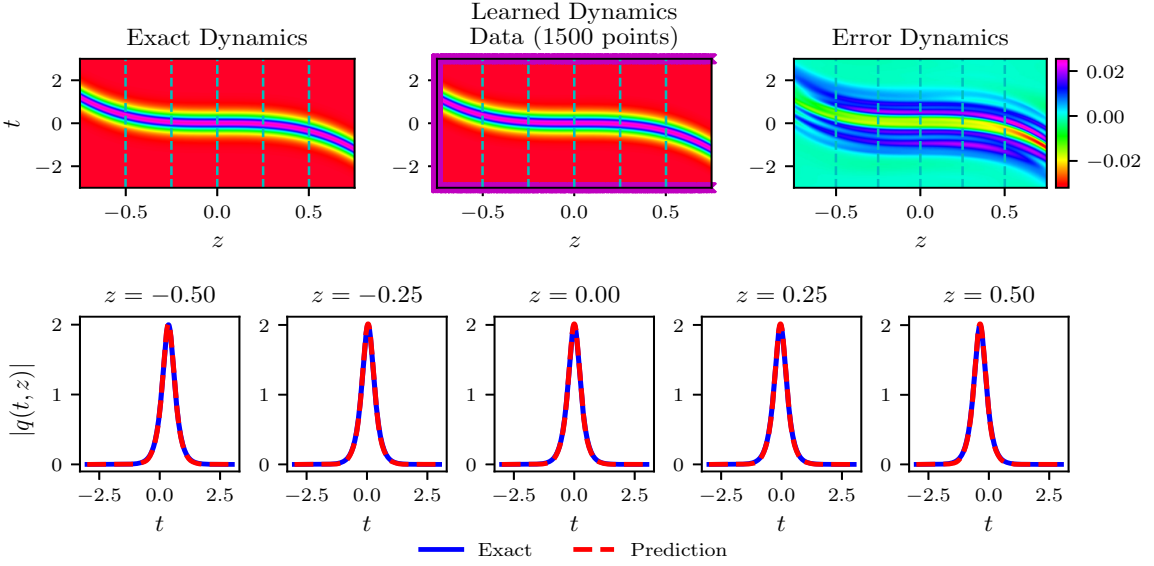}
\centering
\caption{(Color online) The density plots and sectional drawings for the one-soliton $q_{12}(t,z)$:  The one-solitons $q_{12}(t,z)$ resulted from the IPINN with the randomly chosen initial and boundary points $N_q=1500$ which have been shown by using mediumorchid $``\times"$ in learned dynamics, and $N_f = 20000$ collocation points in the corresponding spatiotemporal region. The exact, learned and error dynamics density plots for the one-solitons $q_{12}(t,z)$ with five distinct training moments $z=-0.50, -0.25, 0.00, 0.25$ and $0.50$ (darkturquoise dashed lines), and the sectional drawings which contain the learned and explicit  one-solitons $q_{12}(t,z)$ at the aforementioned five distinct moments.}
\label{F5}
\end{figure}

\begin{figure}[htbp]
\centering
\subfigure[]{\label{F6}
\begin{minipage}[t]{0.3\textwidth}
\centering
\includegraphics[height=4.5cm,width=4.5cm]{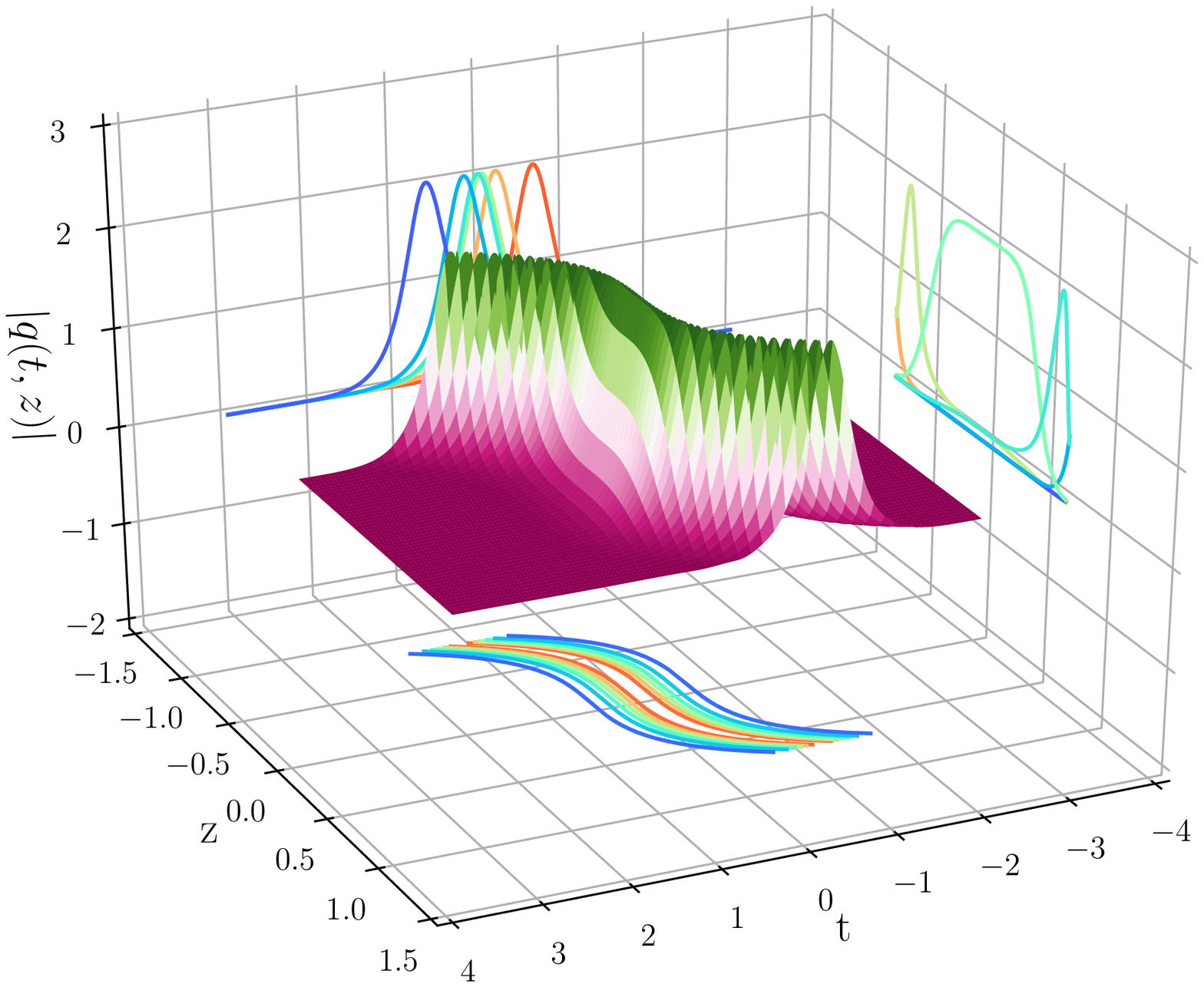}
\end{minipage}
}%
\subfigure[]{\label{F71}
\begin{minipage}[t]{0.3\textwidth}
\centering
\includegraphics[height=4.5cm,width=4.5cm]{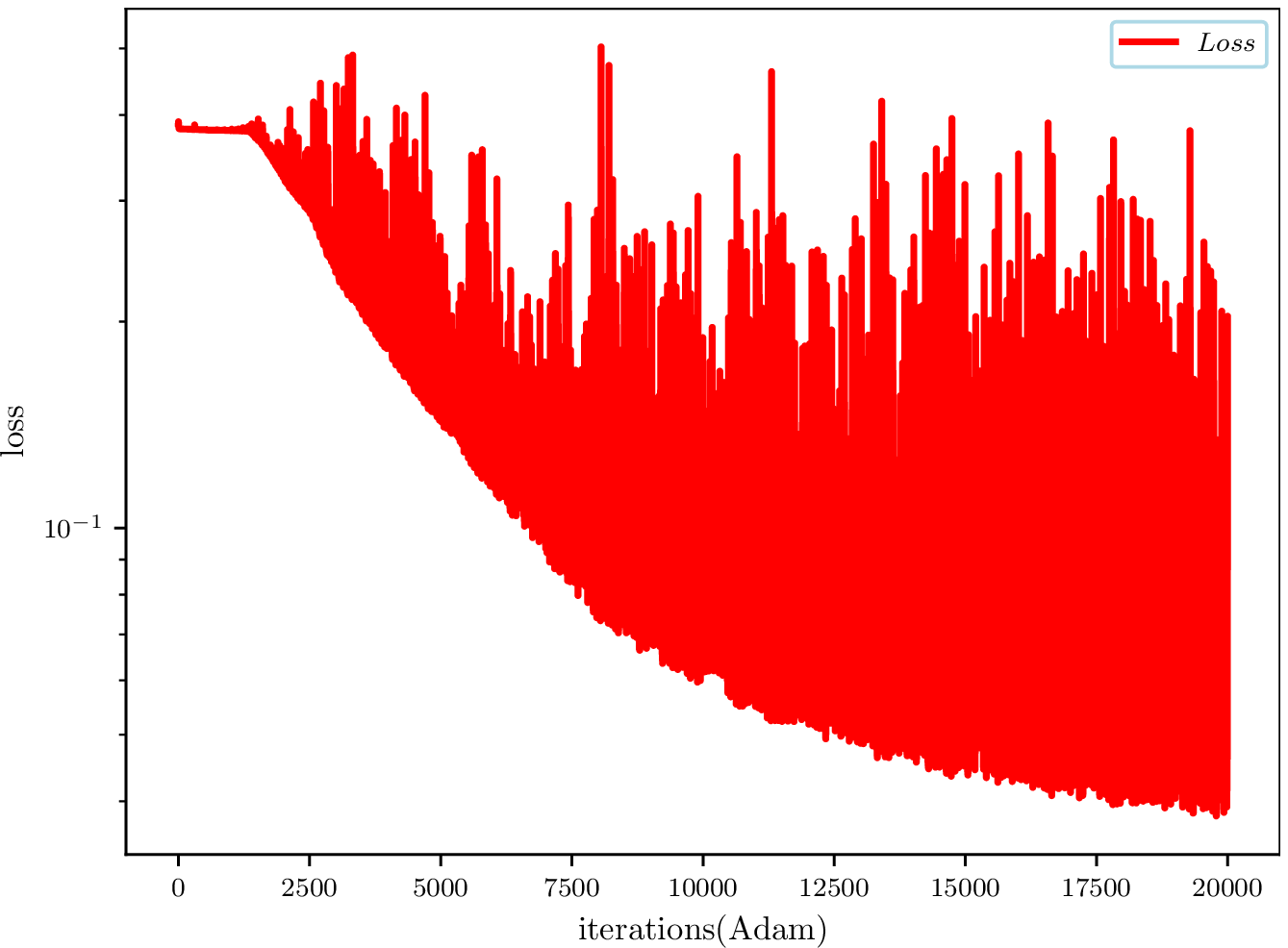}
\end{minipage}
}%
\subfigure[]{\label{F72}
\begin{minipage}[t]{0.3\textwidth}
\centering
\includegraphics[height=4.5cm,width=4.5cm]{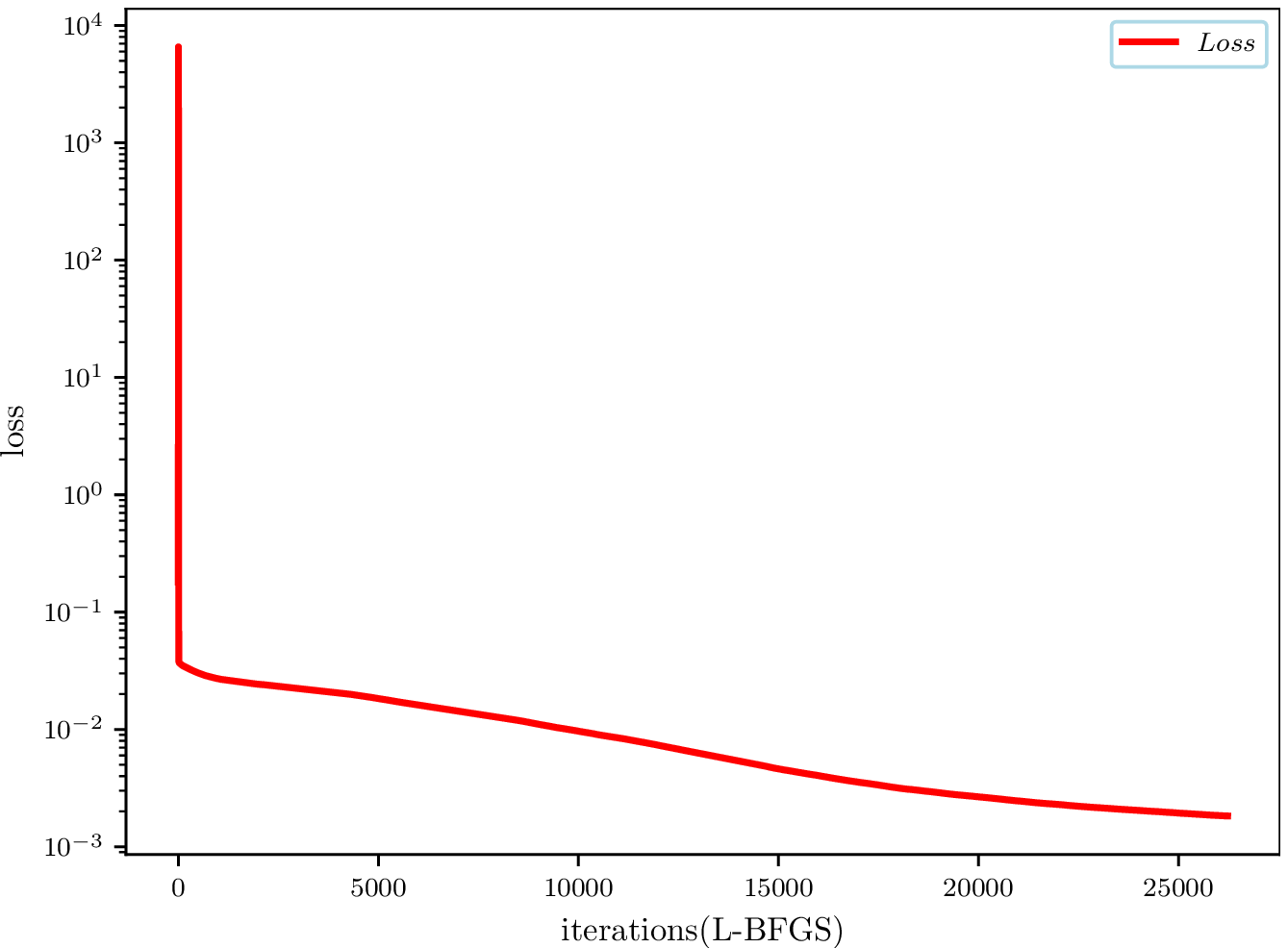}
\end{minipage}
}%
\centering
\caption{(Color online) (a) three-dimensional plot and corresponding contour maps of the predicted one-soliton solutions $\rm\uppercase\expandafter{\romannumeral2}$ stemmed from the IPINN. (b) The loss function curve figures of the predicted one-solitons $\rm\uppercase\expandafter{\romannumeral2}$ with  20000 Adam optimization iterations; (c) The loss function curve figures of the predicted one-solitons $\rm\uppercase\expandafter{\romannumeral2}$ with 16150 L-BFGS optimization iterations.}
\label{F7}
\end{figure}

Figs. \ref{F5} - \ref{F7} exhibit the corresponding training results stemmed from the IPINN with the initial boundary value problem \eqref{E28} and \eqref{E29}. In Fig. \ref{F5}, the exact, learned and error dynamics density plots have been given out.  Fig. \ref{F6} displays the three-dimensional plots with contour map on three planes of the predicted one-soliton $\rm\uppercase\expandafter{\romannumeral2}$ based on the IPINN. From the $(z,t)$ plane of the soliton prediction graph, we can see its center trajectory takes the shape of a "S". Fig. \ref{F71} and \ref{F72} exhibit the loss function curve figures of the data driven one-soliton $\rm\uppercase\expandafter{\romannumeral2}$ arising from the IPINN with the 20000 steps Adam and 16150 steps L-BFGS optimizations on the loss function $\mathscr{L}(\bar{\Theta})$, respectively.

$\bullet$ \textbf{one-soliton solution $\rm\uppercase\expandafter{\romannumeral3}$}

$|q_{1}|^{2}=4\left|\frac{sin(kz)}{\alpha_{2}(z)}\right|sech^{2}(\sqrt {2\delta}(\frac{1}{k}(\frac{4}{3}-\delta)cos(kz)+t))$ is obtained when considering periodic functions  $\alpha_{1}(z)=sin(kz)$ as excitation function in \eqref{q1}. Now the center trajectory of the solution $q_{1}$ is a cosine wave, where $k$ determines the period and $\delta$ has an influence on the shape of the trajectory.  Taking $k=5$, $\delta=6$ and $\xi=\eta$=1, then the explicit one-soliton solution is as follows:

\begin{equation}\label{q13}
q_{13}(t,z)=\frac{4e^{\frac{1}{15}\cos(5z)(72i-(28-44i)\sqrt{3})
-2t(3i-(1-i)\sqrt{3})}}{e^{-\frac{4}{15}\sqrt{3}(14\cos(5z)-15t)}+1}.
\end{equation}

In order to obtain the data-driven one-soliton $q_{13}(t,z)$ of the VC-Hirota equation by using the 9-layer IPINN with 40 neurons per layer, we are committed to studying the initial value and the Dirichlet boundary conditions in the following.
\begin{align}\label{E30}
&q_{13}^0(t)=q_{\mathrm{13}}(t,-1.0),\quad t\in[-3.0,3.0],
\end{align}
\begin{align}\label{E31}
q_{13}^{\mathrm{lb}}(t)=q_{\mathrm{13}}(-3.0,z),\quad q_{13}^{\mathrm{ub}}(t)=q_{\mathrm{13}}(3.0,z),\quad z\in[-1.0,1.0].
\end{align}

Using the same method and procedure as above to obtain the original train date and smaller train data set of the data-driven one-soliton  $\rm\uppercase\expandafter{\romannumeral3}$. After that, introducing the dataset of initial and boundary points into the IPINN, and taking 20000 Adam iterations and 21755 L-BFGS iterations to optimize loss function $\mathscr{L}(\bar{\Theta})$, the predicted one-soliton have been successfully learned by tuning all learnable parameters of the IPINN, and the corresponding network achieved relative $\mathbb{L}_2$ error is 1.885151$\rm e^{-2}$ and the total number of iterations is 41755 with the 10671.1341 seconds training time.

\begin{figure}[htbp]
\centering
\includegraphics[height=7.5cm,width=15cm]{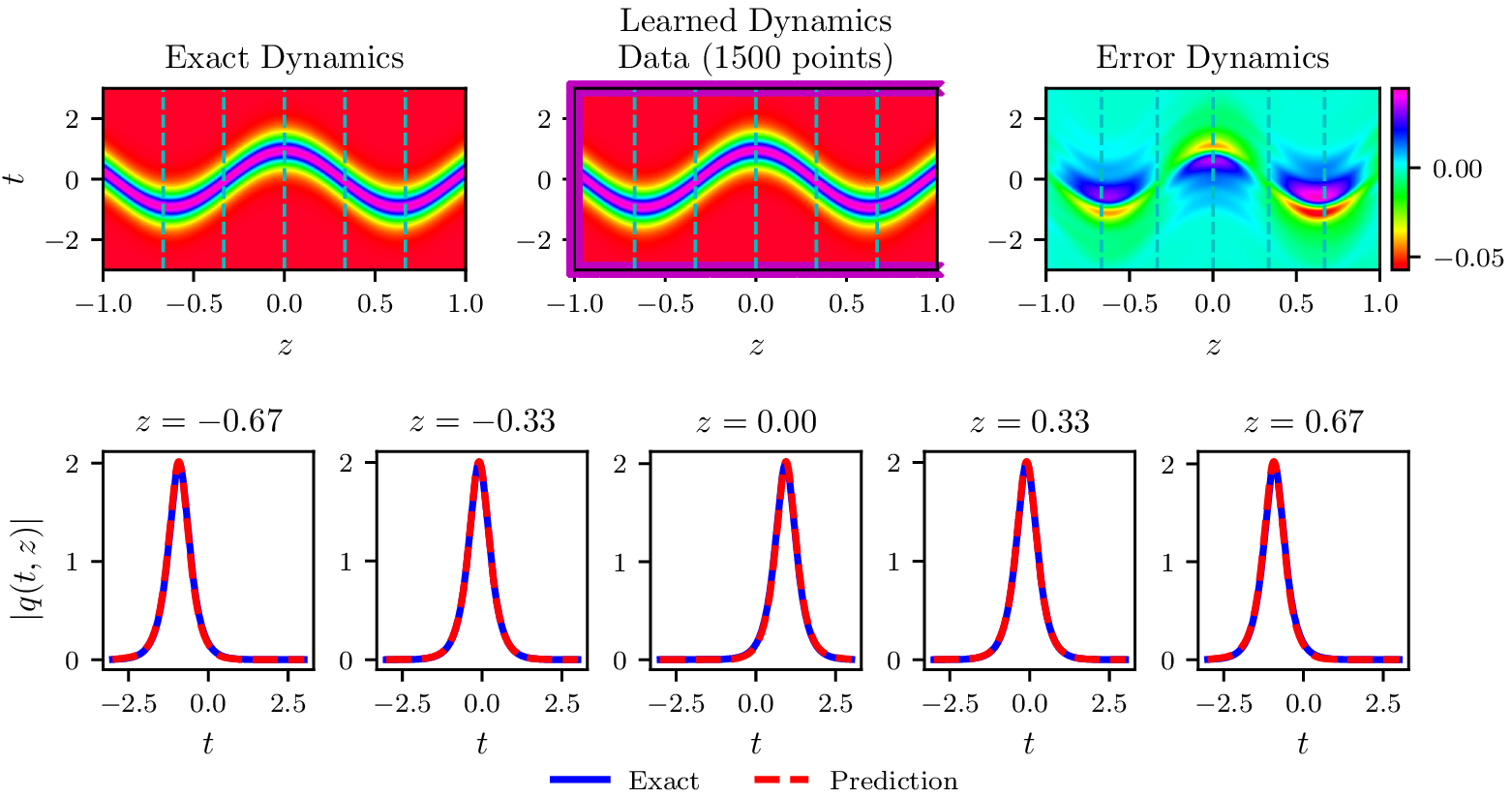}
\centering
\caption{(Color online) The density plots and sectional drawings for the one-soliton: The one-solitons $\rm\uppercase\expandafter{\romannumeral3}$ resulted from the IPINN with the randomly chosen initial and boundary points $N_q=1500$ which have been shown by using mediumorchid $``\times"$ in learned dynamics, and $N_f = 20000$ collocation points in the corresponding spatiotemporal region. The exact, learned and error dynamics density plots for the one-solitons with five distinct training moments $t=-0.67$, $-0.33$, $0.00$, $0.33$ and $0.67$ (darkturquoise dashed lines), and the sectional drawings which contain the learned and explicit one-solitons at the aforementioned five distinct moments.}
\label{F8}
\end{figure}

\begin{figure}[htbp]
\subfigure[]{\label{F9}
\begin{minipage}[t]{0.3\textwidth}
\centering
\includegraphics[height=4.5cm,width=4.5cm]{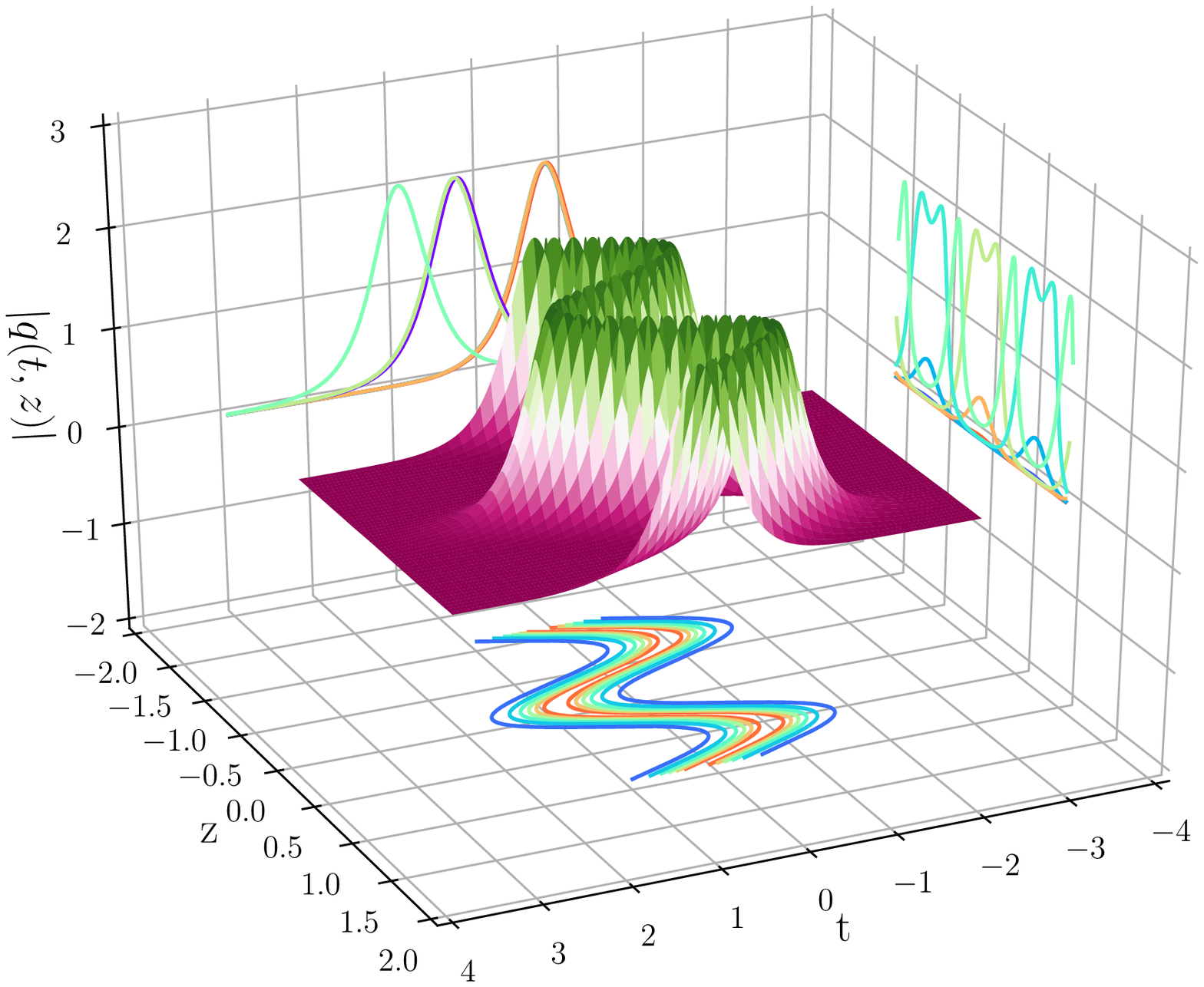}
\end{minipage}}
\centering
\subfigure[]{\label{F101}
\begin{minipage}[t]{0.3\textwidth}
\centering
\includegraphics[height=4.5cm,width=4.5cm]{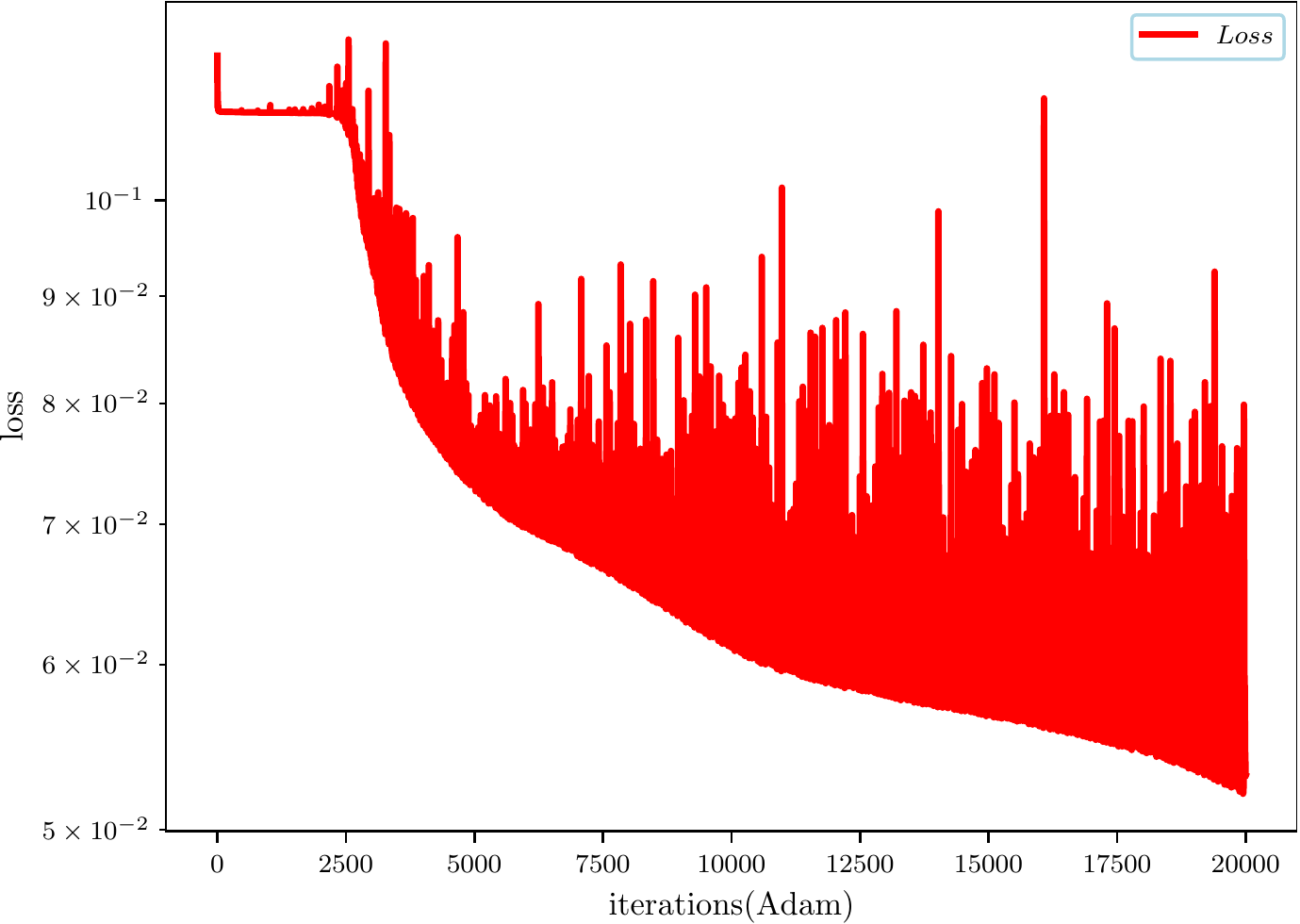}
\end{minipage}
}%
\subfigure[]{\label{F102}
\begin{minipage}[t]{0.3\textwidth}
\centering
\includegraphics[height=4.5cm,width=4.5cm]{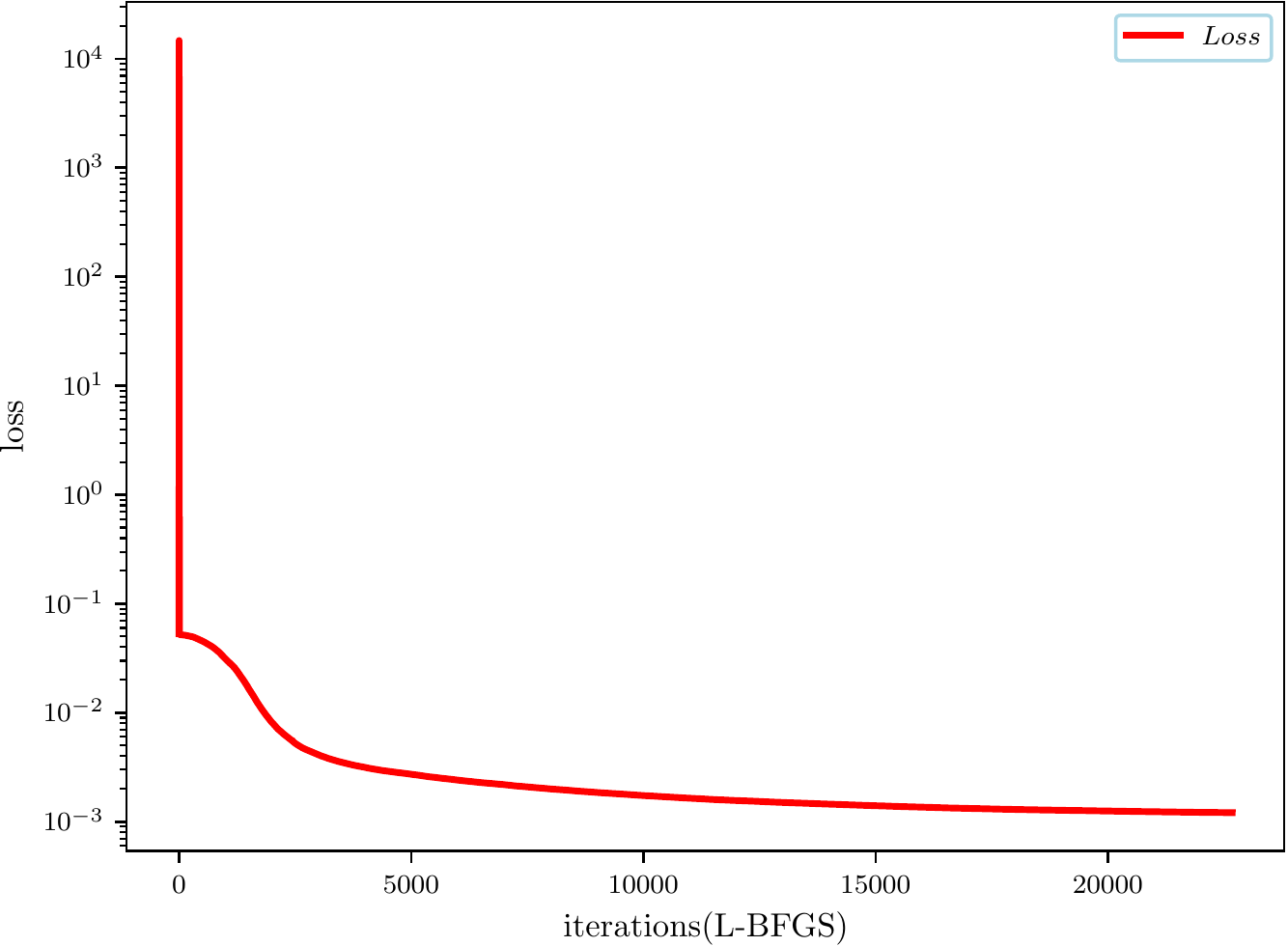}
\end{minipage}%
}%
\centering
\caption{(Color online) (a) three-dimensional plot and corresponding contour maps of the predicted one-soliton solutions $\rm\uppercase\expandafter{\romannumeral3}$ stemmed from the IPINN. (b) The loss function curve figures of the one-solitons $\rm\uppercase\expandafter{\romannumeral3}$ with 20000 Adam optimization iterations; (b) The loss function curve figures of the one-solitons $\rm\uppercase\expandafter{\romannumeral3}$ with 21755 L-BFGS optimization iterations.}
\label{F10}
\end{figure}

Figs. \ref{F8}-\ref{F10} provide the training results arising from the IPINN for the one-soliton \eqref{q13} of the VC-Hirota equation with the initial boundary value problem \eqref{E30} and \eqref{E31}. In Fig. \ref{F8}, the exact, learned and error dynamics density plots have been exhibited, it is worth mentioning that the $N_q=1500$ training data points involved in the initial-boundary condition are marked by mediumorchid symbol $``\times"$ in the learned density plots. Meanwhile, the sectional drawings which include the learned and exact one-soliton \eqref{q13} have been shown at the five distinct moments pointed out in the exact, learned and error dynamics density plots by using darkturquoise dashed lines in the bottom panel of Fig. \ref{F8}. The  three-dimensional plots with contour map on three planes of the predicted one-soliton $\rm\uppercase\expandafter{\romannumeral3}$ based on the IPINN is drawn in Fig. \ref{F9}.  From the $(z,t)$ plane of the soliton prediction graph, the center trajectory of one-soliton $\rm\uppercase\expandafter{\romannumeral3}$ takes the shape of cosine wave. Fig. \ref{F101} and \ref{F102}, respectively,  display the loss function curve figures of the one-soliton  $\rm\uppercase\expandafter{\romannumeral3}$ arising from the IPINN with the 20000 steps Adam and 21755 steps L-BFGS optimizations. 

\subsection{Data-driven two-solitons of the VC-Hirota equation}

In the formula for $N$-solitons formula \eqref{VC-Hirotas}, take $N=2$, $\alpha=0$, $c_{1}=c_{2}=1$, $\zeta_{1}=1+i$ and $\zeta_{2}=2+2i$, then the exact expression of the corresponding two-soliton solution is
\begin{equation}
\label{q2}
q_{2}(t,z)=\frac{\Delta_{1}}{\Delta_{2}},
\end{equation}
\begin{equation}\notag
\begin{split}
\Delta_{1}&=-8\sqrt{\frac{\alpha_{1}(z)}{\alpha_{2}(z)}}
((i-2)e^{-\sqrt{2\delta}(((i-4)\delta+\frac{40}{3}+20i)\int\alpha_{1}(z)dz+(i-4)t)}
-(2+i)e^{-\sqrt{2\delta}(((i-2)\delta+\frac{32}{3}+20i)\int\alpha_{1}(z)dz+(i-2)t)}\\
&+(\frac{1}{2}-i)e^{-\frac{32}{3}\sqrt{2\delta}((-\frac{15}{32}\delta+\frac{17}{8}+i)\int\alpha_{1}(z)dz-\frac{15}{32}t)}
+(i+\frac{1}{2})e^{-\frac{32}{3}\sqrt{2\delta}((-\frac{3}{32}\delta+i+\frac{1}{8})\int\alpha_{1}(z)dz-\frac{3}{32}t)})e^{-\frac{1}{3}i\delta(2\delta\int\alpha_{1}(z)dz+3t)},\\
\Delta_{2}&=e^{\sqrt{2\delta}
((-24-\frac{28}{3}i+(6+i)\delta)\int\alpha_{1}(z)dz+(6+i)t)
}
+5e^{\sqrt{2\delta}((-\frac{64}{3}-\frac{28}{3}i+(4+i)\delta)\int\alpha_{1}(z)dz+(4+i)t)}
-4e^{-\frac{56}{3}\sqrt{2\delta}((\frac{9}{14}+i-\frac{9}{56}\delta)\int\alpha_{1}(z)dz-\frac{9}{56}t)}\\
&-4e^{2\sqrt{2\delta}((-6+(\frac{3}{2}+i)\delta)\int\alpha_{1}(z)dz+(\frac{3}{2}+i)t)}
+5e^{\sqrt{2\delta}((-\frac{8}{3}-\frac{28}{3}i+(2+i)\delta)\int\alpha_{1}(z)dz+(2+i)t)}
+e^{\frac{1}{3}i\sqrt{2\delta}(3\delta\int\alpha_{1}(z)dz-28\int\alpha_{1}(z)dz+3t)}.
\end{split}
\end{equation}

$\bullet$ \textbf{two-soliton solution $\rm\uppercase\expandafter{\romannumeral1}$}

First, fixed coefficient $\alpha_{1}(z)=\alpha_{2}(z)=z$ and $\delta=10$, then an explicit express of two-soliton solution \eqref{q2} is as follows:
\begin{equation}\label{q21}
q_{21}(t,z)=\frac{\Delta_{11}}{\Delta_{21}},
\end{equation}
\begin{equation}\notag
\begin{split}
\Delta_{11}&=-4e^{-\frac{10i}{3}(10z^{2}+3t)}
((2i-4)e^{-\frac{2\sqrt{5}}{3}(45iz^{2}+3it-40z^{2}-12t)}
-(2i+4)e^{-\frac{2\sqrt{5}}{3}(45iz^{2}+3it-14z^{2}-6t)}
-(2i-1)e^{-\frac{2\sqrt{5}}{3}(16iz^{2}-41z^{2}-15t)}\\
&+(2i+1)e^{-\frac{2\sqrt{5}}{3}(16iz^{2}-13z^{2}-3t)}),\\
\Delta_{21}&=e^{\frac{2\sqrt{5}}{3}(iz^{2}
+3it+54z^{2}+18t)}
+5e^{\frac{2\sqrt{5}}{3}(iz^{2}+3it+28z^{2}+12t)}
-4e^{-\frac{2\sqrt{5}}{3}(28iz^{2}-27z^{2}-9t)}
-4e^{2\sqrt{5}(10iz^{2}+2it+9z^{2}+3t)}
+5e^{\frac{2\sqrt{5}}{3}(iz^{2}+3it+26z^{2}+6t)}\\
&+e^{\frac{2\sqrt{5}}{3}i(z^{2}+3t)}.
\end{split}
\end{equation}

In the following, the initial-boundary problem for the VC-Hirota equation is considered by applying an 8-layer IPINN with 50 neurons per layer to obtain a data-driven two-soliton. Similarly, taking $[L_0,L_1]\times [T_0,T_1]$= $[-1.5,1.0] \times [-0.3,0.3]$ in Eq. \eqref{E1}, then the corresponding initial condition $q^0(x)$ and Dirichlet boundary conditions $q^{\mathrm{lb}}(z)$ and $q^{\mathrm{ub}}(z)$ are rewritten as
\begin{align}\label{E17}
\begin{split}
q^0(x)=q_{\mathrm{21}}(t,-0.3),\quad t\in[-1.5,1.0],
\end{split}
\end{align}
and  
\begin{align}\label{E18}
q^{\mathrm{lb}}(z)=q_{\mathrm{21}}(-1.5,z),\quad q^{\mathrm{ub}}(z)=q_{\mathrm{21}}(6.0,z),\quad z\in[-0.3,0.3].
\end{align}

The original training data is obtained by dividing the spatial region $[-1.5,1.0]$ into 1000 points and the temporal region $[-0.3,0.3]$ into 1000 points, the remaining data will be used to obtain training errors by comparing with predicted two-solitons. After that, we generate a smaller training dataset containing initial-boundary data by randomly extracting $N_q=1600$ from original training dataset and $N_f=20000$ collocation points produced via LHS in the corresponding spatiotemporal region. Then, the predicted two-solitons $\rm\uppercase\expandafter{\romannumeral1}$ have been successfully learned by imposing an 8-hidden-layer IPINN with 50 neurons per layer, and the related loss functions are optimized through 20000 Adam iterations and 29359 L-BFGS iterations. The relative $\mathbb{L}_2$ error of the IPINN model is 4.121996$\rm e^{-2}$, the total number of iterations is 49359 with 12990.2661 seconds training time.

\begin{figure}[htbp]
\centering
\includegraphics[height=7.5cm,width=15cm]{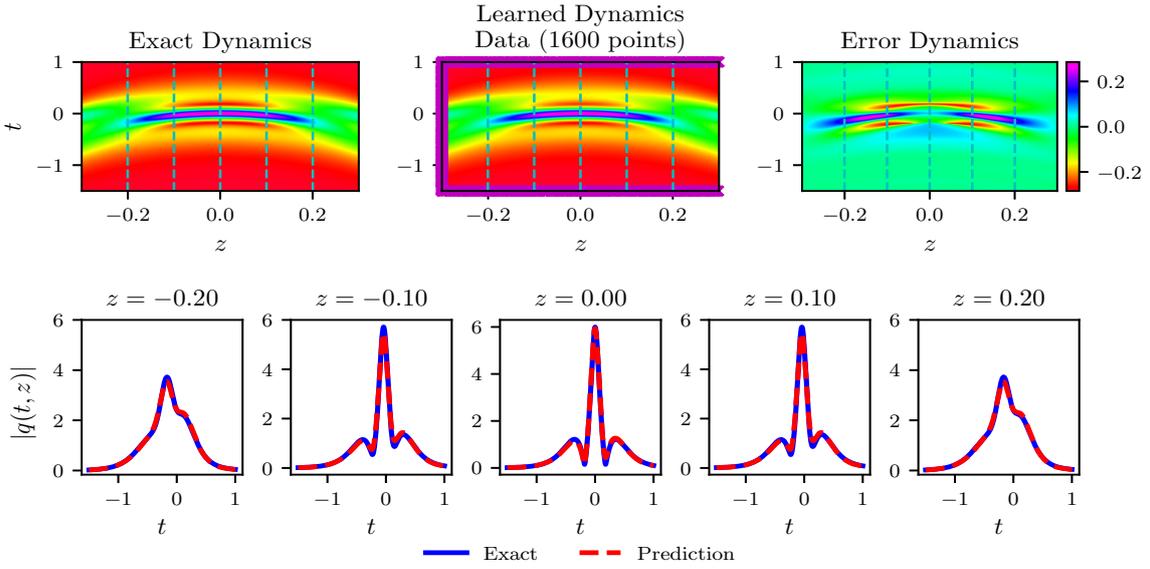}
\centering
\caption{(Color online) The two-solitons $\rm\uppercase\expandafter{\romannumeral1}$ resulted from the IPINN with the randomly chosen initial and boundary points $N_q=1500$ which have been shown by using mediumorchid $``\times"$ in learned dynamics, and $N_f = 20000$ collocation points in the corresponding spatiotemporal region. The exact, predicted and error dynamics density plots for the two-solitons with five distinct training moments $z=-0.2, -0.1, 0.0, 0.1$ and $0.2$ (darkturquoise dashed lines), and the sectional drawings which contain the learned and explicit two-solitons at the aforementioned five distinct moments.}
\label{F11}
\end{figure}

\begin{figure}[htbp]
\subfigure[]{\label{F12}
\begin{minipage}[t]{0.3\textwidth}
\centering
\includegraphics[height=4.5cm,width=4.5cm]{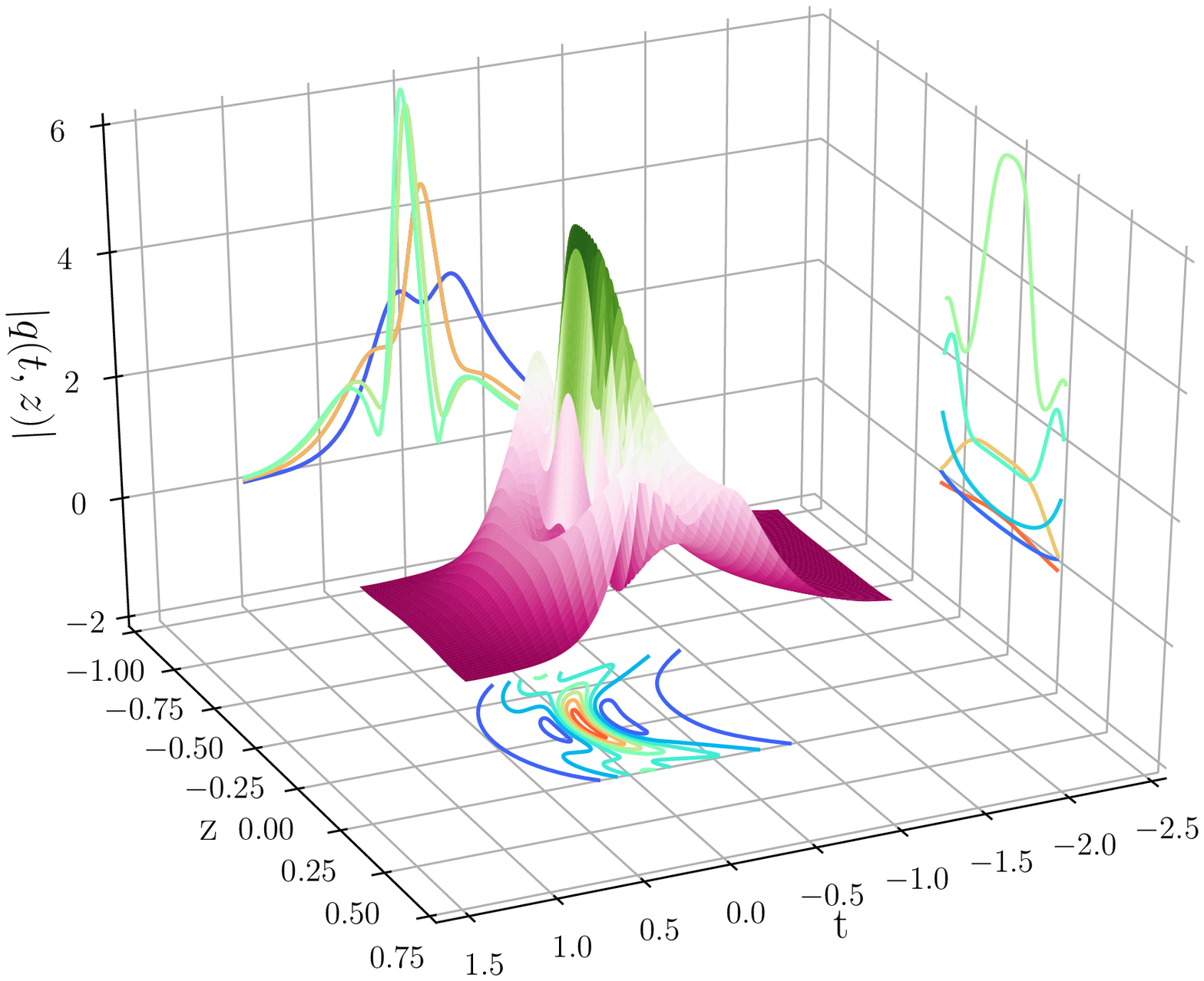}
\end{minipage}%
}%
\centering
\subfigure[]{\label{F131}
\begin{minipage}[t]{0.3\textwidth}
\centering
\includegraphics[height=4.5cm,width=4.5cm]{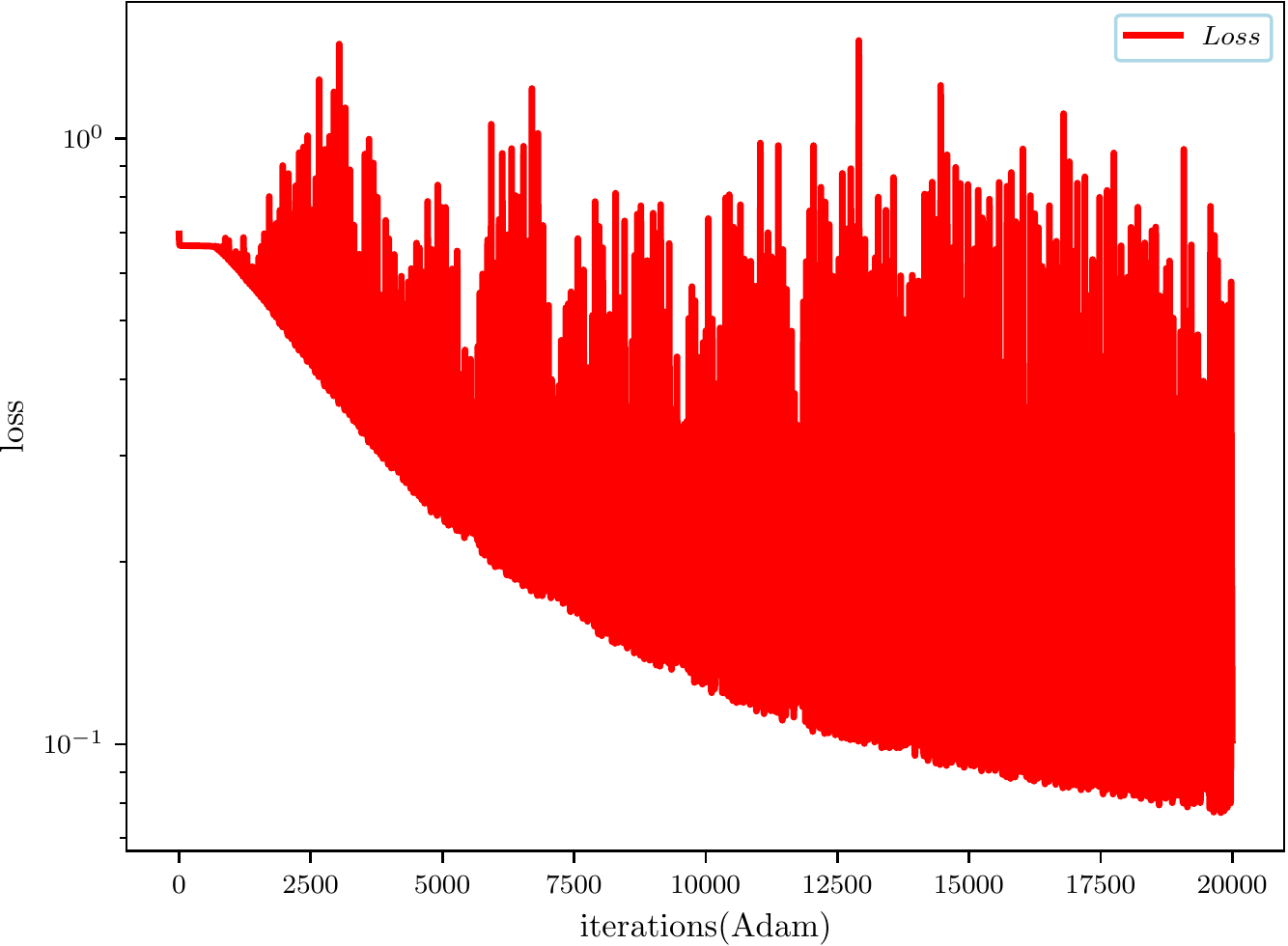}
\end{minipage}
}%
\subfigure[]{\label{F132}
\begin{minipage}[t]{0.3\textwidth}
\centering
\includegraphics[height=4.5cm,width=4.5cm]{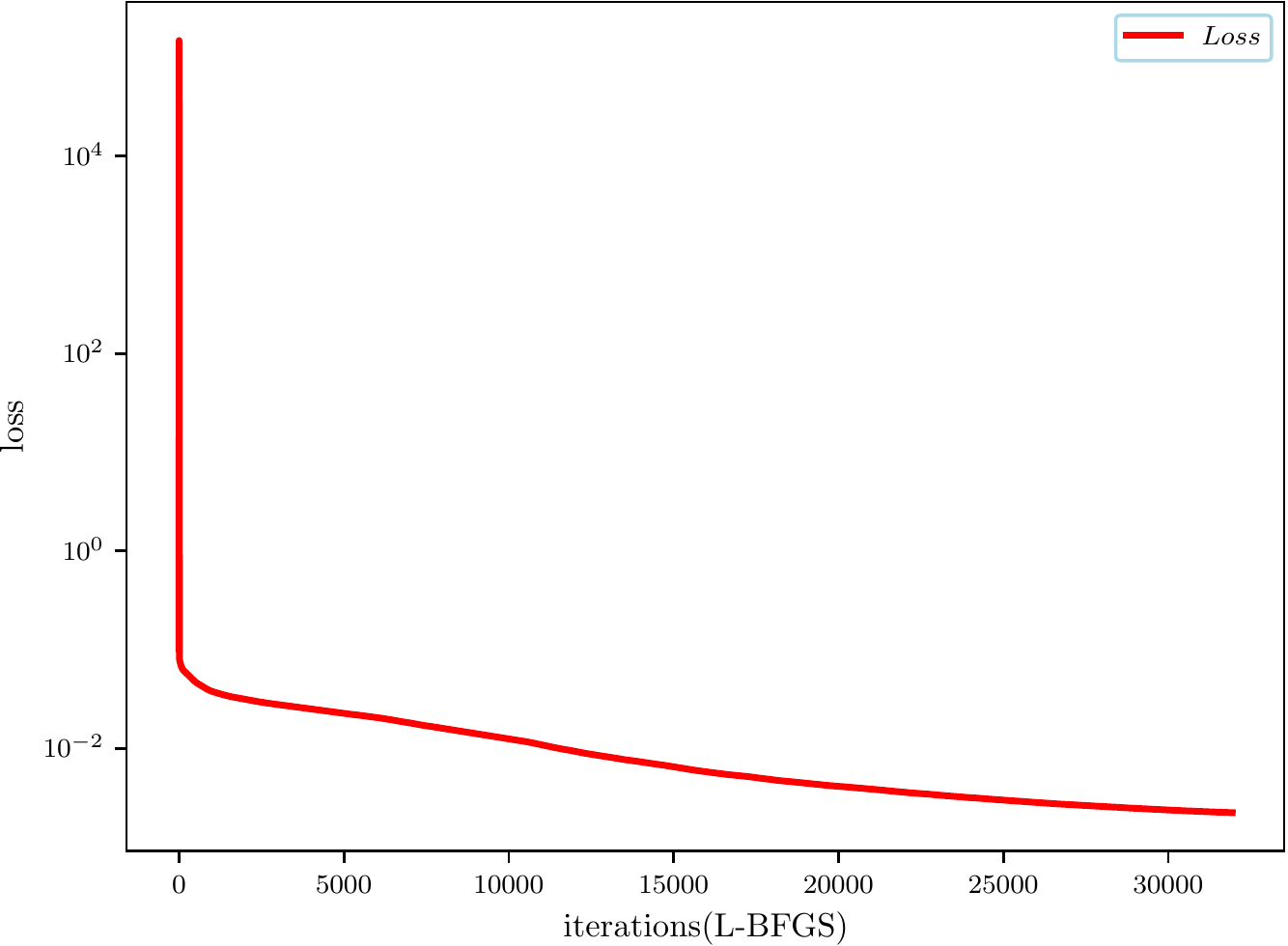}
\end{minipage}%
}%
\caption{(Color online) (a) three-dimensional plot and corresponding contour maps of the predicted two-soliton solutions $\rm\uppercase\expandafter{\romannumeral1}$  stemmed from the IPINN. (b) The loss function curve figures of the two-solitons $\rm\uppercase\expandafter{\romannumeral1}$ with 20000 Adam optimization iterations; (c) The loss function curve figures of the two-solitons $\rm\uppercase\expandafter{\romannumeral1}$ with 26239 L-BFGS optimization iterations.}
\label{F13}
\end{figure}

Figs. \ref{F11}-\ref{F13} display the training results of the two-soliton $q_{21}(t,z)$ based on the IPINN related to the initial-boundary value problem \eqref{E17} and \eqref{E18} of the VC-Hirota equation. Fig. \ref{F11} depicts the exact, learned and error dynamic density plots and sectional drawing at different moments for the exact and prediction two-soliton. The three-dimensional plot and corresponding contour maps of the predicted two-soliton solution   $\rm\uppercase\expandafter{\romannumeral1}$ is shown in Fig. \ref{F12}.  In the IPINN framework, curve plot of the loss function after 20000 Adam optimization iterations is diaplays in Fig. \ref{F131}, the loss function plot with 26239 L-BFGS optimization iterations is showcases in Fig. \ref{F132}.  The shape of the two-soliton solution of variable coefficients  equation shown in this paper is more complex than that of the rogue wave solution of the traditional constant coefficient equation. 

$\bullet$ \textbf{two-soliton solution $\rm\uppercase\expandafter{\romannumeral2}$}

Fix the coefficient $\alpha_{1}(z)=\alpha_{2}(z)=z^{2}$ and $\delta=10$, then the explicit express of the two-soliton solution  is derived form \eqref{VC-Hirotas} as follows:
\begin{equation}\label{q22}
q_{22}=\frac{\Delta_{12}}{\Delta_{22}},
\end{equation}

\begin{equation}\notag
\begin{split}
\Delta_{12}&=-4e^{-\frac{10}{9}i(20z^{3}+9t)}(
(2i-4)e^{-\frac{2}{9}\sqrt{5}((90i-80)z^{3}+(9i-36)t)}
-(2i+4)e^{-\frac{2}{9}\sqrt{5}((90i-28)z^{3}+(9i-18)t)}
+(1-2i)e^{-\frac{2}{9}\sqrt{5}((32i-82)z^{3}-45t)}\\
&+(2i+1)e^{-\frac{2}{9}\sqrt{5}((32i-26)z^{3}-9t)}),\\
\Delta_{22}&=e^{\frac{2}{9}\sqrt{5}((2i+108)z^{3}+(9i+54)t)}+5e^{\frac{2}{9}\sqrt{5}((2i+56)z^{3}+(9i+36)t)}
 -4e^{-\frac{2}{9}\sqrt{5}(56iz^{3}-54z^{3}-27t)}
 -4e^{\frac{2}{3}\sqrt{5}(20iz^{3}+18z^{3}+6it+9t)}+e^{\frac{2}{9}\sqrt{5}i(2z^{3}+9t)}\\
 &+5e^{\frac{2}{9}\sqrt{5}(2iz^{3}+52z^{3}+9it+18t)}.
 \end{split}
\end{equation}

Now, we consider the initial-boundary value problem of the VC-Hirota equation for obtaining the data-driven two-soliton $\rm\uppercase\expandafter{\romannumeral2}$  by applying the multilayer IPINN. Similarly, taking $[L_0,L_1]$ and $[T_0,T_1]$ in Eq. \eqref{E1} as $[-2.0,2.0]$ and $[-0.5,0.5]$ respectively, then the corresponding initial condition $q^0(x)$ and Dirichlet boundary conditions $q^{\mathrm{lb}}(z)$ and $q^{\mathrm{ub}}(z)]$ are shown in the following:
\begin{align}\label{E171}
\begin{split}
&q^0(t)=q_{\mathrm{22}}(t,-3.0),\quad t\in[-2.0,2.0],
\end{split}
\end{align}
and
\begin{align}\label{E181}
q^{\mathrm{lb}}(z)=q_{\mathrm{22}}(-0.5,z),\quad q^{\mathrm{ub}}(z)=q_{\mathrm{22}}(2.0,z),\quad z\in[-0.5,0.5].
\end{align}
Dividing the spatial region $[-2.0,2.0]$ into 1000 points and the temporal region $[-0.5,0.5]$ into 1000 points, and setting $N_q=1500$  and $N_f=20000$. Then, the predicted two-solitons $\rm\uppercase\expandafter{\romannumeral2}$ have been successfully learned by imposing a 8-hidden-layer IPINN with 50 neurons per layer, and the related loss functions are optimized through 20000 Adam iterations and 26239 L-BFGS iterations. The relative $\mathbb{L}_2$ errors of the IPINN model are  1.770931$\rm e^{-2}$ and the total number of iterations is  23461 with training time of 11106.1975 seconds.

\begin{figure}[htbp]
\centering
\includegraphics[height=7.5cm,width=15cm]{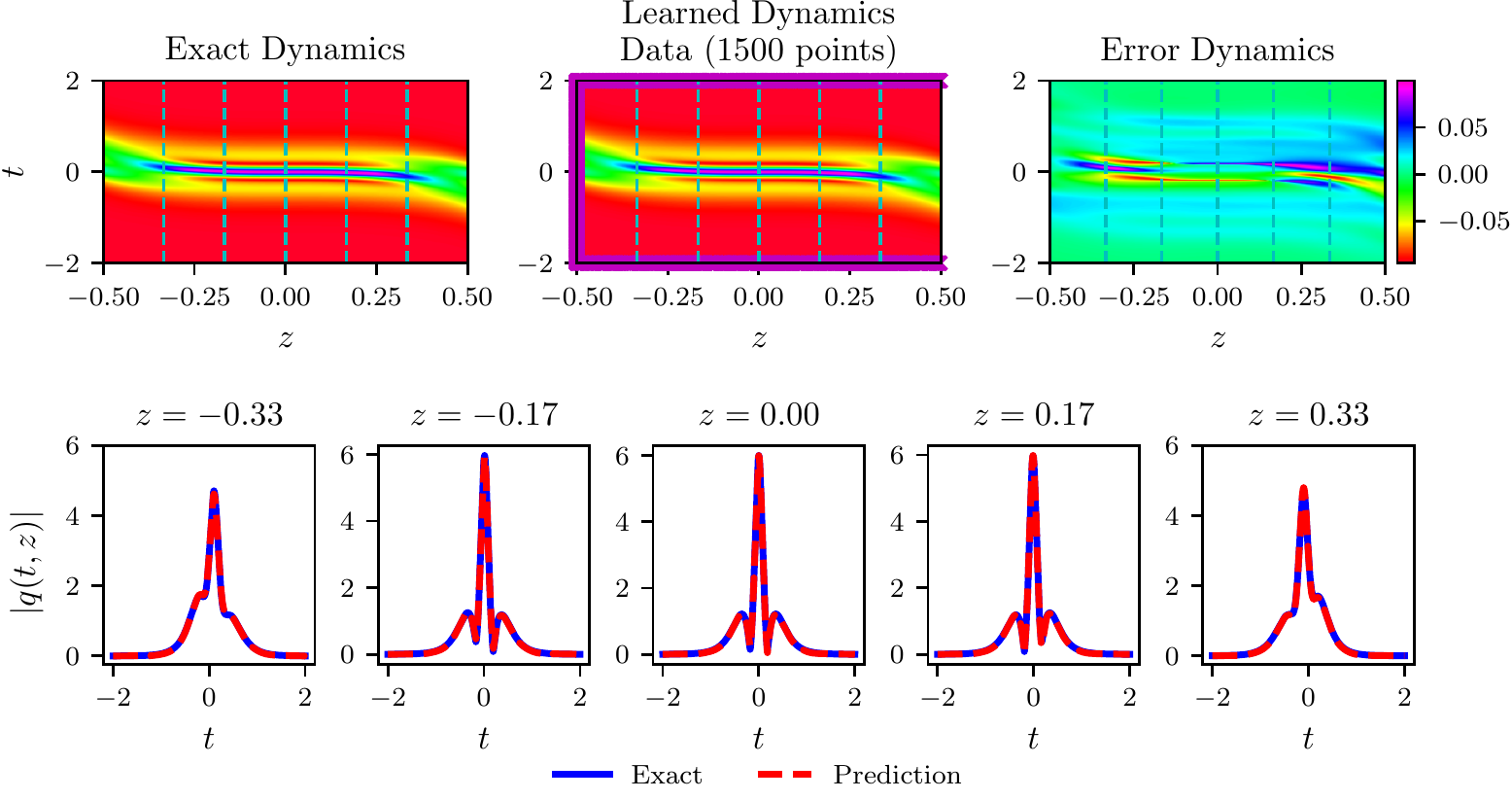}
\centering
\caption{(Color online)  The density plots and sectional drawings for the two-soliton: the two-solitons  resulted from the IPINN with the randomly chosen initial and boundary points $N_q=1500$ which have been shown by using mediumorchid $``\times"$ in learned dynamics, and $N_f = 20000$ collocation points in the corresponding spatiotemporal region. The exact, predicted and error dynamics density plots for the two-solitons $q_{22}(t,z)$ with five distinct training moments $z=-0.33, -0.17, 0.00, 0.17$ and $0.33$ (darkturquoise dashed lines), and the sectional drawings which contain the learned and exact two-solitons $q_{22}(t,z)$ at the aforementioned five distinct moments.}
\label{F14}
\end{figure}

\begin{figure}[htbp]
\centering
\subfigure[]{\label{F15}
\begin{minipage}[t]{0.3\textwidth}
\centering
\includegraphics[height=4.5cm,width=4.5cm]{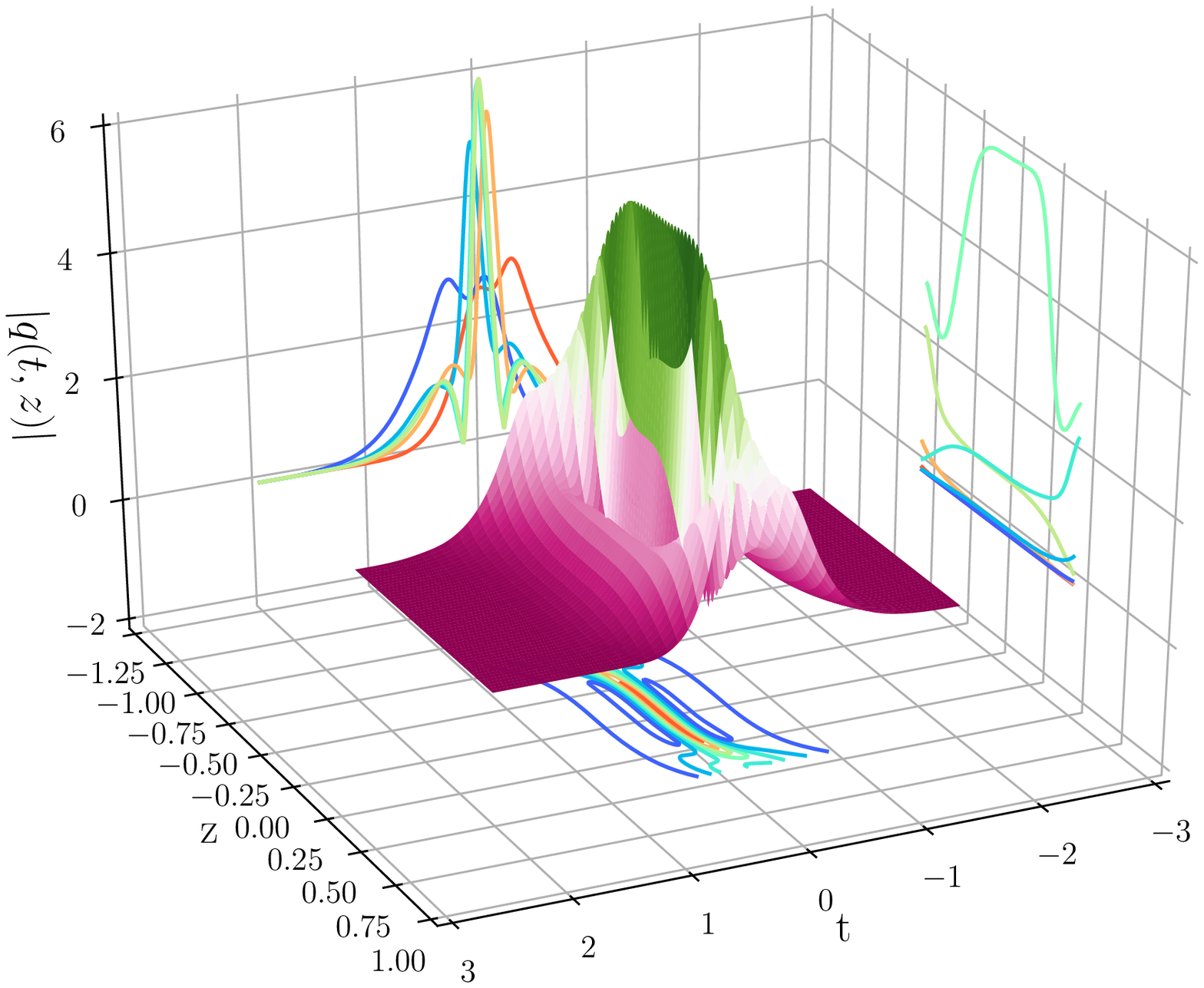}
\end{minipage}%
}%
\subfigure[]{\label{F161}
\begin{minipage}[t]{0.3\textwidth}
\centering
\includegraphics[height=4.5cm,width=4.5cm]{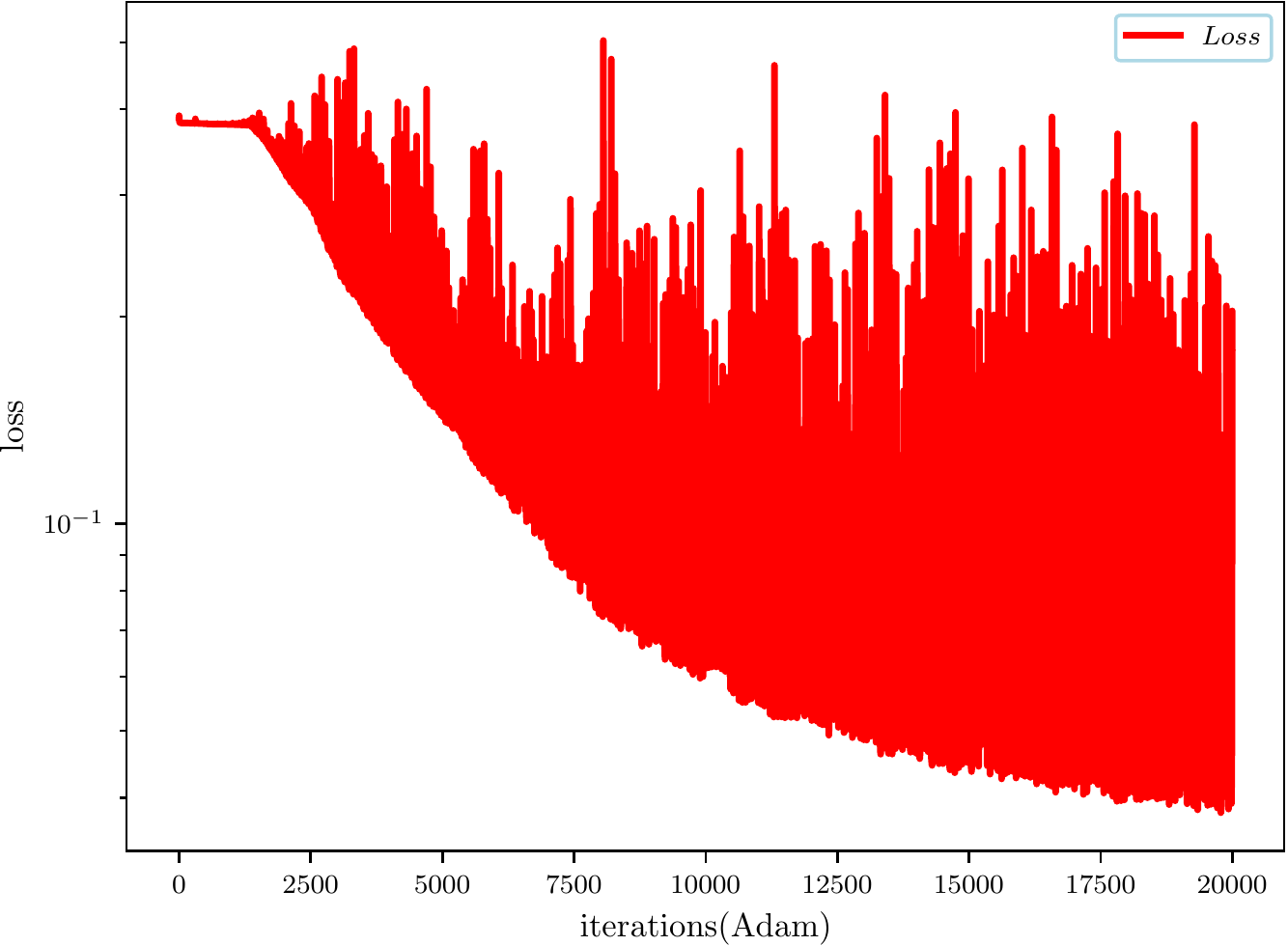}
\end{minipage}
}%
\subfigure[]{\label{F162}
\begin{minipage}[t]{0.3\textwidth}
\centering
\includegraphics[height=4.5cm,width=4.5cm]{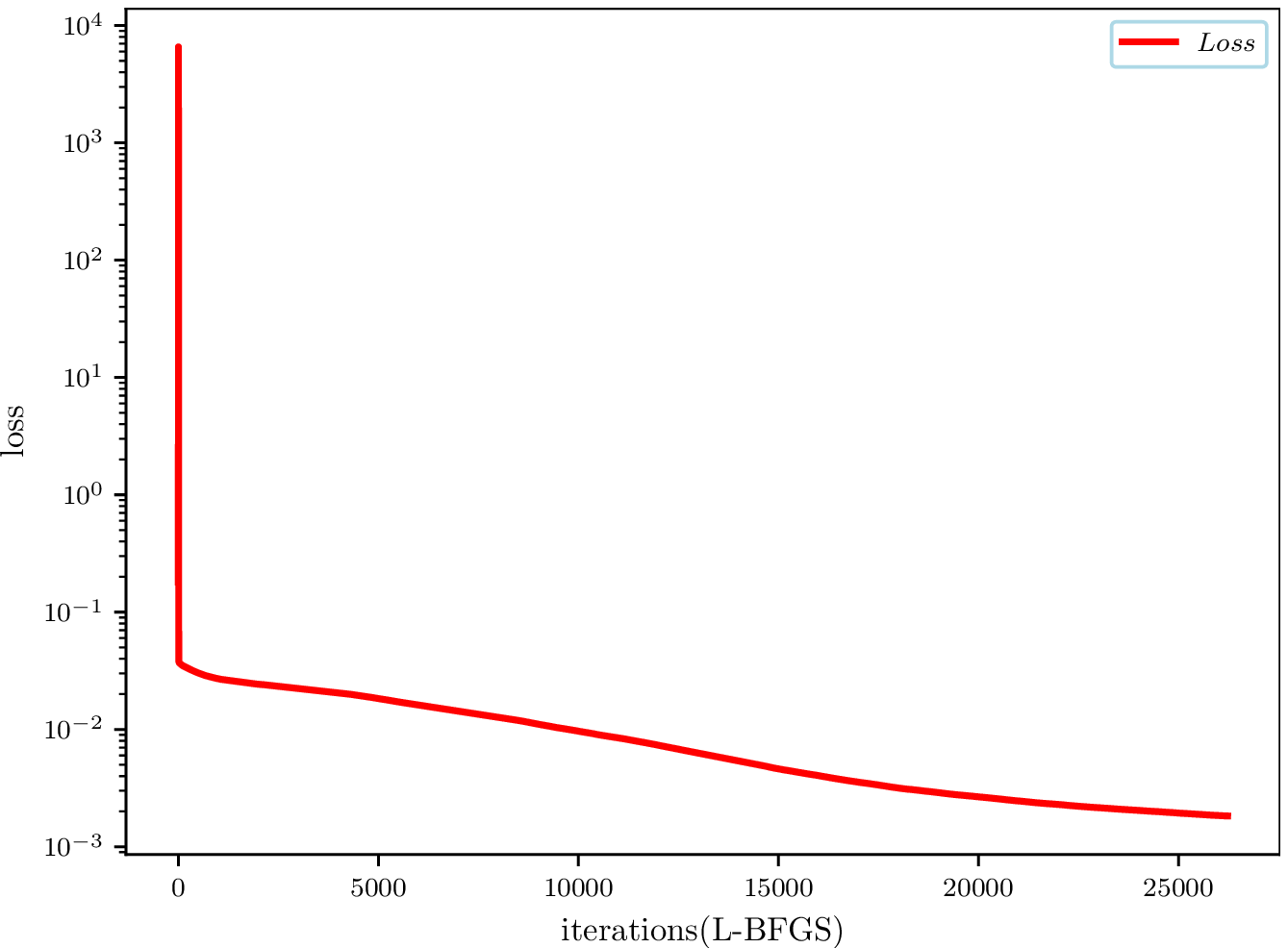}
\end{minipage}%
}%
\centering
\caption{(Color online) (a) three-dimensional plot and corresponding contour maps of the predicted two-soliton solutions $\rm\uppercase\expandafter{\romannumeral2}$ stemmed from the IPINN. (b) The loss function curve figures of the two-solitons $\rm\uppercase\expandafter{\romannumeral2}$ with 20000 Adam optimization iterations; (b)  The loss function curve figures of the two-solitons $\rm\uppercase\expandafter{\romannumeral2}$ with 26239 L-BFGS optimization iterations.}
\label{F16}
\end{figure}

Figs. \ref{F14}-\ref{F16} display the training results of the 
two-solitons $q_{22}(t,z)$ based on the IPINN related to the
 initial-boundary value problem \eqref{E171} and \eqref{E181} of the VC-Hirota equation \eqref{E1}. Fig. \ref{F14} depicts various dynamic density plots and sectional drawing at different moments for the VC-Hirota equation \eqref{E1}. The three-dimensional plot of predicted
  two-soliton solutions $\rm\uppercase\expandafter{\romannumeral2}$ is shown in Fig. \ref{F15}. Fig. \ref{F161} and \ref{F162} showcase curve plots of the loss function after 20000 Adam optimization iterations and 26239 L-BFGS optimization iterations in IPINN framework, respectively. 

From the figure above, it can be seen intuitively that the dynamic behavior of the two-soliton solution of the VC-Hirota equation is more complex than that of the traditional rogue wave solutions of the constant coefficients integrable equation. This brings a greater challenge to the deep learning process of two-solitons. In this paper, an 8-layer IPINN algorithm with 50 neurons in each layer is adopted to successfully train the 2-solitons of the VC-Hirota equation after several experiments to optimize the relevant parameters.

\subsection{Data-driven higher-order solitons of the VC-Hirota equation}

In complex systems, the research of high-order soliton of the IVC model has more appropriate practical significance. In this section, we will focus on the data-driven  high-order soliton of the VC-Hirota equation with $\zeta_{1}=1+i$ and $\zeta_{2}=1-i$ based on the IPINN method. The formula of the high-order solitons for the VC-Hirota equation is obtained by means of Riemann-Hilbert method in \cite{zhj-arxiv-2022}. For simple, we just use a second-order soliton as an example to verify the effective of the IPINN method. The express of the second-order soliton is as follows.

\begin{equation}\label{qh}
q_{h}=\frac{\Theta_{1}}{\Theta_{2}},
\end{equation}

\begin{equation}\notag
\begin{split}
\Theta_{1}&=
-8i\sqrt{\frac{\alpha_{1}(z)}{\alpha_{2}(z)}}
(((-i\sqrt{2}\delta^{\frac{3}{2}}+4\sqrt{2\delta})\int\alpha_{1}(z)dz
-i\sqrt{2\delta}t+i)
e^{(2\sqrt{2}\delta^{\frac{3}{2}}+\frac{1}{3}((-8-8i)\sqrt{2\delta}-2i\delta^{2}))\int\alpha_{1}(z)dz
-(i\delta-2\sqrt{2\delta})t}+
(i\sqrt{2\delta}t\\
&+(i\sqrt{2}\delta^{\frac{3}{2}}
+4\sqrt{2\delta})\int\alpha_{1}(z)dz+4\sqrt{2\delta}i)e^{-\frac{i}{3}((8\sqrt{2\delta}+2\delta^{2})\int\alpha_{1}(z)dz+3\delta t)}
,\\
\Theta_{2}&=(8\delta^{3}+128\delta)
((\int\alpha_{1}(z)dz)^{2}+16t\delta^{2}\int\alpha_{1}(z)dz+8t^{2}\delta+2)
e^{(\frac{1}{3}+\frac{1}{3}i)\sqrt{2\delta}
((3\delta-4)\int\alpha_{1}(z)dz+3t)}
+e^{(1+\frac{1}{3}i)\sqrt{2\delta}((3\delta-4)\int\alpha_{1}(z)dz+3t)}\\
&+e^{(-\frac{1}{3}+\frac{1}{3}i)\sqrt{2\delta}((3\delta-4)\int\alpha_{1}(z)dz+3t)}).
\end{split}
\end{equation}

From the express of \eqref{qh}, we can construct abundant second-order solitons by adjustment $\alpha_{1}(z)$ and $\alpha_{2}(z)$. Next, we give an expression of the second-order soliton with $\alpha_{1}=\alpha_{2}=sinh(z)$ as follows:

\begin{align}\label{E21}
\begin{split}
q_{h1}(x,t)=\frac{\Theta_{h1}}{\Theta_{h2}},
\end{split}
\end{align}

\begin{equation}\notag
\begin{split}
\Theta_{h1}&=
8i( (5\cosh(z)+t)i\sqrt{10}e^{(\frac{22}{3}\sqrt{10}-\frac{8}{3}i\sqrt{10}-\frac{50}{3}i)\cosh(z)+2\sqrt{10}t-5it}
-(5\cosh(z)+1)i\sqrt{10}e^{-\frac{1}{3}i((8\sqrt{10}+50)\cosh(z)+15t)}\\
&-(4\cosh(z)\sqrt{10}+i)(e^{-\frac{1}{3}i((8\sqrt{10}+50)\cosh(z)+15t)}
+e^{(\frac{22}{3}\sqrt{10}-\frac{8}{3}i\sqrt{10}
-\frac{50}{3}i)\cosh(z)+2\sqrt{10}t-5it})
),\\
\Theta_{h2}&=(
1640\cosh(z)^{2}
+400t\cosh(z)
+40t^{2}
+2)e^{\frac{\sqrt{10}}{3}(i+1)(11\cosh(z)+3t)}
+e^{\frac{\sqrt{10}}{3}(11\cosh(z)+3t)(i+3)}
+e^{\frac{\sqrt{10}}{3}(11\cosh(z)+3t)(i-1)}.
\end{split}
\end{equation}

In order to recover the data-driven second-order soliton solution via eq.\eqref{E21}, we commit to introducing the initial boundary value conditions of the VC-Hirota equation to the 9-layer IPINN with 40 neurons per layer. Selecting $[L_0,L_1]$ and $[T_0,T_1]$ in Eq. \eqref{E1} as $[-10.0,0.0]$ and $[-0.75,0.75]$ respectively, then we have the corresponding initial value and Dirichlet boundary conditions:
\begin{align}\label{E26}
\begin{split}
&q^0(t)=q_{\mathrm{h1}}(t,-0.75),\quad t\in[-10.0,0.0],
\end{split}
\end{align}

\begin{align}\label{E27}
q^{\mathrm{lb}}(z)=q_{\mathrm{h1}}(-10.0,z),\quad q^{\mathrm{ub}}(z)=q_{\mathrm{h1}}(0.0,z),\quad z\in[-0.75,0.75].
\end{align}

In order to obtain the original training data set of the above initial boundary value conditions \eqref{E26} and \eqref{E27}, we discretize the exact second-order soliton solution \eqref{E21} based on the finite difference method by dividing the spatial region $[-10.0,0.0]$ into 1000 points and the temporal region $[-0.75,0.75]$ into 1000 points in Matlab. Furthermore, in addition to the data set composed of the aforementioned initial boundary value conditions, the residual data set is used to calculate the $\mathbb{L}_2$ norm error by comparing with the predicted second-order soliton solution \eqref{E21}. After that, a smaller training dataset that containing initial-boundary data is generated by randomly extracting $N_q=1500$ from original dataset and $N_f=20000$ collocation points which are produced by the LHS. According to 20000 Adam iterations and 25117 L-BFGS iterations, the latent second-order soliton $q_{h1}(t,z)$ have been successfully learned by employing the IPINN, and the network achieved relative $\mathbb{L}_2$ error of 1.622958$\rm e^{-2}$ for the second-order soliton $q_{h1}(t,z)$, and the total number of iterations is  20932 with the training time is 11016.8943 seconds.

\begin{figure}[htbp]
\centering
\includegraphics[height=7.5cm,width=15cm]{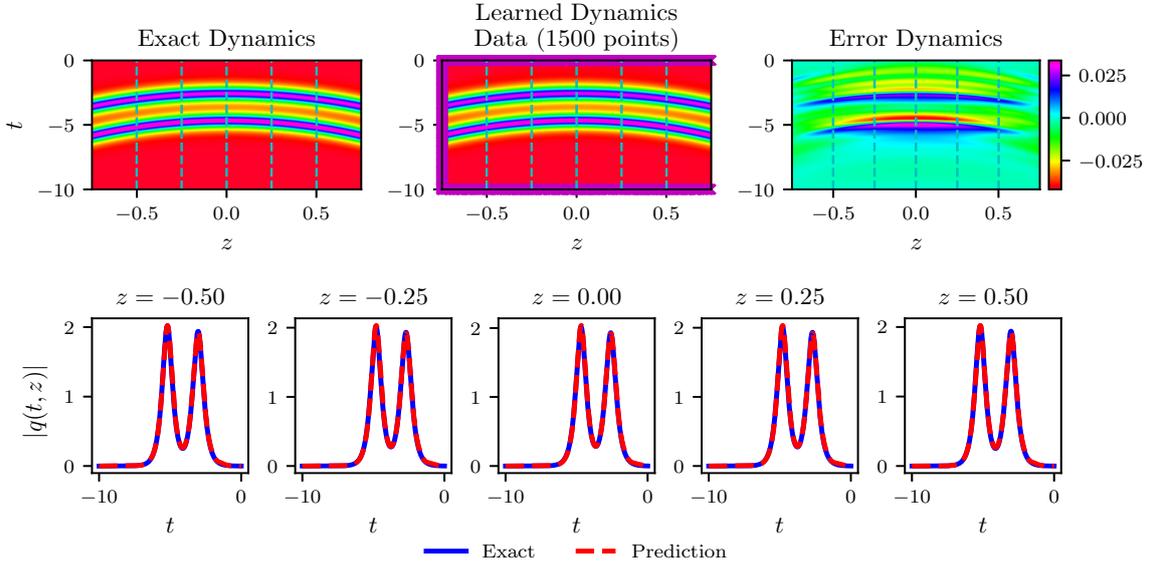}
\centering
\caption{(Color online) The density plots and sectional drawings for the second-order soliton $q_{h1}(t,z)$: The second-order soliton $q_{h1}(t,z)$ resulted from the IPINN with the randomly chosen initial and boundary points $N_q=1500$ which have been shown by using mediumorchid $``\times"$ in learned dynamics , and $N_f = 20000$ collocation points in the corresponding spatiotemporal region. The exact, learned and error dynamics density plots for the second-order soliton $q_{h1}(t,z)$ with five distinct training moments $t=-0.50, -0.25, 0.00, 0.25$ and $0.50$ (darkturquoise dashed lines), and the sectional drawings which contain the learned and explicit second-order soliton $q_{h1}(t,z)$ at the aforementioned five distinct moments.}
\label{F17}
\end{figure}

\begin{figure}[htbp]
\centering
\subfigure[]{\label{F18}
\begin{minipage}[t]{0.3\textwidth}
\centering
\includegraphics[height=4.5cm,width=4.5cm]{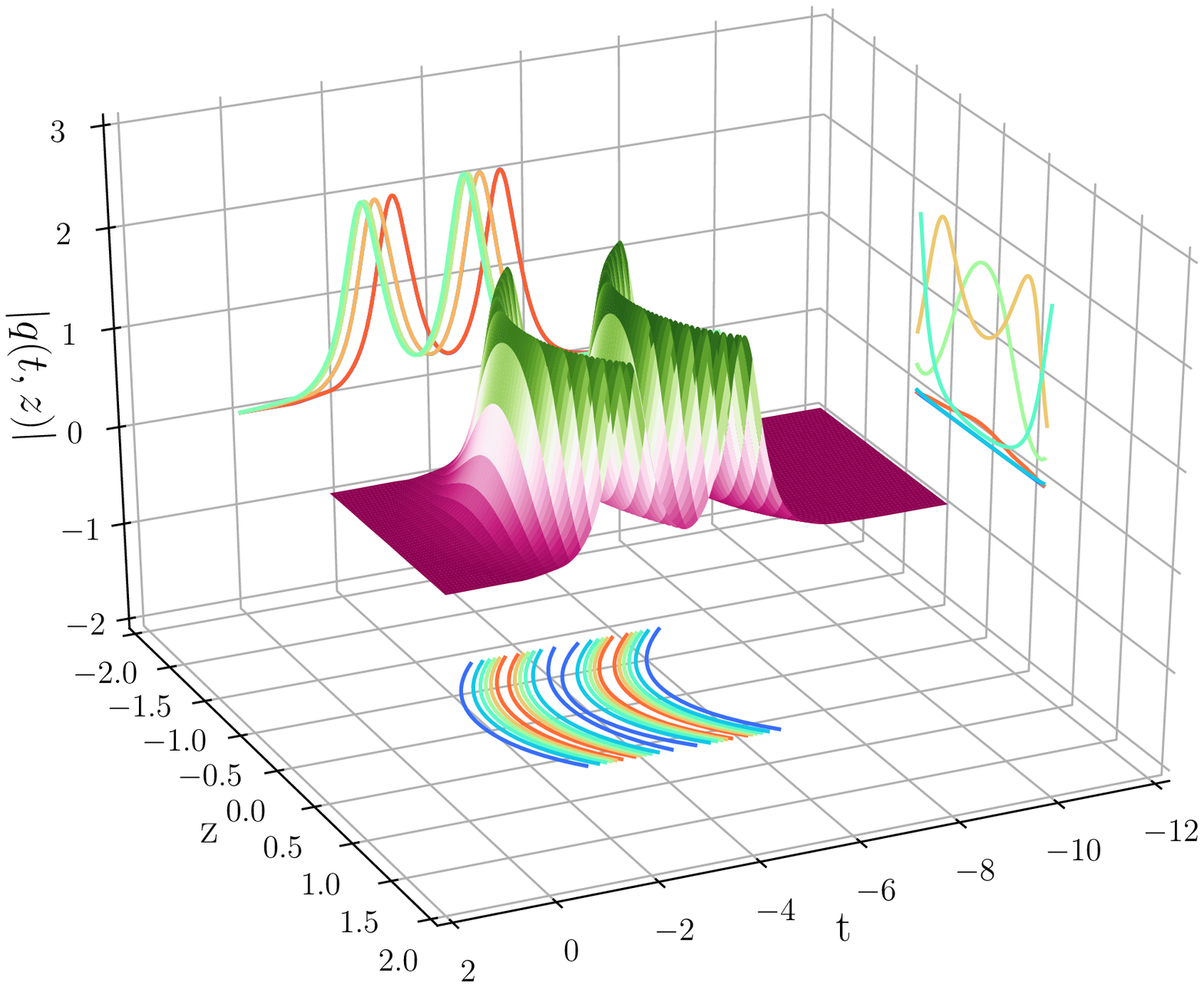}
\end{minipage}}
\subfigure[]{\label{F191}
\begin{minipage}[t]{0.3\textwidth}
\centering
\includegraphics[height=4.5cm,width=4.5cm]{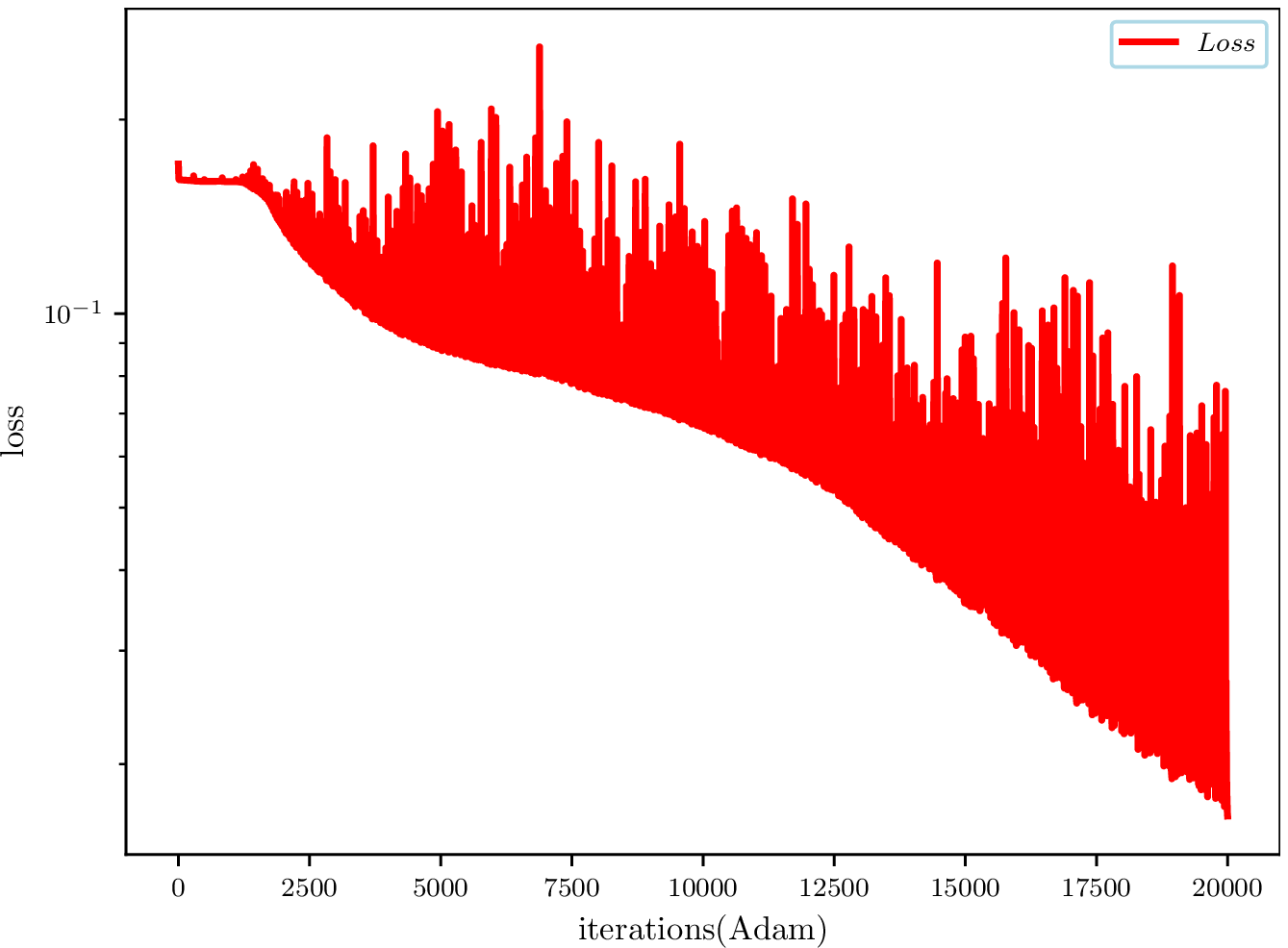}
\end{minipage}
}%
\subfigure[]{\label{F192}
\begin{minipage}[t]{0.3\textwidth}
\centering
\includegraphics[height=4.5cm,width=4.5cm]{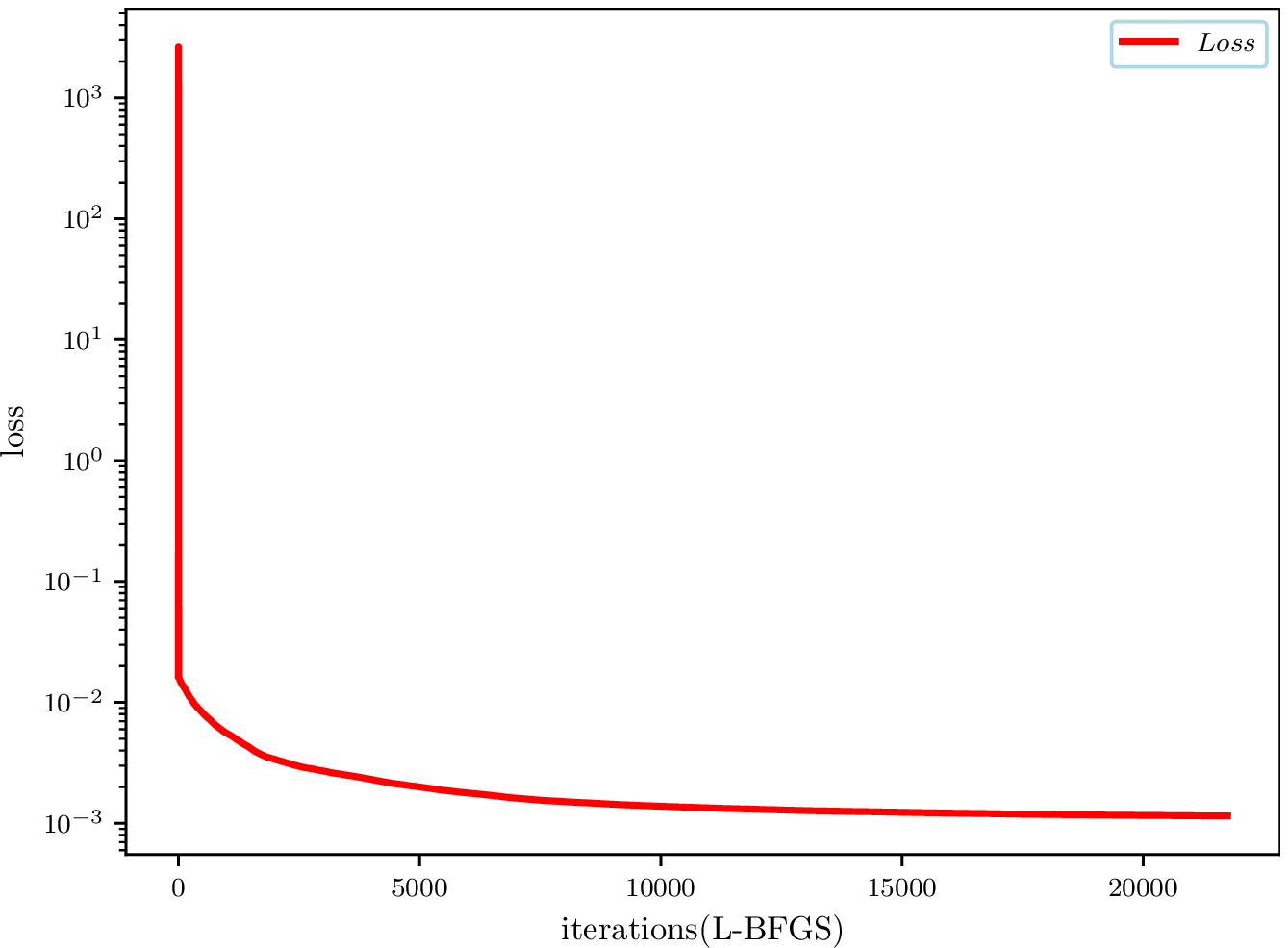}
\end{minipage}%
}%
\centering
\caption{(Color online) (a) three-dimensional plot and corresponding contour maps of the predicted second-order soliton  stemmed from the IPINN. (b) The loss function curve figures of the second-order soliton solutions with 20000 Adam optimization iterations; (c) The loss function curve figures of the second-order soliton solutions with 25117 L-BFGS optimization iterations.}
\label{F19}
\end{figure}

Figs. \ref{F17}-\ref{F19} present the deep learning results of the second-order soliton  based on the IPINN for the VC-Hirota equation with the initial-boundary value problem \eqref{E26} and \eqref{E27}. Fig. \ref{F17} displays the exact, learned and error dynamics of the second-order soliton, and exhibits the sectional drawings which contain the learned and explicit second-order soliton solution at five different moments. From the density plots of learned dynamics and profiles which reveal amplitude and error of exact and prediction second-order soliton in Fig. \ref{F17}.  The 3D plots of the predicted second-order soliton is shown in Fig. \ref{F18}. Graphically, it can be seen that the dynamic behavior of second-order solution is similar to two parallel one solitons $q_{11}$ with different amplitudes. The loss function curve figures of the second-order soliton solutions with 20000 Adam optimization iterations is shown in Fig. \ref{F191}. The loss function curve figures of the second-order soliton solutions with 25117 L-BFGS optimization iterations is displayed in Fig. \ref{F192}.

Through a lot of numerical training, we found that the IPINN algorithm has a good effect when describing relatively smooth solutions in a small range, but when the dynamic behavior of the solution becomes complex, the learning effect will become worse. In order to obtain better numerical results, it becomes extremely dependent on the choice of parameters, which means that the training difficulty will increase with the increase of the complexity of the solution. A summary of the aforementioned data-driven forward problems of the VC-Hirota equation is shown in the following Tab. \ref{Tab-SB22}.
 
 \begin{table}[htbp]
  \caption{Results of the different forward problems for the VC-Hirota equation by IPINN model}
  \label{Tab-SB22}
  \centering
  \scalebox{0.8}{
  \begin{tabular}{|c|c|c|c|c|c|c|}
  \hline
  \diagbox{\textbf{\textbf{Results}}}{\textbf{solution Types}} & one-soliton $\rm\uppercase\expandafter{\romannumeral1}$ & one-soliton $\rm\uppercase\expandafter{\romannumeral2}$ & one-soliton $\rm\uppercase\expandafter{\romannumeral3}$ & two-soliton $\rm\uppercase\expandafter{\romannumeral1}$ & two-soliton $\rm\uppercase\expandafter{\romannumeral2}$ & second-order soliton\\
  \hline
  Relative error & 1.168275$\rm e^{-2}$ &  1.466678$\rm e^{-2}$ & 1.885151$\rm e^{-2}$ & 4.121996$\rm e^{-2}$ & 1.770931$\rm e^{-2}$ & 1.622958$\rm e^{-2}$\\
  \hline
  Training time & 10732.7923s & 9521.1221s & 10671.1341s & 12990.2661s & 11106.1975s & 11016.8943s\\
  \hline
  Iterations   & 39108 &  36150 & 41755  &  49359 & 41755 & 40932\\
  \hline
  \end{tabular}}
\end{table}

\section{Data-driven inverse problems of the VC-Hirota equation}

In the inverse problem of VC-Hirota equation, we mainly consider two cases: first,
data-driven parameters discovery of VC-Hirota equation by using IPINN system when the variable coefficient $\alpha_{1}(z)$, $\alpha_{2}(z)$ and $\alpha_{3}(z)$ of VC-Hirota equation are known; Second, the data-driven variable coefficient equation discovery, namely, data-driven function discovery. 

\subsection{Data-driven parameters discovery of the VC-Hirota equation}
The inverse problems of the VC-Hirota equation with unknown parameters $\delta$ is considered by using the IPINN model in this part. First, initialize the unknown parameters $\delta$ to $\delta_{0}=1$, then we can get the predicted parameter $\hat{\delta}$ from the IPINN algorithm. Define the relative error of unknown parameters as
\begin{align}\label{E36}
RE=\frac{|\hat{\delta}-\delta|}{\delta}\times100\%,
\end{align}
where the $\hat{\delta}$ and $\delta$ represent predicted value and true value, respectively. All noise interference in this section is added to the randomly chosen small data set, the specific form is as below:
\begin{align}\notag
Data_{-}train_{1}= Data_{-}train + noise*np.std(Data_{-}train)*np.random.randn(Data_{-}train.shape[0], Data_{-}train.shape[1]),
\end{align}
where $noise$ and $Data_{-}train$ indicate the noise intensity and a small randomly chosen training data set, respectively. The $np.std(\cdot)$ returns the standard deviation of an array element, and $np.random.randn(\cdot,\cdot)$ returns a set of samples with a standard normal distribution.

In order to accurately and effectively learn unknown parameters, we try to use $L^2$ norm parameter regularization into IPINN to study the parameter discovery problem of VC-Hirota equation. Now we construct a new loss function $\widetilde{\mathscr{L}}(\bar{\Theta})$ with $L^2$ norm weight decay as follows \cite{pu-manokov}
\begin{align}\label{E40}
&\widetilde{\mathscr{L}}(\bar{\Theta})=Loss_{PR}=\mathscr{L}(\bar{\Theta})+\frac{\alpha}{2}\|\textbf{W}\|^2_2,
\end{align}
where $\mathscr{L}(\bar{\Theta})$ and $\textbf{W}$ have been defined in Eq. \eqref{E6} and Eq. \eqref{E-buchong}.

For learning the parameters $\delta$ in Eq. \eqref{E-pi} with the aid of one-solitons solution \eqref{q11}, we reset the initial conditions and Dirichlet boundary conditions as follows:
\begin{align}\label{E38}
\begin{split}
&q^0(t)=q_{\mathrm{11}}(t,-0.5),\quad t\in[-3.0,1.0],
\end{split}
\end{align}
\begin{align}\label{E39}
q^{\mathrm{lb}}(t)=q_{\mathrm{11}}(-3.0,z),\quad q^{\mathrm{ub}}(t)=q_{\mathrm{11}}(1.0,z),\quad z\in[-0.5,0.5].
\end{align}

The original data set can be obtained by dividing the spatial region $[-3.0,1.0]$ into 1000 points and temporal region $[-0.5,0.5]$ into 1000 points. Then a smaller training dataset that containing initial-boundary data \eqref{E38} and \eqref{E39} by randomly extracting $N_q=1700$ from original dataset and $N_f=20000$ collocation points which is generated by the LHS method.

\begin{figure}[htbp]
\centering
\subfigure[]{
\begin{minipage}[t]{0.48\textwidth}
\centering
\includegraphics[height=6cm,width=7.5cm]{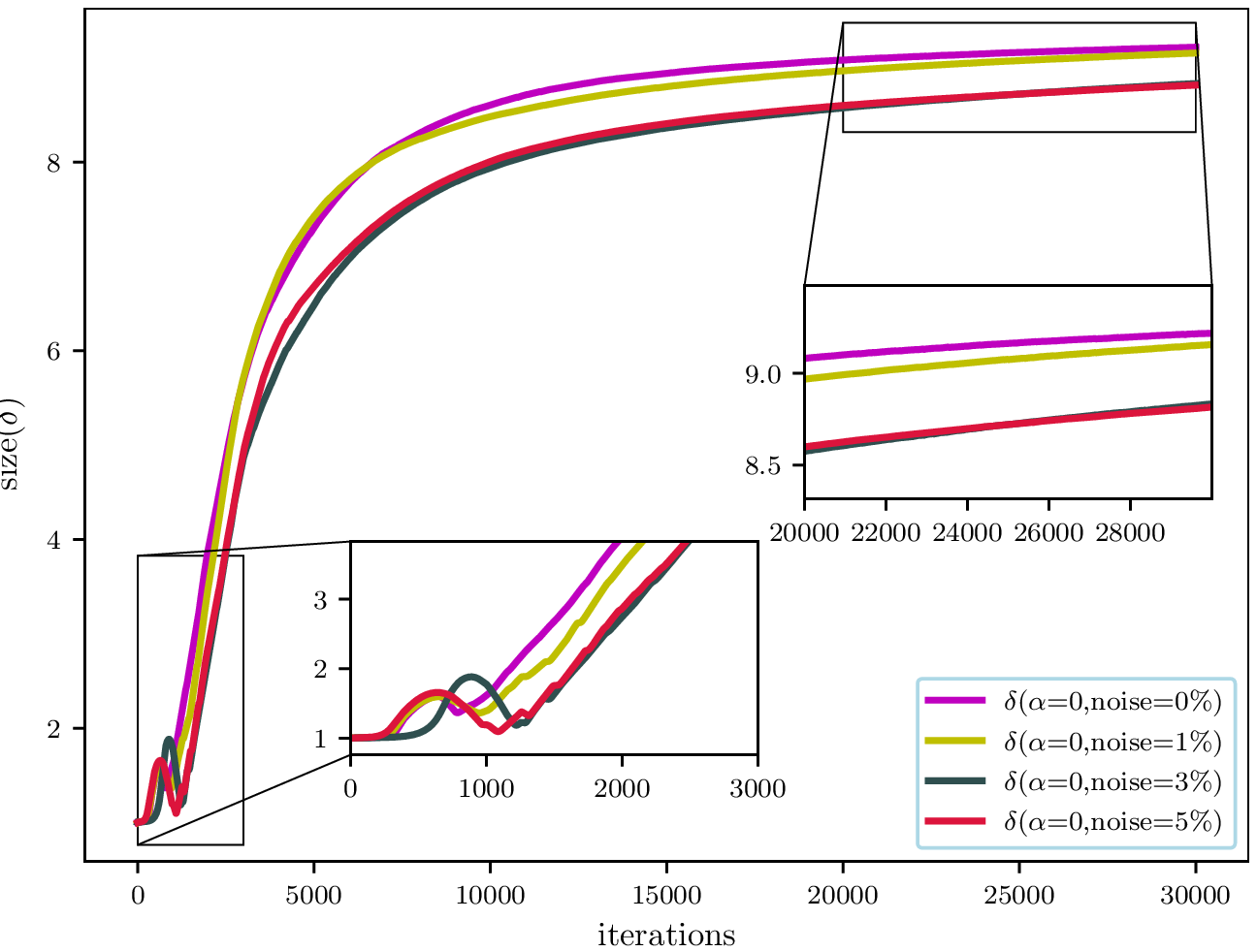}
\end{minipage}
}%
\subfigure[]{
\begin{minipage}[t]{0.48\textwidth}
\centering
\includegraphics[height=6cm,width=7.5cm]{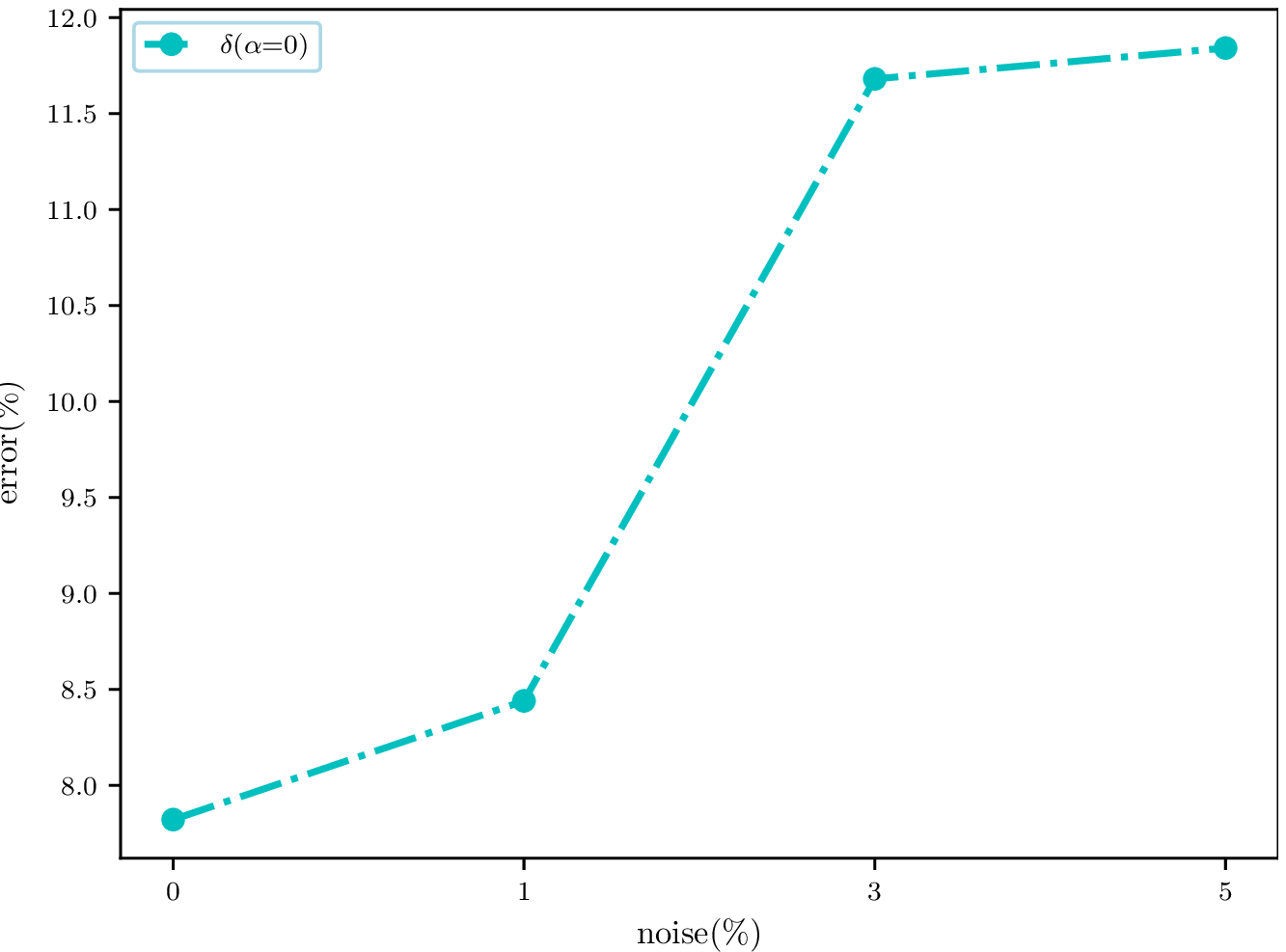}
\end{minipage}
}%
\centering
\caption{(Color online) Training results of parameter discovered by means of the IPINN with 30000 Adam iterations: (a) the variation curves of unknown coefficients $\delta$ with different noise intensity;  (b) unknown coefficients $\delta$ error variation plots under different interference noise.}
\label{F26}
\end{figure}

When we set $\alpha=0$, namely, recover the unknown parameter $\delta$  under the IPINN without $L^2$ norm parameter regularization. We find that the parameter $\delta$ error learned under the IPINN without $L^2$ norm parameter regularization is not very satisfactory by tuning various parameters. As an example here we only show a relatively good result in the following. The training results of unknown parameters $\delta$ under the above initial boundary value conditions by using the IPINN with 30000 Adam iterations are shown in Fig. \ref{F26}. Fig. \ref{F26}(a) exhibit the variation curves of unknown coefficients $\delta$ with different noise intensity. When we use the clean initial-boundary data (noise=$0\%$) in IPINN, the relative error of $\delta$ is about $7.821617\%$. With the increase of noise, the relative error of unknown parameter values also increases. When the noise is 3$\%$, the relative error reaches $11.841373\%$. The error variation plots of unknown coefficients $\delta$  under different interference noise are revealed in Fig. \ref{F26} (b). 

\begin{figure}[htbp]
\centering
\subfigure[]{
\begin{minipage}[t]{0.48\textwidth}
\centering
\includegraphics[height=6cm,width=7.5cm]{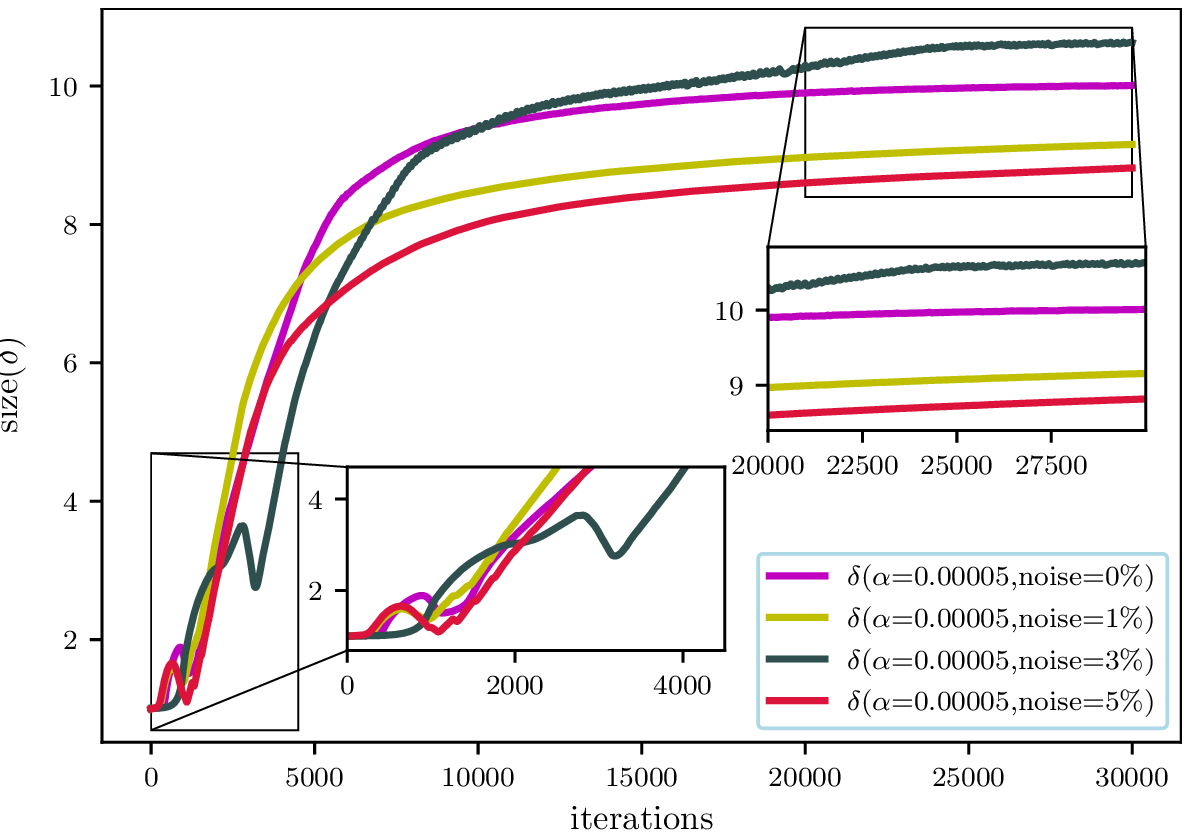}
\end{minipage}
}%
\subfigure[]{
\begin{minipage}[t]{0.48\textwidth}
\centering
\includegraphics[height=6cm,width=7.5cm]{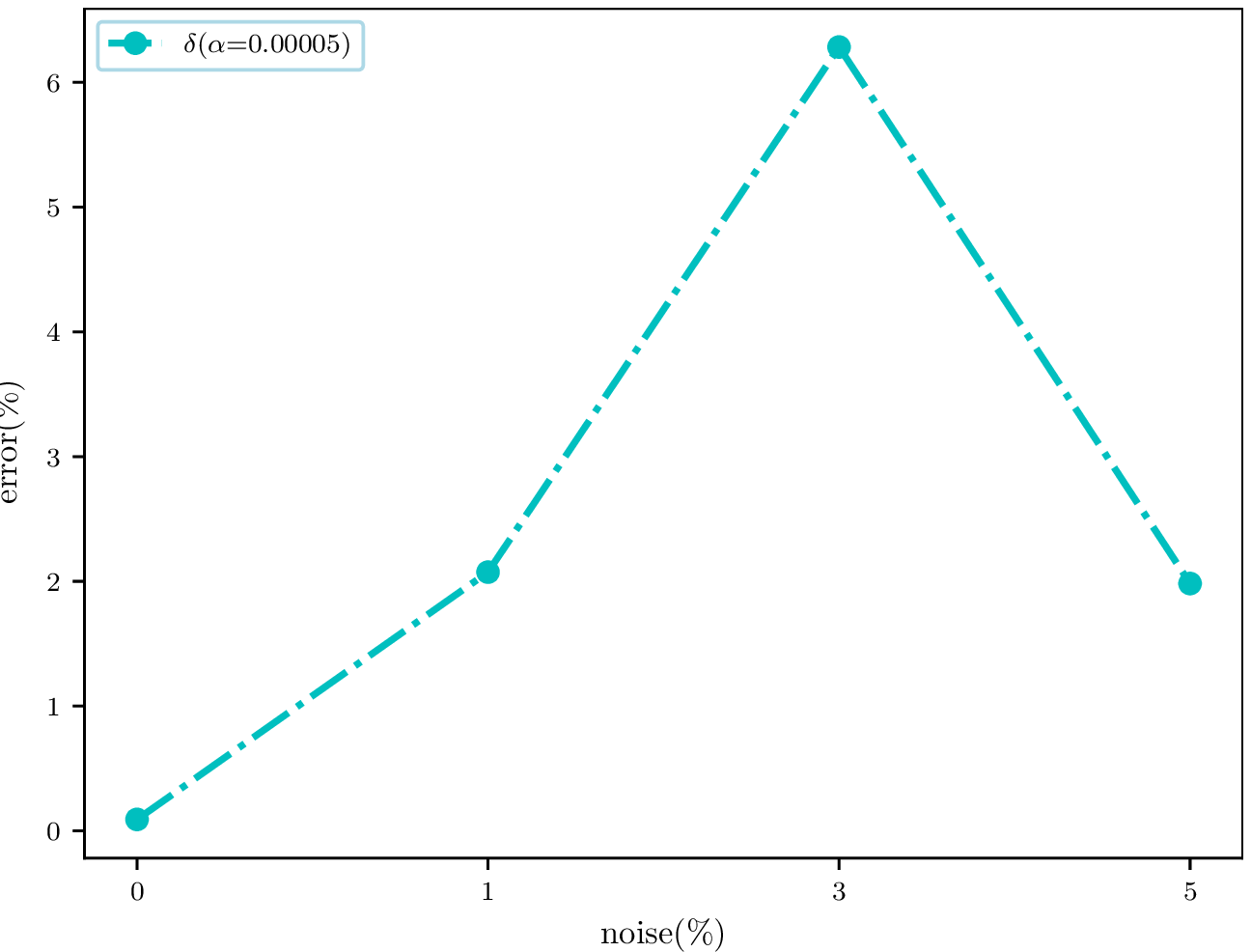}
\end{minipage}
}%
\centering
\caption{(Color online) Training results of parameter discovered by means of the IPINN with $\alpha=0.00005$ weight decay: (a) the variation curves of unknown coefficients $\delta$  with different noise intensity; (b) unknown coefficients $\delta$ error variation plots under different interference noise.}
\label{F27}
\end{figure}

Recall the smaller training dataset that containing initial-boundary data \eqref{E38} and \eqref{E39} by randomly extracting $N_q=1700$ from the original dataset and $N_f=20000$ collocation points which generated by the LHS method. We successfully used IPINN method with parameter regularization to learn the value of the unknown parameter $\alpha$ with a small relative error by adjusting the iteration number of Adam, increasing or decreasing the weight $\alpha$ and other tuning methods. 
When we set the 30000 times  Adam iterations, Fig. \ref{F27} (a) displays the variation curves of unknown coefficients $\alpha$ with different noise intensity by using the IPINN with $\alpha=0.00005$ weight decay. 
The noise intensity and relative error plots are shown in Fig. \ref{F27} (b). The specific numerical results show that $\alpha$ obtains the minimum relative error of 0.091810$\%$ as the noise intensity is $0\%$, and maximum relative error 6.2832007$\%$ as the noise intensity is 3$\%$. Compared with the numerical results obtained by IPINN without parameter regularization in Fig. \ref{F26}, the relative error of parameters learned by IPINN with $L^2$ norm parameter regularization are significantly reduced in both clean and noisy data.
In particular, the relative error of the unknown parameter reaches 0.091810$\%$ when the noise is 0$\%$, which is a great training result of the data-driven parameters discovery. 

\begin{figure}[htbp]
\centering
\subfigure[]{
\begin{minipage}[t]{0.48\textwidth}
\centering
\includegraphics[height=6cm,width=7.5cm]{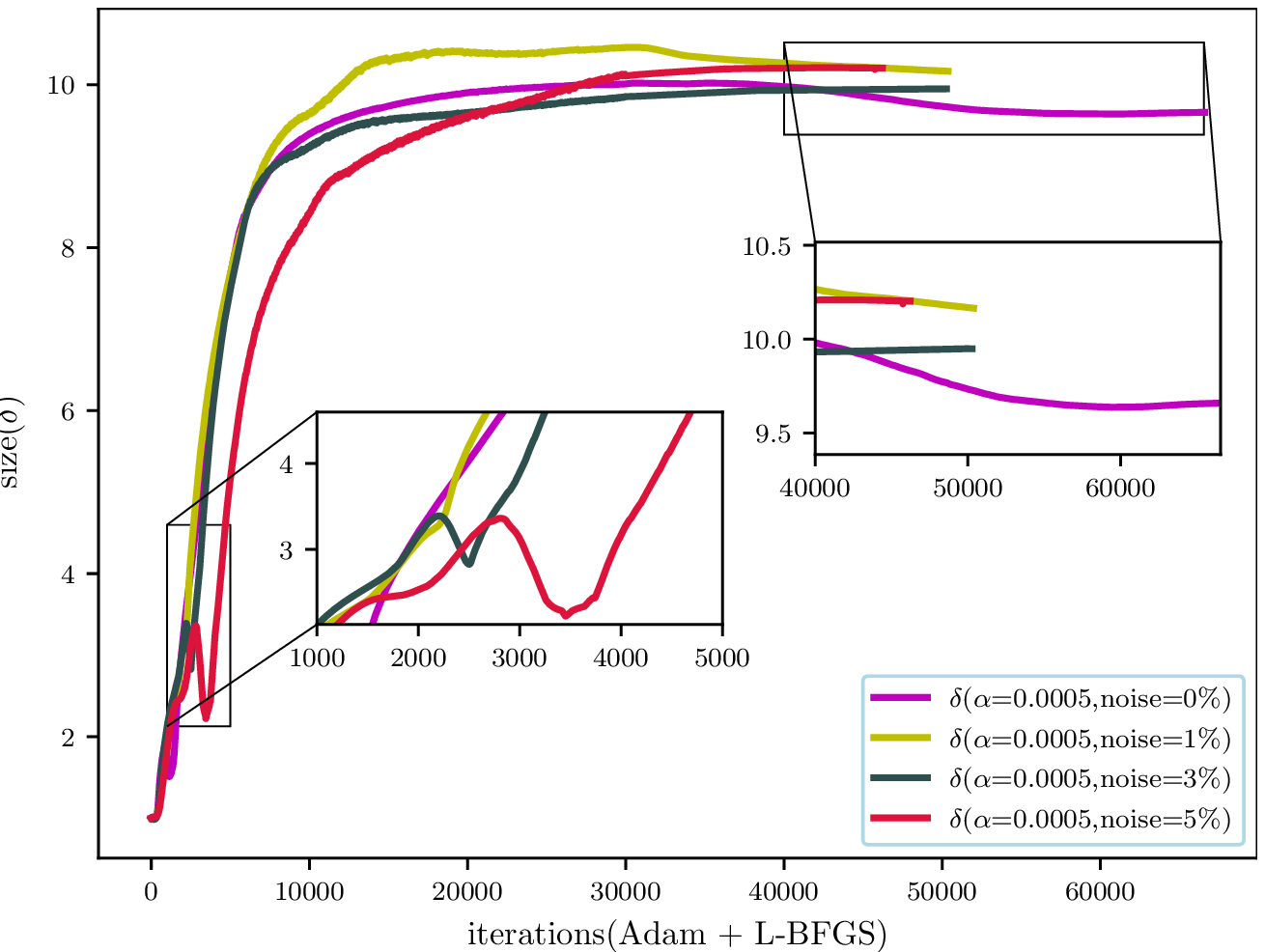}
\end{minipage}
}%
\subfigure[]{
\begin{minipage}[t]{0.48\textwidth}
\centering
\includegraphics[height=6cm,width=7.5cm]{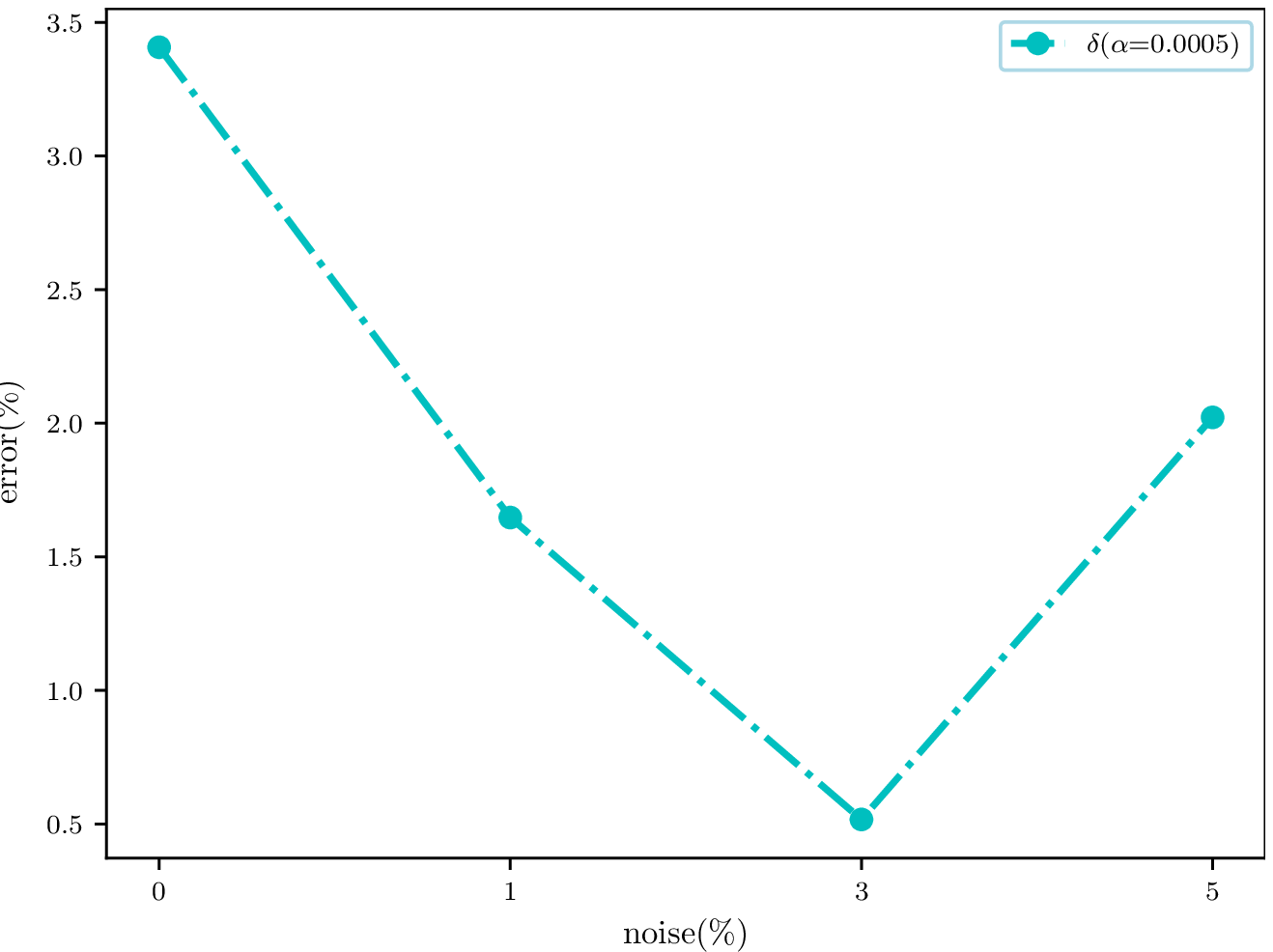}
\end{minipage}
}%
\centering
\caption{(Color online) Training results of parameter discovered by means of the IPINN with $\alpha=0.00005$ weight decay: (a) the variation curves of unknown coefficients $\delta$  with different noise intensity; (b) unknown coefficients $\delta$ error variation plots under different interference noise.}
\label{lbfgs}
\end{figure}

In other words, the training effect has been very good in the case of clean data. However, when using the data with noise interference for training, the relative error of the prediction parameters is larger than that of the training results of clean data. Therefore, we need to reset the weight decay coefficient to improve the training effect of using the data with noise interference.  When we put the L-BFGS iterations into the above method, one can find that the training effect with data of various noise is much better than that with clean data. Compare with the Adam iteration without L-BFGS iterations, we find that the parameter discovery result is worse in the pure data, but significantly better in the 3$\%$ noise, the generalization training shows that IPINN with coefficient $\alpha=0.00005$ of weight decay has excellent noise resistance. The detail can be seen in the Fig.\ref{lbfgs}.

 At the same time, we also consider the influence of  weight $\alpha$ and the number of Adam iterations on the parameter discovery results, and the specific results can be seen in the following Fig. \ref{F28}. From the above figures, we can find that the training effect is better once setting the appropriate weight coefficients, noise intensity data and other associated parameters.
\begin{figure}[htbp]
\centering
\subfigure[]{
\begin{minipage}[t]{0.48\textwidth}
\centering
\includegraphics[height=6cm,width=7.5cm]{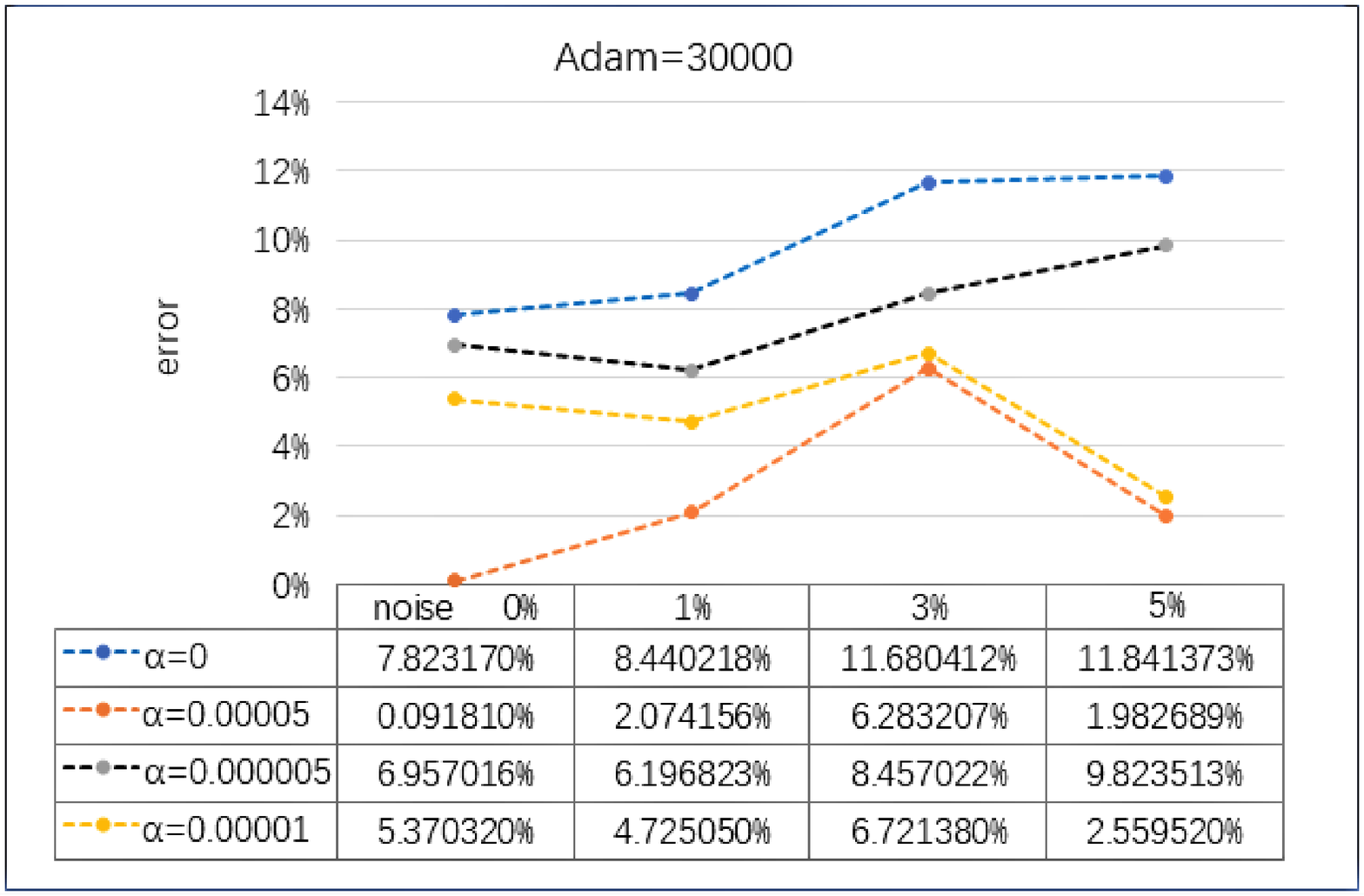}
\end{minipage}
}%
\centering
\subfigure[]{
\begin{minipage}[t]{0.48\textwidth}
\centering
\includegraphics[height=6cm,width=7.5cm]{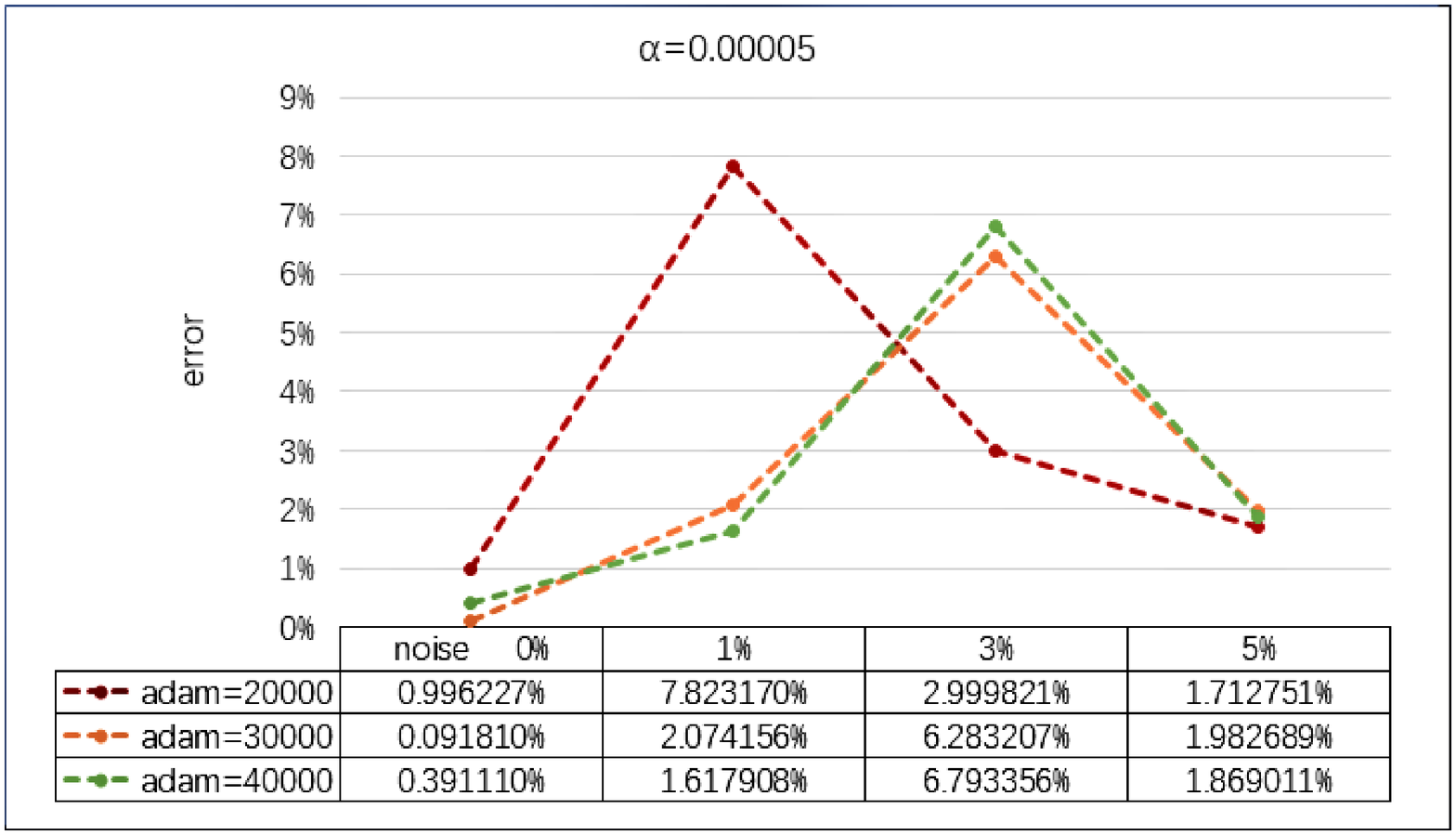}
\end{minipage}
}%
\\
\centering
\subfigure[]{
\begin{minipage}[t]{0.48\textwidth}
\centering
\includegraphics[height=6cm,width=7.5cm]{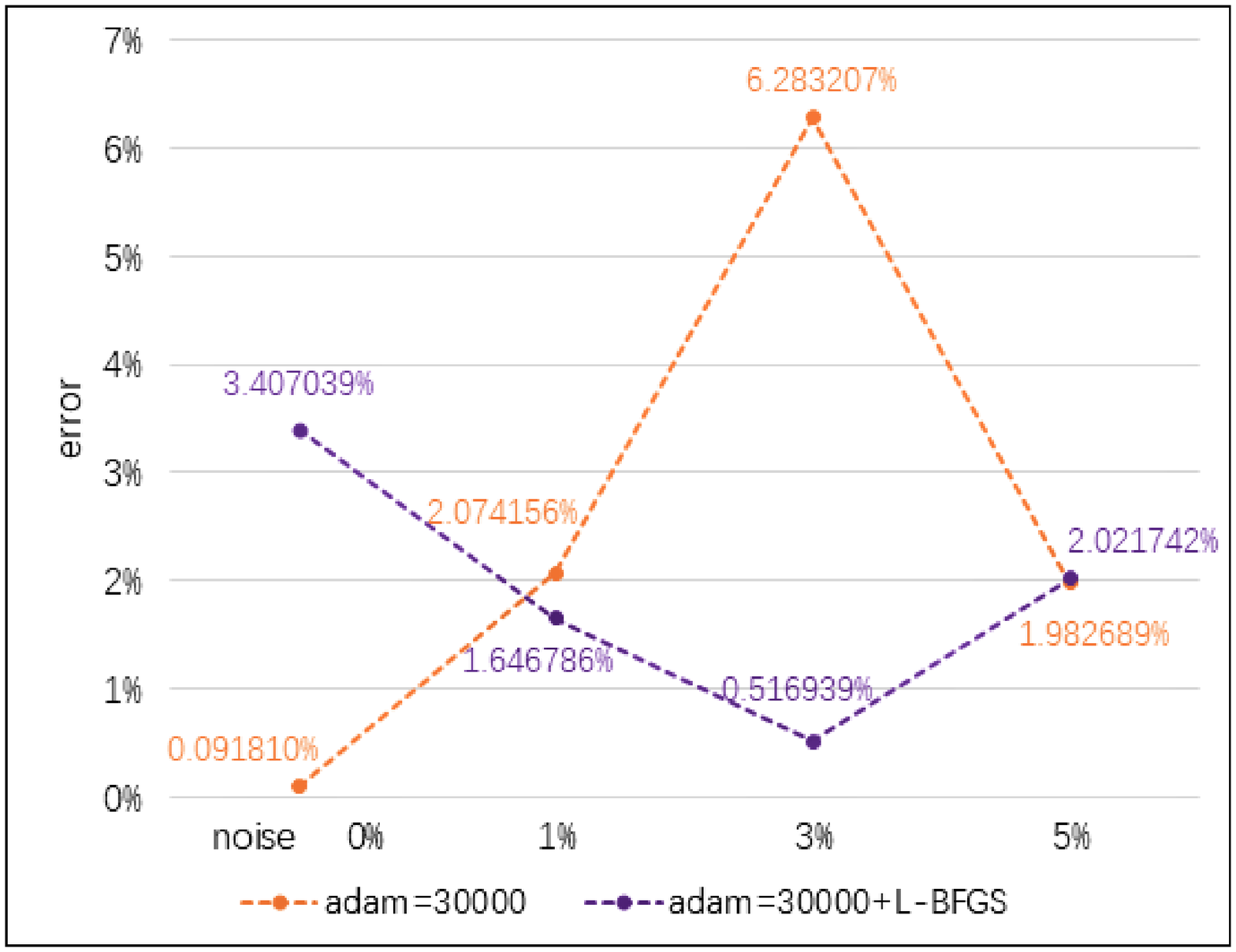}
\end{minipage}
}%
\caption{(Color online) Training results of parameter discovered by means of the IPINN with different parameters: (a) with a fixed number of 30000 Adam iterations, unknown coefficients $\delta$ error variation plots under different weight decay; 
(b) with a fixed weight decay $\alpha=0.00005$, the variation curves of unknown coefficients $\delta$  with different iteration numbers of Adam; (c) the unknown coefficients $\delta$ error variation plots under different interference noise after using only 30000 Adam iterations and adding L-BFGS algorithm.}
\label{F28}
\end{figure}

\subsection{Data-driven function discovery of the VC-Hirota equation}

\begin{figure}[htbp]
\centering
\begin{minipage}[t]{0.99\textwidth}
\centering
\includegraphics[height=9cm,width=15cm]{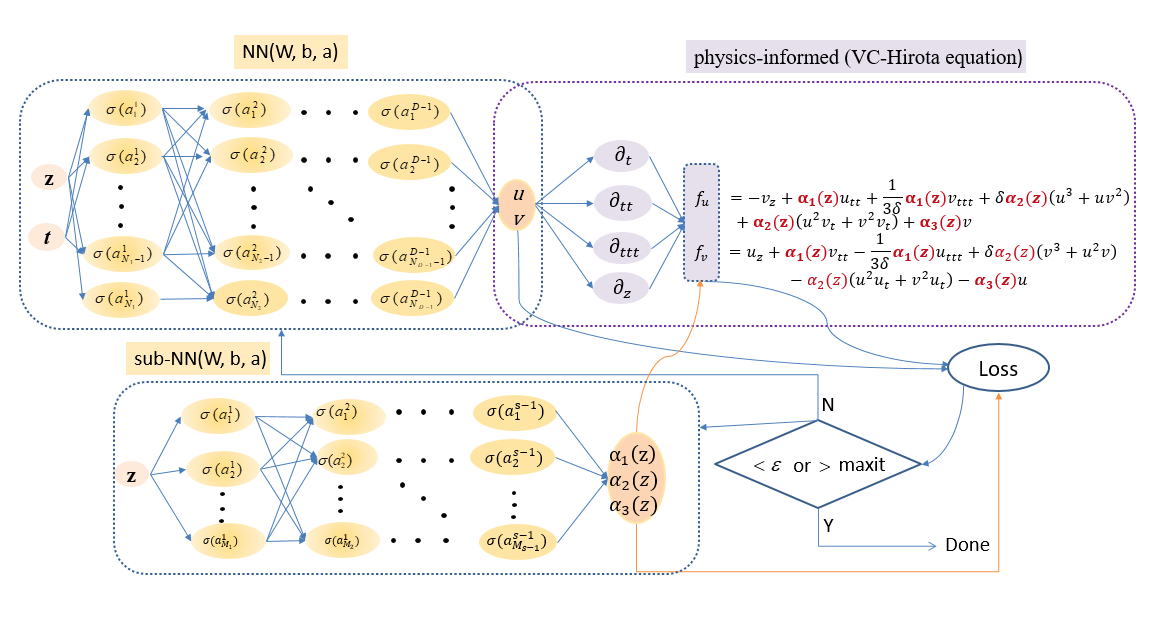}
\end{minipage}
\centering
\caption{(Color online) Schematic of the PINN algorithm with sub-neural networks for study the unknown function of VC-Hirota equation.}
\label{double-NN}
\end{figure}

In this subsection, we will use a PINN algorithm with sub-neural networks to study the unknown function for the VC-Hirota equation. The flow chart of the PINN algorithm with sub-neural networks is given in Fig. \ref{double-NN} to describe it more intuitively.

Compared with the flow chart of the algorithm \ref{F1}, in the NNs part, we not only use a main  neural network to train the data-driven solutions $u$ and $v$, but also add a sub-neural network to train the unknown functions $\alpha_{1}(z)$, $\alpha_{2}(z)$ and $\alpha_{3}(z)$ in the equation \eqref{VC-Hirotaeq}. At this point, $Loss_{\alpha_{1}(z)}$, $Loss_{\alpha_{2}(z)}$ and $Loss_{\alpha_{3}(z)}$ are added to the loss function$\mathscr{L}(\bar{\Theta})$ where $Loss_{\alpha_{1}(z)}$, $Loss_{\alpha_{2}(z)}$ and $Loss_{\alpha_{3}(z)}$ are defined as below:

\begin{align}\label{E71}
Loss_{\alpha_{i}(z)}=\big|\hat{\alpha_{i}(T_{0})}-\alpha_{i}(T_{0})\big|^{2},
\end{align}
where the $\hat{\alpha_{i}(T_{0})}$ and $\alpha_{i}(T_{0})$ (i=1, 2 and 3) represent predicted value and true value, respectively.

Without loss of generality, we can consider a simple case: $\alpha_{1}(z)=\alpha_{2}(z)=\Gamma(z)$ and $\delta=10$ in the VC-Hirota equation \eqref{VC-Hirotaeq}, then  $\alpha_{3}=0$ and the physics-informed parts of the PINN algorithm with sub-neural networks for VC-Hirota equation \eqref{VC-Hirotaeq} can be rewritten as:
\begin{align}\label{E-pi1}
\begin{split}
&f_u:=-v_z+\Gamma(z)u_{tt}+\frac{1}{30}\Gamma(z)v_{ttt}+10\Gamma(z)(u^{3}+uv^{2})+\Gamma(z)(z)(u^{2}v_{t}+v^{2}v_{t}),\\
&f_v:=u_z+\Gamma(z)v_{tt}-\frac{1}{30}\Gamma(z)u_{ttt}+10\Gamma(z)(vu^{2}+v^{3})-\Gamma(z)(u^{2}u_{t}+v^{2}u_{t}).
\end{split}
\end{align}

Now $Loss_{\Gamma(z)}=Loss_{\alpha_{1}(z)}=Loss_{\alpha_{2}(z)}=\big|\hat{\Gamma(T_{0})}-\Gamma(T_{0})\big|^{2}$.

In this section, we fixed $a_i^d=1$, which has the advantage of ensuring good training results while training fewer parameters. Without loss of generality, we can take $N_a=0$, Now the loss function which used in the PINNs is shown as follows:
\begin{align}\label{E6}
\mathscr{L}(\bar{\Theta})=Loss=Loss_{u}+Loss_{v}+Loss_{f_u}+Loss_{f_v}+Loss_{\Gamma(z)},
\end{align}
and rewritten the $Loss_{u}$ and $Loss_{v}$ as  
\begin{align}\label{E7}
\begin{split}
Loss_{u}&=\frac{1}{N_{q_{in}}}\sum^{N_{q_{in}}}_{j=1}\big|\hat{u}(t^j,z^j)-u^j\big|^{2},
\\
Loss_{v}&=\frac{1}{N_{q_{in}}}\sum^{N_{q_{in}}}_{j=1}\big|\hat{v}(t^j,z^j)-v^j\big|^{2},
\end{split}
\end{align}
 where $N_{q_{in}}$ represents the internal grid point from original dataset.
 
For learning the unknown function $\Gamma(z)$, $z \in [z_0,z_1]=[-1.0,1.0]$ in Eq. \eqref{E-pi} with the aid of one-solitons solution \eqref{q11}, we reset the initial conditions  as follows:

\begin{align}\label{E401}
\Gamma(z_0)=c,
\end{align}
where $c=1$. The temporal region $[-1.0,1.0]$ of function $\Gamma(z)$ is divided into 500 points. In the main neural network, the original data set  can be obtained by dividing the spatial region $[-3.0,1.0]$ into 512 points and temporal region $[-1.0,1.0]$ into 200 points. Then a smaller training dataset is obtained by randomly extracting $N_{q_{in}}=2000$ and $N_f=40000$ which is generated by the random sampling method. 

Define $\mathbb{L}_2$ norm error of unknown function as:
\begin{align}\label{E361}
\mathrm{Error}=\frac{\sqrt{\sum\limits_{k=1}^{N}\big|\hat{\Gamma(z_{k})}-\Gamma(z_{k})\big|^{2}}}{\sqrt{\sum\limits_{k=1}^{N}\big|\Gamma(z_{k})\big|^{2}}},
\end{align}
where the $\hat{\Gamma(z_{k})}$ and $\Gamma(z_{k})$ represent predicted value and true value,respectively. $z_{k}$ denotes the equidistant point after dividing the interval 
$[-1,1]$ into 500 equal parts.

\begin{figure}[htbp]
\centering
\subfigure[]{
\begin{minipage}[t]{0.48\textwidth}\label{gammat1}
\centering
\includegraphics[height=6cm,width=7.5cm]{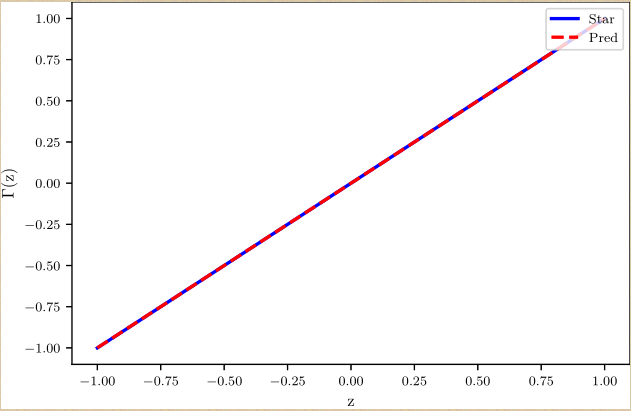}
\end{minipage}
}
\subfigure[]{
\begin{minipage}[t]{0.48\textwidth}\label{gammat2}
\centering
\includegraphics[height=6cm,width=7.5cm]{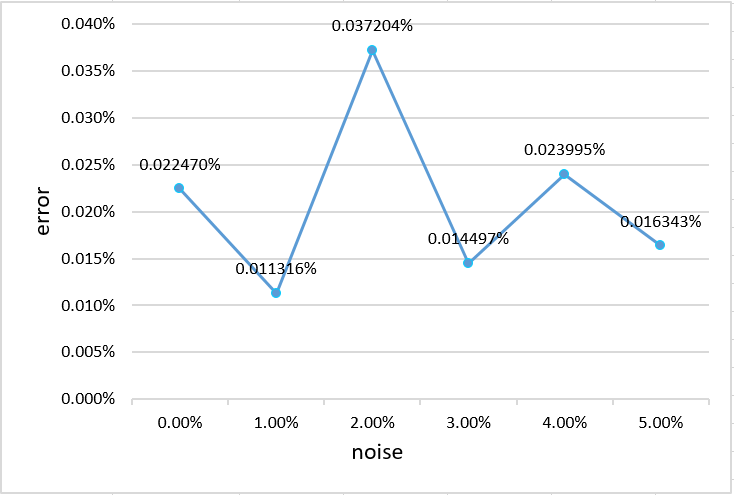}
\end{minipage}
}
\centering
\caption{(Color online) Training results of function discovery by means of the PINN algorithm with sub-neural networks: (a) the plot of predicted and true  function $\Gamma(z)$;  (b) error variation plots of predicted function $\Gamma(z)$ under different interference noise.}
\label{gammat}
\end{figure}

The training results of unknown function $\Gamma(z)$ under the above conditions by using the PINN algorithm with sub-neural networks with 5000 times Adam iterations and a number of L-BFGS iterations are shown in Fig. \ref{gammat}. Since unknown function $\Gamma(z)$ can obtain very good training results under both pure data and noisy conditions, the images of predictive solutions under different
noises are visually indistinguishable.  Therefore, in Fig.\ref{gammat1}, we only present the images of predictive function solution and real function solution $\Gamma(z)$ under pure data. When we use the clean data (noise$=0\%$) in the PINN algorithm with sub-neural networks, the $\mathbb{L}_2$ norm error of 
 $\Gamma(z)$ is about $1.124902e^{-4}$. With the addition of different noise, we find that 
  the PINN algorithm with sub-neural networks has excellent anti-noise ability. For example, the $\mathbb{L}_2$ norm error 
 of $\Gamma(z)$ is about $1.449694e^{-4}$ when we add $3\%$ noise into the data. Error 
variation plots of unknown function $\Gamma(z)$ under different interference noise is revealed in Fig.\ref{gammat2}. In error variation plots Fig.\ref{gammat2} we convert scientific counting to percentages.

\section{Conclusions and discussions}

The conclusion drawn from the numerical evidence presented in this paper is the IPINN framework with neuron-wise locally adaptive activation function and slope recovery term is an efficient method for solving the data-driven solutions including the one-solitons, two-solitons and second-order soliton of the VC-Hirota equation with a small sample data set. Meanwhile, we also study the parameter discovery problem of VC-Hirota equation through the IPINN method and found that the PINN with neuron-wise locally adaptive activation function and $L^2$ norm parameter regularization shows amazing effect in studying the inverse problem of VC-Hirota equation. Furthermore, we learn the data-driven unknown function discovery stably and efficiently in the variable coefficient equation via the PINN algorithm with sub-neural networks, and find that the network has superior anti-noise ability. The data-driven forward and inverse problems of variable coefficients equation are given for the first time by using the PINN methods mentioned above in this work, which provides an essential theoretical basis and experimental experience for variable coefficients model to conduct more professional research.

Due to the soliton solutions of the variable coefficient equation present  more complex shapes including parabola type, "S-shape" and cosine wave type etc., which brings more challenges in the training process. The IPINN method displays  well training effect and convergence speed in training the one-soliton solution of VC-Hirota equation. Even though the two-soliton solution of variable coefficients equation, the dynamic behavior shown in this paper is more complex than that of the rogue wave solution of the traditional constant coefficient equation, we also obtain the data driven two-solitons of the VC-Hirota successful  through a large number of training experiments. The experiments result that in addition to the number of network layers and neurons, other influencing factors, such as randomly extracting from original dataset and collocation points may have a great impact on the learning results. During the parameter discovery, we find that the training error is not ideal either using clean data or noisy data based on earlier IPINN method. However, the training results improve greatly under different parameters when the parameter regularization strategy is added into the IPINN framework. In the progress of function discovery, we can obtain the data-driven function solution with relatively small error whether using clean data or data with different levels of noise by means of the PINN algorithm with sub-neural networks. 
 
The IPINN algorithm has good effect when describing relatively smooth solutions in a small range, but as the solution range becomes larger or the dynamic behavior of the solution is steeper, the learning effect becomes worse. In future work, we will focus on overcoming this problem by optimization and improvement of the PINN algorithms. Since the variable coefficient model used in this paper is relatively complex, in the process of data-driven function discovery in this paper, we only give the example of linear function. In future work, we will also consider how to construct an appropriate PINN network to solve the data-driven problem of nonlinear unknown function.


\end{document}